\documentclass[journal=jacsat,manuscript=article]{achemso}

\usepackage{amsmath,amssymb,graphicx,epstopdf,bm,bbm,hyperref}

\usepackage[version=3]{mhchem} 

\author{Michael S. Dimitriyev}
\affiliation{Department of Polymer Science and Engineering, University of Massachusetts, Amherst, MA 01003}
\author{Abhiram Reddy}
\affiliation{Department of Polymer Science and Engineering, University of Massachusetts, Amherst, MA 01003}
\author{Gregory M. Grason}
\affiliation{Department of Polymer Science and Engineering, University of Massachusetts, Amherst, MA 01003}
\email{grason@umass.edu}

\title[Medial packing, frustration and competing network phases]
  {Medial packing, frustration and competing network phases in strongly-segregated block copolymers}

\usepackage{xcolor}
\newcommand{\note}[1]{#1}

\begin{document}

\begin{abstract}
Self-consistent field theory (SCFT) has established that for cubic network phases in diblock copolymer melts, the double-gyroid (DG) is thermodynamically stable relative to the competitor double-diamond (DD) and double-primitive (DP) phases, and exhibits a window of stability intermediate to the classical lamellar and columnar phases. This competition is widely thought to be controlled by ``packing frustration'' -- the incompatibility of uniformly filling melts with a locally preferred chain packing motif. Here, we reassess the thermodynamics of cubic network formation in strongly-segregated diblock melts, based on a recently developed \emph{medial strong segregation theory} (``mSST'') approach that directly connects the shape and thermodynamics of chain packing environments to the \emph{medial geometry} of tubular network surfaces. We first show that medial packing significantly relaxes prior SST upper bounds on the free energy of network phases, which we attribute to the spreading of terminal chain ends within network nodal regions.  Exploring geometric and thermodynamic metrics of chain packing in network phases, we show that mSST reproduces effects dependent on the elastic asymmetry of the blocks that are consistent with SCFT at large $\chi N$.
We then characterize geometric frustration in terms of the spatially-variant distributions of local entropic and enthalpic costs throughout the morphologies, extracted from mSST predictions.
Analyzing these distributions, we find that the DG morphology, due to its unique medial geometry in the nodal regions, is stabilized by the incorporation of favorable, quasi-lamellar packing over much of its morphology, motifs which are inaccessible to DD and DP morphologies due to ``interior corners'' in their medial geometries.  
Finally, we use our results to analyze ``hot spots'' of chain stretching and discuss implications for network susceptibility to the uptake of guest molecules.
\end{abstract}

\section{Introduction}

Amphiphilic molecules, from lyotropic liquid crystals to macromolecular block copolymer (BCP) analogs, are known to assemble into a wide range of morphologies upon microphase separation, from the classical lamellar, columnar, and spherical phases, to a variety of intercatenated, triply-periodic network phases \cite{Isrealachvili2011,Hyde1997_ch4}.  
While the propensity of such molecular building blocks to assemble into the classical phases can be largely understood from simple packing arguments relating interfacial curvature to asymmetry of molecular architecture, the stability of network phases over the classical phases for certain molecular architectures is difficult to rationalize using such arguments. 
Such networks are ``in-between'' phases that typically form in conditions where layers and columns are in close competition, exhibiting interfaces that are almost cylindrical in some regions, almost flat in others.  
Initial heuristic pictures of bicontinuous phase structure formation focused on the likely role of area optimization at the intermaterial dividing surface (IMDS) and the geometric connection to triply-periodic, area-minimizing surfaces \cite{Scriven1976,Thomas1988}.  
By far, the most commonly considered network phases are those related to the associated family of cubic minimal surfaces: Primitive (P), Diamond (D), and Gyroid (G), which are infinite surfaces of every zero mean curvature and negative Gaussian curvature that divide space into two inter-catenated, equal-volume regions\cite{Schoen2012}.

\begin{figure}[h!]
\centering
\includegraphics[width=3.5in]{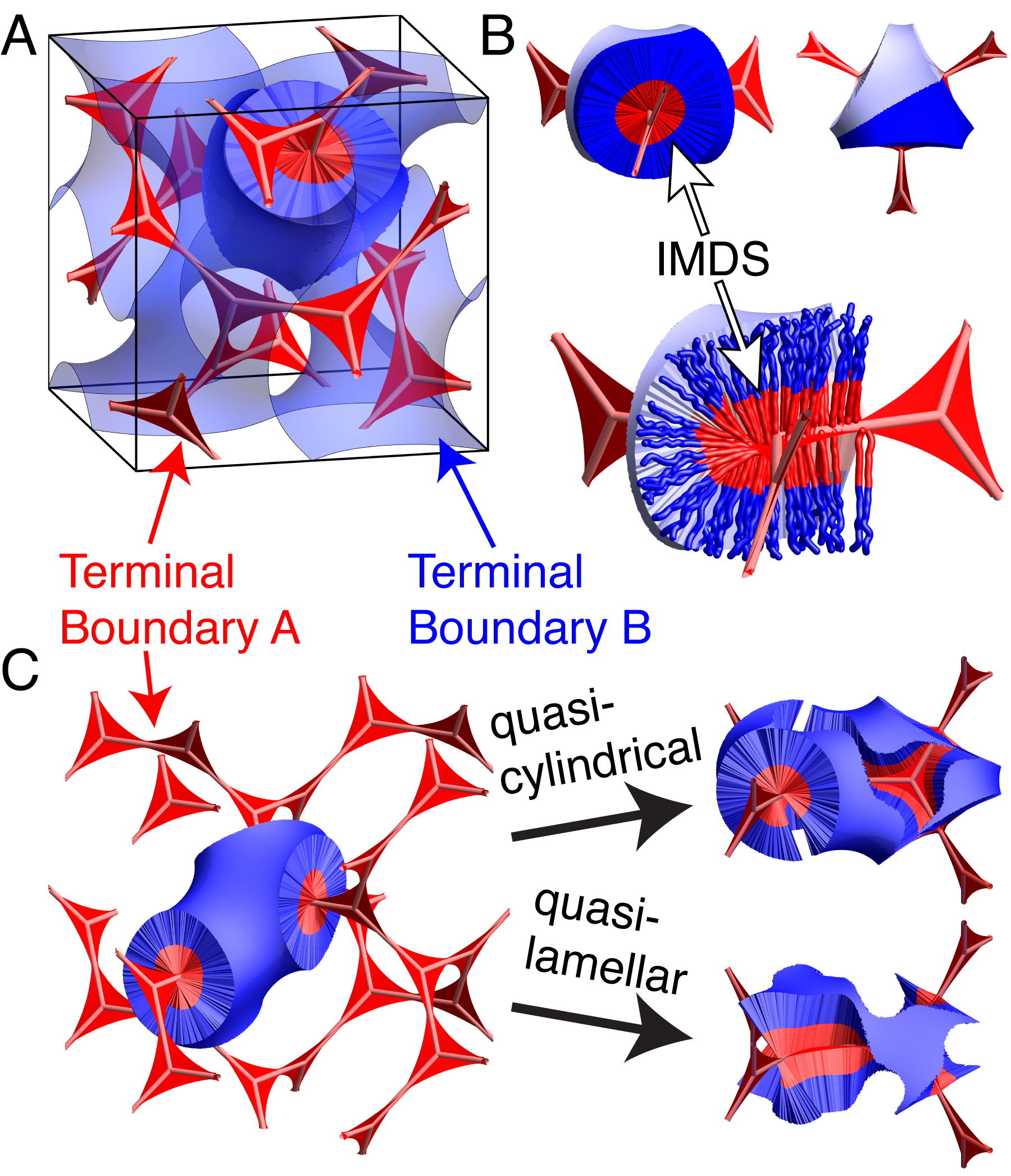}
\caption{\label{fig:dg_overview} Overview of the (A) DG unit cell structure and (B) subdomain structure within a DG mesoatom (here, $\epsilon = 1.0$ and $f \approx 0.29$). In (A), the terminal surface of the A-block domain is modeled by a web-like medial surface whereas the terminal surface of the B-block domain is shown as a gyroid minimal surface. The highlighted mesoatom unit in (B) is cut open to show the wedge packing motifs in greater detail and a selection of representative chain conformations is shown. (C) shows a pair of DG mesoatoms exhibiting hybrid packing motifs, separately highlighting wedges in regions of cylinder-like packing and lamellar-like packing, distinguished by the degree of alignment of individual packing environments with the local orientation of the tubular medial surface.}
\end{figure}

\begin{figure}[h!]
\centering
\includegraphics[width=\textwidth]{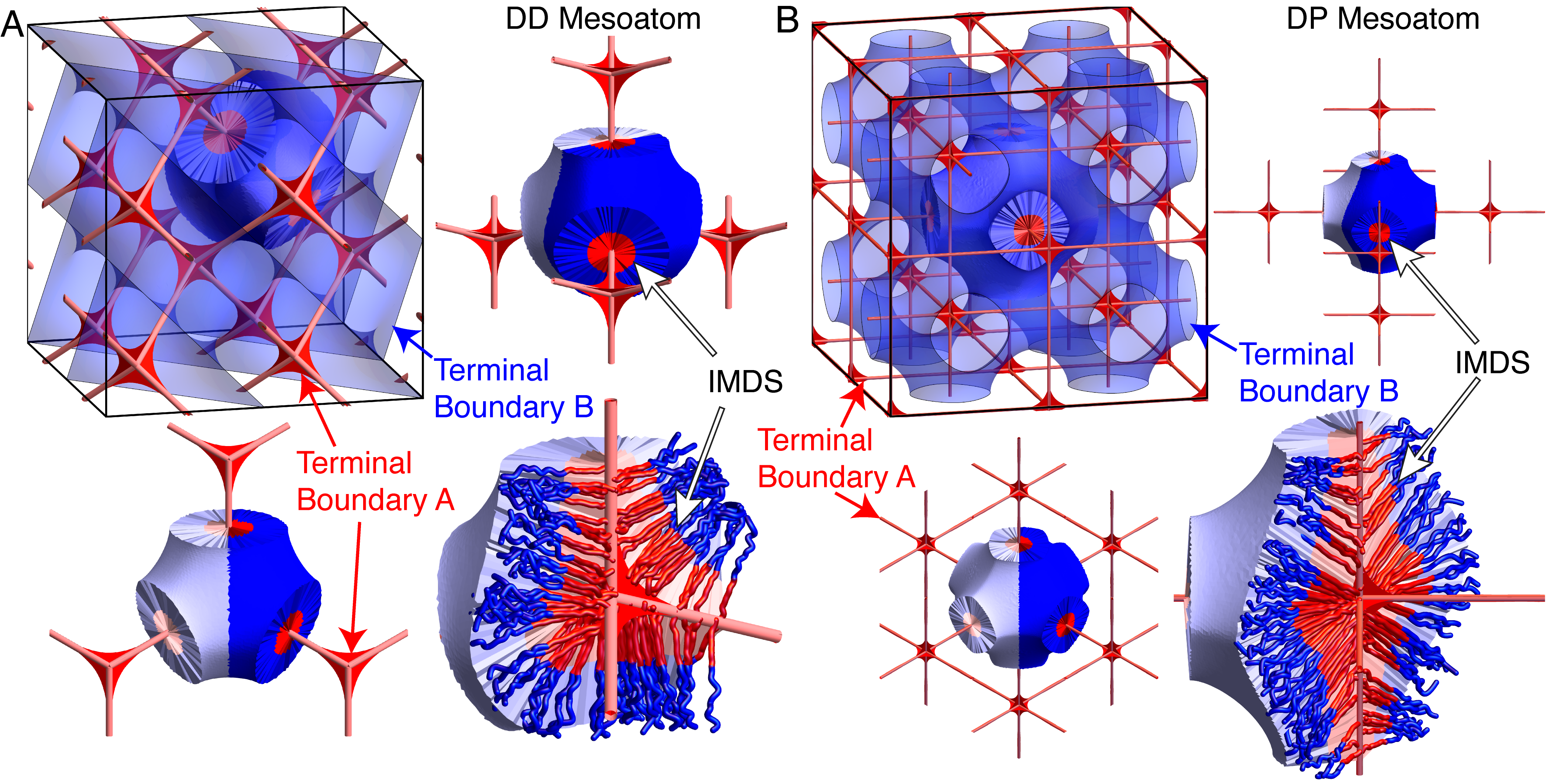}
\caption{\label{fig:network_geometry} Depictions of the (A) DD and (B) DP structures computed using medial strong segregation theory (mSST) at $\epsilon = 1$ and $f = 0.29$. For both networks, a unit cell as well as single mesoatoms (in two different perspectives) are shown. Sliced versions of the unit cell show the internal structure about the medial set approximation of the A-block terminal surface (red fins), with a representative selection of polymers.}
\end{figure}

Unlike in lyotropic or solvated systems, understanding network formation in BCP melts presents a particular challenge due to the requirement of molecules to fill all of space, with chains stretching to fill to the center of the network domains, as well as the entire matrix domain.  
The basic antagonism between the usual rules that shape the intermaterial dividing surface (IMDS) -- the enthalpy of mixing and entropic rigidity of different blocks -- and the constraint that chains must fill all of space is loosely referred to as ``packing frustration.''\cite{Anderson1988,Matsen1996,Grason2006,ACShi2021,Duesing1997} 
That is, the thermodynamic competition selects for a preferred local arrangement of molecules, which can be described as occupying a certain geometric motif.
These motifs, which can be approximated as wedge-like volumes of particular sizes and shapes, cannot, in general, tile space at uniform density.
Hence, physical BCP assemblies must be composed of ensembles of spatially distorted variants of the preferred motif, and consequently incur an additional free energy penalty associated with that distortion.  
Elementary considerations show that all BCP morphologies, with the exception of lamellae, are subject to some measure of frustration\cite{ACShi2021}.  
Perhaps the most obvious example derives from the fact that spheres and cylinders do not fill space without gaps.  
Hence convex, quasi-spherical or cylindrical domains arranged in 3D crystalline or 2D columnar packings are distorted away from perfect rotational symmetry.  
Packing frustration has long been cited to be particularly vexing for the formation of bicontinuous networks in BCP melts\cite{Matsen1996,MatsenBates1997,Schroder-Turk2007}, leading to narrow equilibrium windows in the diblock copolymer phase diagram (if they are predicted at all), as well as the stability of the double-gyroid (DG) network phase (shown in Fig.~\ref{fig:dg_overview}) over competitor double-diamond (DD) and double-primitive (DP) phases, whose structures are displayed in Fig.~\ref{fig:network_geometry}.   

Efforts to understand the connections between the thermodynamics and complex geometries of bicontinuous networks were stimulated by experiments that reported apparently equilibrium cubic networks in diblock melts at compositions intermediate to where columnar and lamellar morphologies are observed. 
Early observations of star diblocks suggested the formation of a DD structure\cite{Thomas1986}, but it was later concluded that in linear diblock melts, under conditions approaching equilibrium, cubic network formation is predominantly DG\cite{Hajduk1994,Foerster1994,Khandpur1995}.
The thermodynamic stability of bicontinuous networks in BCP has a complex history, owing in part to a long-standing and apparent discrepancy between strong segregation theory (SST) and numerical studies of self-consistent field theory (SCFT).  
One one hand, early SST models of network phases in linear diblocks by Olmsted and Milner \cite{Olmsted1994,Olmsted1995,Olmsted1998}, and separately by Lihktman and Semenov \cite{Likhtman1994,Likhtman1997}, predicted that network phases are never thermodynamically stable in the $\chi N \to \infty$ limit due to a large free-energy gap relative to competitor hexagonal cylinder (Hex) and lamellar (Lam) phases.  
This was consistent with initial SCFT calculations which suggested that the DG phase might become unstable at very large segregation\cite{Matsen1996b}.  
However, as resolution of SCFT algorithms improved sufficiently to address very strong segregation regimes, calculations showed that DG phase retains a finite, albeit narrowing window of equilibrium up to at least $\chi N \gtrsim 100$ \cite{Cochran2006}, which is also consistent with experimental studies of this regime \cite{Davidock2003}.

These original SST models of chain packing in network phases are based on what might be called a ``skeletal ansatz," which assumes that the 1D skeletal graph that connects the nodal centers of the tubular network domains constitutes what has been dubbed a {\it terminal boundary}\cite{Reddy2021}, i.e.~the region of maximal extension of chain trajectories away from the IMDS within a brush-like subdomain composed of one polymer block.  
While the skeletal ansatz is a seemingly intuitive proxy for the tubular network regions\cite{Prasad2018}, it has recently been understood that this approximation severely overestimates the entropic penalty of stretching in those domains.
Recently, we\cite{Reddy2022} showed that a more realistic and thermodynamically favorable packing ansatz for DG networks formed by diblocks in the SST limit considers terminal boundaries that spread well away from the 1D skeletal graph, in web-like surface patches derived from the medial map of gyroid morphologies. 
In short, the {\it medial map} provides the shortest-distance map of points within a volume \emph{onto} points on its bounding surface as well as a corresponding {\it medial set}, which is the set of maximally-distant points at the general ``center'' of a domain of arbitrary shape\cite{Schroder2003,Reddy2021}.   
Importantly, we showed that the stability of the DG phase relies on the ability of chain ends to spread out over a web-like surface (shown in Fig.~\ref{fig:dg_overview}) in each tubular domain, rather than being forced to stretch to the 1D skeletal graph, thus lowering the entropic penalty for tubular-domain filling to the point where the free-energy gap between competitor morphologies is eliminated.
As a result, we predicted DG stability windows, between Hex and Lam phases for diblocks, that open up and widen as the conformational asymmetry between blocks is increased.

In this article, we apply this medial strong segregation theory (mSST) approach to a comparative study of DG, DD, and DP phases.  
In neat BCP systems, DG is almost always the thermodynamically favored network structure.   
This has long begged the question, what aspects of the DG morphology make it favorable over its apparently structurally-similar DD and DP cousins?  
Long-standing heuristic pictures point to costs of packing frustration associated with the center of the nodes\cite{Matsen1996}.  
The DG, DD and DP networks differ in terms of their functionality (i.e.~the number of nearest-neighbor nodes within a contiguous tubular domain): 3, 4 and 6, respectively.  
It has been suggested that the relatively high free energy of DD and DP relative to DG can be attributed to a frustration cost that increases with node functionality, due to an increasing distance of the domain center from the IDMS for nodal regions with additional tubular interconnections.  
While seemingly intuitive, this heuristic picture makes it unclear why any network morphology composed of such ``problematic'' nodal regions should be favored over the classical cylindrical or layered structures, and whether and when conditions exist such that the competition between network phases can be shifted to higher-functionality structures (e.g.~DD over DG). 

Here we employ the mSST framework to assess the role of chain packing and its connection to complex domain shape in detail.  
We analyze the specific relationships between the \emph{shapes} of the IMDS and terminal boundaries and the \emph{thermodynamic costs} of variable stretching and IMDS enthalpy in the competing networks.  
Additionally, this medial packing perspective reveals what is uniquely advantageous about DG among its competitor morphologies, a feature that can be associated with terminal boundary geometry, shown in Figs.~\ref{fig:dg_overview} and \ref{fig:network_geometry} for all three network phases.  
The DG (Fig.~\ref{fig:dg_overview}A,B) terminal boundaries are composed of twisted, ribbon-like webs that thread through tubular subdomains and a Gyroid-like surface that subdivides the matrix domain.  
While the DD and DP (Fig.~\ref{fig:network_geometry}) matrix regions are also divided by minimal surface-like terminal boundaries, their tubular domains possess multiple leaves, or ``fins,'' that meet in abrupt angles.  
This can intuitively be understood as a consequence of the fact that the terminal boundaries are composed of web-like surfaces that span between portions of the network skeletons.   
Hence, the terminal boundaries of DD and DP possess \emph{corners}, which are analogous to singular geometric features that are typically found in the boundaries between matrix brushes for cylinder- and sphere-like domain assemblies (i.e.~the corners of Voronoi cells).   
While the struts of individual DG nodes are coplanar and can be spanned by a single, roughly triangular surface (appropriately twisting between nodes), the respective tetrahedral and octahedral geometries spanned by the DD and DP struts are not coplanar, and pairs of struts must be joined by distinct webs. 
The special ``corner-free'' nature of DG terminal boundaries facilitates a hybridized morphology, as shown in Fig.~\ref{fig:dg_overview}C, which is composted of local regions of quasi-lamellar packing coexistent with curved, cylinder-like regions.  
As we demonstrate below, this sharp distinction in terminal boundary geometry between DG relative to DD and DP significantly impacts the nature and thermodynamics of chain packing underlying network phases.

While ``packing frustration'' is a seemingly intuitive and widely invoked notion in amphiphile self-assembly\cite{Templer1998,Kulkarni2010}, a central perspective of this article is that it is difficult, if not impossible, to characterize it by a single quantitative measure of the morphology\cite{Schroder-Turk2006}, beyond possibly the free energy of a morphology.  
On one hand, packing frustration is associated with the \textit{variability} of BCP packing and the resulting entropic and enthalpic costs\cite{Schroder-Turk2007,Matsen1996,Reddy2021,Chen2022}.  
Alternatively, packing frustration is often connected to certain {\it locations} in a morphology that can be identified as the ``sources" of frustration.  
Such a perspective focuses on where is frustration coming from in the packing, and in turn, how this relates to especially costly ``hot spots" in the resulting morphology.   
Related to this latter concept is a third notion of the \textit{susceptibility} of a morphology to filling the hot spots by blending in guest molecules (e.g.~homopolymers, nanoparticles, or solvent) that can can relieve some, if not all, of the frustration\cite{Matsen1995,Likhtman1997,Martinez-Veracoechea2009,Martinez-Veracoechea2009b,Cheong2020,Xie2021,Lai2021,Takagi2021}.  
These distinct, yet interrelated, facets of packing frustration challenge a simplified, overarching theoretical understanding of what it is, and more importantly, the development of rational principles for manipulating it via chemical design of supramolecular systems.  
In what follows, we make no particular attempt to resolve these multiple, at at times countervailing, perspectives on packing frustration.  
Instead, we exploit medial SST to translate fine features of complex morphology into explicit thermodynamic terms in order to assess formation of distinct cubic double-networks in diblock melts.  
The primary focus of our analysis will be on the first two facets of frustration: the statistical variability of molecular packing and its spatial correlations with geometric aspects of the morphology.  
While blended systems are beyond the primary scope of this article, we include a brief analysis of the ``hot-spot" distributions within the distinct tubular vs.~matrix domains of networks and discuss the likely implications of hot-spot distributions for susceptibility of frustrated network assembly to incorporate ``guest'' molecules (e.g.~homopolymers) in blends. 

The remainder of this article is outlined as follows. 
We first introduce the medial strong segregation theory (mSST) and our variational approach to finding optimal network morphologies.
Next we explore the geometry of each of the cubic network phases and identify ways in which the packing environments of the DG phase are optimal in comparison with the DD and DP.
We then examine how these variable packing environments give rise to significant spatial variations in the free energy per chain, with optimal morphologies forming regions of low chain entropy at the cost of high interfacial enthalpy, and \emph{vice versa}.
Building on our explorations of packing geometry and variations in free energy per chain, we present a multifaceted and generalized picture of \emph{packing frustration}.
Finally, we consider the distribution of chain stretching free energy in tubular and matrix brush subdomains.  
We analyze the shapes and locations of so-called ``hot spots'' in the chain stretching and discuss potential implications for systems blended with guest molecules (e.g.~homopolymers).

\section{Medial SST Methodology}

\subsection{Strong Segregation Theory}

Expanding our attention beyond linear diblocks, we consider the more general situation of A$_{n_{\rm A}}$B$_{n_{\rm B}}$ equal-arm starblocks, where $n_{\rm A}$ and $n_{\rm B}$ are the number of branches.
Each branch of a single block has the same number of segments, $N_{\rm A}/n_{\rm A}$ and $N_{\rm B}/n_{\rm B}$, where $N_{\rm A}$ and $N_{\rm B}$ are the total numbers of A-block and B-block segments, respectively, and $N = N_{\rm A} + N_{\rm B}$ is the total number of segments in the chain.
It has been shown by Milner \cite{Milner1994,Grason2004,Grason2006} that the entropic stiffness of such starblock copolymers can be encoded in a single elastic asymmetry parameter $\epsilon = n_{B}a_{A}/n_{A}a_{B}$, where $a_{\rm A}$ and $a_{\rm B}$ are the block segment lengths.
Note that this elastic asymmetry parameter encodes both architectural asymmetry ($n_{\rm A} \neq n_{\rm B}$) as well as conformational asymmetry ($a_{\rm A} \neq a_{\rm B}$), and that a starblock can be represented as an equivalent linear diblock with asymmetric segment lengths.
Thus, we consider a parameter space of chains whose properties are encoded in a pair of parameters -- the elastic asymmetry $\epsilon$ and the A-block fraction $f = N_{\rm A}/N$.

In the SST limit, $\chi N \to \infty$, the total free energy $F$ is given by\cite{Semenov1985,Goveas1997}
\begin{equation}
    F = H + S_{\rm A} + S_{\rm B} \ ,
\end{equation}
where $H$ is the interfacial enthalpy and $S_{\alpha}$ for $\alpha \in \{{\rm A},{\rm B}\}$ are the free energy contributions arising from reductions in the entropy of stretched chain conformations.
The enthalpy is given by $H = \gamma A$, where $A$ is the area of the IMDS and the interfacial energy density $\gamma$ is expressed as\cite{Helfand1975}
\begin{equation}
\gamma = k_{\rm B}T\rho \overline{a} \sqrt{\frac{\chi}{6}}\left(\frac{2}{3}\frac{\epsilon_0^{3/2} - \epsilon_0^{-3/2}}{\epsilon_0 - \epsilon_0^{-1}}\right) = k_{\rm B}T\rho \overline{a} \sqrt{\frac{\chi}{6}} \overline{\gamma} \, ,
\end{equation}
where $\rho$ is the segment density, $\overline{a} \equiv \sqrt{a_{\rm A} a_{\rm B}}$ is the geometric mean of the two statistical segment length $a_{\rm A}$ and $a_{\rm B}$, $\epsilon_0 = a_{\rm A}/a_{\rm B}$ is a parameter describing conformational asymmetry, which is folded into a dimensionless number $\overline{\gamma}$.
The stretching free energy of each subdomain $\alpha$ is given by 
\begin{equation}
S_{\alpha} = \frac{1}{2}\kappa_{\alpha}I_{\alpha}\, ,
\end{equation}
where the entropic stiffness parameters $\kappa_\alpha$ are given by
\begin{equation}
\begin{split}
    \kappa_{\rm A} &= \frac{3\pi^2 \rho k_{\rm B}T}{4 N^2 \overline{a}^2}\frac{ \overline{n}^2}{f^2 \epsilon} = \frac{3\pi^2 \rho k_{\rm B}T}{4 N^2 \overline{a}^2} \overline{\kappa}_{\rm A}\, , \\
    \kappa_{\rm B} &= \frac{3\pi^2 \rho k_{\rm B}T}{4 N^2 \overline{a}^2}\frac{ \overline{n}^2\epsilon}{(1-f)^2} = \frac{3\pi^2 \rho k_{\rm B}T}{4 N^2 \overline{a}^2} \overline{\kappa}_{\rm B} \, ,
\end{split}
\end{equation}
where $\overline{n} \equiv \sqrt{n_{\rm A} n_{\rm B}}$ is the geometric mean of A- and B-block branch numbers $n_{\rm A}$ and $n_{\rm B}$, $\epsilon = (n_{\rm B}/n_{\rm A})\epsilon_0$ is the elastic asymmetry parameter, and $\overline{\kappa}_\alpha$ are numerical parameters that contain all dependence on architectural and conformational asymmetry between the two blocks.
The geometric cost of chain stretching is well-approximated by the results of parabolic brush theory\cite{Milner1988,Likhtman1994} and encoded in the total second moments of volume 
\begin{equation}
I_{\alpha} \equiv \int_{V_{\alpha}}{\rm d}V\, z^2
\end{equation}
for each subdomain $\alpha$, where $z$ is a brush height coordinate extending from the IMDS at $z = 0$.
While parabolic brush theory accounts for the statistics of chains whose free ends are allowed to exist anywhere within a brush in a manner consistent with melt conditions, it fails to be self-consistent for brushes on surfaces whose curvatures force the free chain ends to splay apart.\cite{Ball1991,Belyi2004}
However, we have previously shown\cite{Dimitriyev2021,Reddy2022} that the necessary ``end-exclusion zone'' corrections to parabolic brush theory are typically negligible for network phases, increasing the free energy per chain by $\lesssim 0.01\%$, with appreciable increases in free energy predicted for highly-curved sphere phases, justifying our use of parabolic brush theory for the results presented in this manuscript.

\begin{figure}[h!]
\centering
\includegraphics[width=3in]{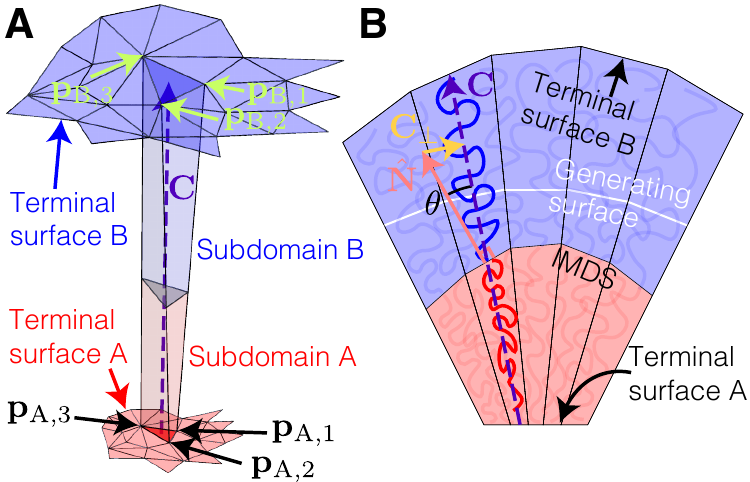}
\caption{\label{fig:wedges_schematic} (A) Schematic of a wedge built off of two medial surface patches. Vertices $\{\mathbf{p}_{{\rm A},i}\}$ and $\{\mathbf{p}_{{\rm B},i}\}$ lie on terminal surface A and B, respectively. (B) Multiple volume-balanced wedges are shown with continuous IMDS patches whose normals $\hat{\mathbf{N}}$ are tilted relative to the local chain orientations, provided by the centroidal vector $\mathbf{C}$.}
\end{figure}

Next, we develop an ansatz for how chains fill space.
We imagine that the average chain conformations follow trajectories that join points $\mathbf{p}_{\rm A}$ on the terminal surface $\mathbf{T}_{\rm A}$ in the tubular A-block subdomain to points $\mathbf{p}_{\rm B}$ on the terminal surface $\mathbf{T}_{\rm B}$ in the matrix B-block subdomain.
Since all of space is occupied by chains, any point in space can be mapped to a pair of terminal points on each of the terminal surfaces along such trajectories, the collection of which foliates space and defines the \emph{association map}.
In SST, chains are in a strongly-stretched limit, which suggests that straight-line association maps $\mathbf{p}(t) = (1-t)\mathbf{p}_{\rm A} + t\mathbf{p}_{\rm B}$ for $t \in [0,1]$ are reasonable approximations of chain trajectories.

Triplets of points on each terminal surface, $\{\mathbf{p}_{\alpha,1},\mathbf{p}_{\alpha,2},\mathbf{p}_{\alpha,3}\}$ (for $\alpha = {\rm A},\,{\rm B}$), form triangular faces and lines joining corresponding pairs of these points $\mathbf{p}_i(t) = (1-t)\mathbf{p}_{{\rm A},i} + t\mathbf{p}_{{\rm B},i}$ (for $i = 1,\,2,\,3$) provide the long edges of slender, pentahedral ``wedge'' volumes that approximate local packing environments for collections of chains, as depicted in Fig.~\ref{fig:wedges_schematic}A.
Each wedge can be constructed as a stack of triangular regions for fixed values of $t$, terminating at the triangular facets on the two terminal surfaces, so that an arbitrary point $\mathbf{X}$ within any triangular slice can be represented as
\begin{equation}
    \mathbf{X}(u,v,t) = (1-v)\mathbf{p}_1(t) + (1 - u)v\mathbf{p}_2(t) + u v \mathbf{p}_3(t) \, ,
\end{equation}
where $u,v \in [0,1]$ provide local coordinates for each triangle at fixed $t$.
If the wedge is sufficiently thin then the centroidal vector $\mathbf{C} \equiv \sum_{i=1}^3\Delta\mathbf{p}_i/3$, where $\Delta \mathbf{p}_i \equiv \mathbf{p}_{{\rm B},i} - \mathbf{p}_{{\rm A},i}$, provides an average trajectory of chains within the wedge and a definition of wedge height, $h \equiv |\mathbf{C}|$ and a local ``wedge height'' coordinate $\zeta \equiv ht$.
The area $A(\zeta)$ of each triangular cross-section of the wedge is given by a version of Steiner's formula\cite{Hyde1997_ch4}, 
\begin{equation}
    A(\zeta) = \hat{\mathbf{C}}\cdot\mathbf{A}_{\rm A}\left(1 + 2\mathcal{H}\zeta + \mathcal{K}\zeta^2\right)\, ,
\end{equation}
where $\mathbf{A}_{\rm A} = (\mathbf{p}_{{\rm A},3} - \mathbf{p}_{{\rm A},2})\times(\mathbf{p}_{{\rm A},1} - \mathbf{p}_{{\rm A},2})/2$ is the area vector of the triangular facet on the terminal surface $\mathbf{T}_{\rm A}$ and
\begin{equation}
    \begin{split}
        \mathcal{H} &\equiv \frac{\hat{\mathbf{C}}}{4h\hat{\mathbf{C}}\cdot\mathbf{A}_{\rm A}}\cdot\left[(\Delta \mathbf{p}_3 - \Delta\mathbf{p}_2)\times(\mathbf{p}_{{\rm A},1} - \mathbf{p}_{{\rm A},2}) + (\mathbf{p}_{{\rm A},3} - \mathbf{p}_{{\rm A},2})\times(\Delta \mathbf{p}_1 - \Delta\mathbf{p}_2)\right] \\
        \mathcal{K} &\equiv \frac{\hat{\mathbf{C}}}{2h^2\hat{\mathbf{C}}\cdot\mathbf{A}_{\rm A}}\cdot\left[\Delta\mathbf{p}_3\times\Delta\mathbf{p}_1 + \Delta\mathbf{p}_2\times\Delta\mathbf{p}_3 + \Delta\mathbf{p}_1\times\Delta\mathbf{p}_2\right]
    \end{split}
\end{equation}
are analogous to the mean and Gaussian curvatures, respectively.
In the strong segregation limit, each wedge is sharply divided into two sub-wedges by a dividing triangle at height $\zeta = h_b$, representing a small surface patch of the IMDS.
The location of the IMDS patch is determined by local volume balance, $V(h_b) = fV(h)$, where the volume of a sub-wedge with one end on $\mathbf{T}_{\rm A}$ and the other at height $\zeta$ is given by
\begin{equation}
    V(\zeta) = \hat{\mathbf{C}}\cdot\mathbf{A}_{\rm A}\,\zeta\left(1 + \mathcal{H}\zeta+\frac{\mathcal{K}}{3}\zeta^2\right)\, ,
\end{equation}
so that $h_b$ is determined by solving a cubic equation for each wedge.

It is important to emphasize that while this representation of local packing environments allows for local volume balance to be satisfied everywhere by solving for the distance $h_b$ of the IMDS from the A-block terminal surface $\mathbf{T}_{\rm A}$, the parametrization that we employ fixes the local surface normal $\hat{\mathbf{N}}$ of the IMDS to lie along the centroidal direction $\hat{\mathbf{C}}$.
However, this enforcement of IMDS orientation over-constrains local packing and since two neighboring wedges generally have different balanced heights $h_b$ and centroidal directions $\hat{\mathbf{C}}$, the resulting IMDS is generally discontinuous.
To fix this, we individually adjust the vertices of each IMDS patch so that neighboring patches are continuous across each shared edge.
This is done by averaging over the vertex positions IMDS patches at shared edges, ensuring that any error in local volume balance due to this adjustment is kept minimal; in our calculations, the error in volume balance is typically less than 1\% of the total volume.
As a result of these local adjustments between neighboring wedges, there is a local tilt between chain orientations and IMDS normal (illustrated in Fig.~\ref{fig:wedges_schematic}B), the implications of which are discussed in a later section.

In terms of wedge geometry, the second moments of volume of a single wedge (indexed by $\mu$) are given by
\begin{equation}
    \begin{split}
        I_{{\rm A},\mu} &= \frac{h_{b,\mu}^3\hat{\mathbf{C}}_\mu\cdot\mathbf{A}_{{\rm A},\mu}}{3}\left[1 + \frac{\mathcal{H}_\mu}{2}h_{b,\mu} + \frac{\mathcal{K}_\mu}{10}h_{b,\mu}^2\right] \\
        I_{{\rm B},\mu} &= \frac{\hat{\mathbf{C}}_\mu\cdot\mathbf{A}_{{\rm A},\mu}(h_\mu - h_{b,\mu})^3}{3}\left[1 + \frac{\mathcal{H}_\mu}{2}(h_{b,\mu} + 3h_\mu) + \frac{\mathcal{K}_\mu}{10}(h_{b,\mu}^2 + 3h_{b,\mu}h_\mu+6h_{\mu}^2)\right]
    \end{split} \, ,
\end{equation}
where the subscript $\mu$ indicates that the corresponding parameters are defined according to individual wedges.
The total free energy $F$ is then given as a sum over individual wedge contributions, i.e.~$F = \sum_\mu F_\mu$; similarly, the total enthalpy $H = \sum_{\mu} H_\mu$ and the total costs of stretching each block are $S_{\alpha} = \sum_{\mu} S_{\mu, \alpha}$.

\subsection{Medial ansatz and variational calculation}

We propose that the terminal surfaces $\mathbf{T}_{\rm A}$ and $\mathbf{T}_{\rm B}$ are well-approximated by the \emph{medial sets} of a suitable family of generating surfaces.
This is based on the observation that the cost of stretching chains in a domain is related to the thickness of the domain.
Thus, this cost is minimized when the thickness of the domain is minimized. 
Given a certain surface $\mathbf{G}$ (which we shall call a ``generating surface''), any point $\mathbf{x}$ has a corresponding point $\mathbf{p}$ on $\mathbf{G}$ that minimizes the Euclidean distance, which is found via the \emph{medial map} $\mathbf{m}(\mathbf{x})$ \cite{Blum1967,Nackman1985,Schroder2003,SiddiqiPizer2008}.
The medial map can be determined from the geometry of $\mathbf{G}$: the Euclidean distance $|\mathbf{p} - \mathbf{x}|$ is minimized when $\mathbf{p} - \mathbf{x} \propto \hat{\mathbf{N}}(\mathbf{p})$, where $\hat{\mathbf{N}}(\mathbf{p})$ is the surface normal evaluated at $\mathbf{p}$; since $\mathbf{x}$ may in general lie on many different lines that lie along different surface normals of $\mathbf{G}$, the medial map $\mathbf{m}(\mathbf{x})$ selects the global minimizer of the Euclidean distance from all of these possibilities.
If $\mathbf{x}$ is near the center of the region bounded by $\mathbf{G}$ then it is potentially equidistant from multiple points on $\mathbf{G}$: the collection of these points whose medial map $\mathbf{m}$ has multiple solutions defines the \emph{medial set} and gives a rigorous definition of the ``center'' of a region.
Since the medial map lies along local surface normals, it falls into the family of straight-line association maps.

To generate medial surfaces that are consistent with the symmetries of the three network phases, we specify generating surfaces $\mathbf{G}$ as level sets of the form $\Psi(\mathbf{r}) = \pm1$, where the sign selects a generating surface for one of the two bicontinuous domains.
These level sets have symmetries given by the non-centrosymmetric subgroups $I4_1 3 2$ (DG), $F d \overline{3} m$ (DD), and $P m \overline{3} m$ (DP) of the network phases' crystallographic groups, $I a \overline{3} d$ (DG), $P n \overline{3} m$ (DD), and $I m \overline{3} m$ (DP) \cite{Wohlgemuth2001}.
As such, the level sets can be represented by a set of basis functions that are adapted to each space group, i.e.~$\Psi = \sum_n c_{n} \psi_{n}(\mathbf{r}/D)$, where $\psi_{n}(\mathbf{r}/D)$ represent the symmetry-adapted basis functions, labeled by $n$, and $c_n$ are expansion coefficients, and $D$ represents the periodicity of the unit cell.
We truncate this expansion at the first four modes; for a list of the basis functions, please refer to the \note{supplementary text}.
It is important to emphasize that the generating surface $\mathbf{G}$ is in fact \emph{distinct} from the volume-balanced IMDS: while we propose that medial packing optimizes the stretching cost of chains, it generally fails to yield environments that satisfy the local volume balance constraint.
Therefore, we use generating surfaces that are close to a reasonable IMDS shape as an initial guess, with the expectation that the corresponding volume-balanced IMDS will be distinct while preserving features of the generating surface, such as its symmetries, topology, and coarse features.

Given a generating surface specified by coefficients $\{c_n\}$ and a fixed set of symmetry-adapted basis functions, we calculate the corresponding medial sets and, treating them as suitable terminal surfaces, construct volume-balanced wedges.
The total free energy of a given network structure depends on the value of the $D$-spacing.
Instead of fixing the $D$-spacing, we allow the network to adjust in size, changing the total number of chains within a unit cell.
Under isotropic scaling, the IMDS area is $A = D^2 \tilde{A}$, the total volume is $V = D^3 \tilde{V}$, and the total second moments of volume are $I_{\alpha} = D^5 \tilde{I}_{\alpha}$, where $\tilde{A}$, $\tilde{V}$, and $\tilde{I}_{\alpha}$ are the nondimensional forms of IMDS area, total volume, and total second moments of volume.
For a fixed morphology, the $D$-spacing can then be found by minimizing the free energy per chain $\hat{F} \equiv F/n_{\rm ch}$ (where the number of chains is given by $n_{\rm ch} = \rho V/N$) with respect to $D$, yielding
\begin{equation}
    D = \sqrt{\frac{2}{3\pi^{4/3}}}(\chi N)^{1/6} N^{1/2}\overline{a}\overline{\lambda}
\end{equation}
where 
\begin{equation}\label{eq:scaling_factor2}
\overline{\lambda} \equiv (\overline{\gamma}\tilde{A}/(\sum_\alpha\overline{\kappa}_\alpha \tilde{I}_\alpha))^{1/3}
\end{equation}
is a dimensionless scale factor that depends only on the wedge geometries and chain architecture.
The free energy per chain is then given by
\begin{equation}\label{eq:free_energy_per_chain}
\hat{F} = \frac{3 \pi^2}{4 \tilde{V}}(\overline{\kappa}_{\rm A}\tilde{I}_{\rm A} + \overline{\kappa}_{\rm B}\tilde{I}_{\rm B})^{1/3}(\overline{\gamma}\tilde{A})^{2/3}\left[(\chi N)^{1/3}k_{\rm B} T\right]  \, ,
\end{equation}
recovering the expected $\sim k_{\rm B}T(\chi N)^{1/3}$ scaling as $\chi N \to \infty$.

Finally, we minimize the free energy per chain (Eq.~\ref{eq:free_energy_per_chain}) over the set of generating surfaces.
This is done by choosing an initial set of basis coefficients $c_n$, meshing the resulting generating surface $\mathbf{G}$ (we used $\sim 10^4$ facets per nodal IMDS), calculating the free energy per chain $\hat{F}$, and then performing a search for values of $c_n$ that minimize $\hat{F}$ using a Nelder-Mead algorithm .
Since changes in A-block fraction $f$ and elastic asymmetry $\epsilon$ result in different equilibrium morphologies, this minimization is performed for each fixed value of the two parameters $(f,\epsilon)$.

Our choice of minimizing the free energy over a set of four basis functions rather than the two of our previous study\cite{Reddy2022} was in order to assess (i) whether the addition of further Fourier modes to the generating surfaces had a significant impact on the results of the calculation and (ii) the generalizability of the variational calculation to further degrees of freedom.
As shown in \note{SI Fig.~S1}, these additional modes lead to slight reductions (on the order of $10^{-5}$ for DG, $10^{-3}$ for DD and DP) in the calculated free energy (within the constraints of the convergence criteria supplied to the Nelder-Mead algorithm) that seemingly diminish with each added mode.
Nonetheless, this demonstrates the potential applicability of the variational form of mSST to calculations involving additional degrees of freedom.

\begin{figure}[h]
\centering
\includegraphics[width=\textwidth]{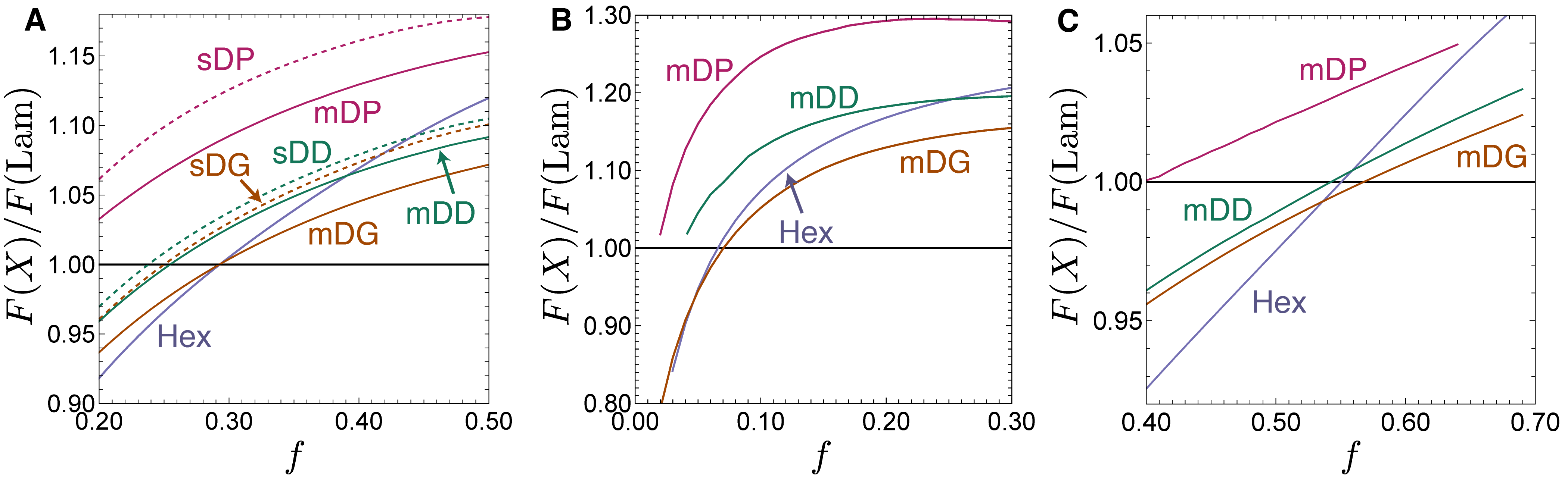}
\caption{\label{fig:free_energy_fixed_eps} (A) Free energy comparison between the cubic double networks computed via mSST (mDG, mDD, and mDP; solid curves) with the prior sSST calculations of Olmsted and Milner\cite{Olmsted1998} (sDG, sDD, and sDP; dashed curves) in the case of elastic symmetry ($\epsilon = 1$). These free energies normalized by the lamellar (Lam) free energy and further compared with the hexagonal cylinder (Hex) free energy, as calculated via the kinked path model. Free energy comparisons using the mSST calculation for $\epsilon = 0.3$ and $\epsilon = 3.0$ are shown in (B) and (C).}
\end{figure}

\begin{figure}[h!]
\centering
\includegraphics[width=3in]{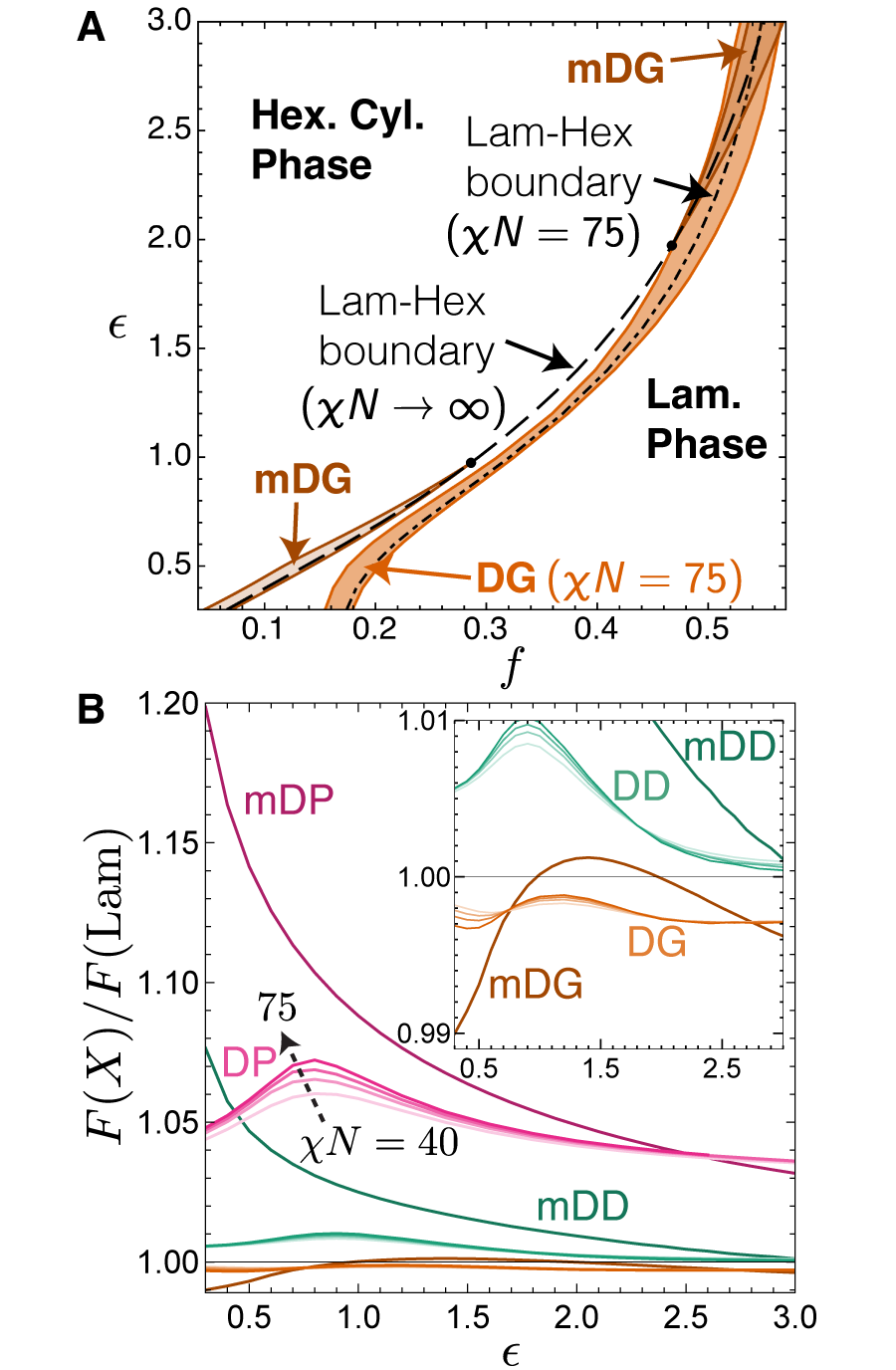}
\caption{\label{fig:fe_compare} (A) Phase diagram showing a stable DG phase at $\chi N = 75$ (computed via SCFT) and re-entrant stability of mDG ($\chi N \to \infty$) compared with lamellar (high $f$) and hexagonal cylinder (low $f$) phases. Dashed lines represent regions of the $(f,\epsilon)$ parameter space for which the free energy of the lamellar and hexagonal cylinder phases are equal, additionally providing a region of direct lam/hex transition between the triple points (shown as dots) at $(f \sim 0.29, \epsilon \sim 0.98)$ and $(f \sim 0.47, \epsilon \sim 1.97)$ in the $\chi \to \infty$ limit. (B) Comparison of mSST calculations (dark curves) with SCFT calculations at finite segregation ($\chi N = 40 - 75$, light to darker). Inset shows a magnification that highlights DG and DD.}
\end{figure}

\subsection{Large $\chi N$, self-consistent field theory}

We use self-consistent field theory (SCFT) calculations\cite{Matsen2002} as a point of comparison for the equilibrium network morphologies predicted by mSST and to demonstrate how our results at $\chi N \to \infty$ translate to finite (but large) $\chi N$.
The SCFT calculations were performed using the open-source PSCF software \cite{Arora2016} at $\chi N = 40,\, 50,\, 60$, and 75.
To simplify comparisons, we chose parameters relative to unit chain length, $N = 1$, and B block statistical segment length, $a_{\rm B} = 1$.
In these units, we tuned the elastic asymmetry $\epsilon$ via the A block statistical segment length, $a_{\rm A} = \epsilon$.
Finally, in order to extract information about chain trajectories, we computed the polar order parameter field $\mathbf{p}(\mathbf{x})$ from the segment flux\cite{Prasad2017}.

\section{Thermodynamics of competing networks}

We first analyze the relative free energies of DG, DD and DP network phases in comparison to competitor Lam and Hex phases, and their dependence on composition and elastic asymmetry between the two blocks.

As shown in Fig.~\ref{fig:free_energy_fixed_eps}A, the four-mode mSST calculations of the free energy for each network is consistently smaller than predictions using the skeletal terminal surface model (here referred to as sSST) of Olmsted and Milner \cite{Olmsted1998} for the case of elastically symmetric diblocks ($\epsilon = 1$).
Here, the SST calculation of the the hexagonal cylinder (Hex) phase is performed using the kinked path ansatz \cite{Grason2004}.
Once again, we find that the free energy of the mSST-constructed DG (mDG) coincides with that of the lamellar (Lam) and Hex phases at $f \approx 0.29$, dramatically relaxing the conditions for stability compared with previous sSST predictions (sDG).
Moreover, the mDG is consistently $\sim 2-3\%$ lower in free energy to sDG, compared with the $\sim 1\%$ of mDD to sDD and $\sim 3\%$ of mDP to sDP.  
While those changes in free energy for DD and DP are significant on the scale of SST thermodynamics in general, they are nevertheless small compared to the much larger gaps of those networks relative to Lam and Hex (of order 3\% and 12\% for DD and DP respectively at $f = 0.29$).  
This demonstrates that relaxation from skeletal to medial terminal boundary geometry accounts for a much greater, and more significant, reduction of the entropic penalty associated with nodal packing for DG relative to its DD and DP competitors.  
We discuss geometrical interpretations of this ``more effective'' medial packing for DG in the following section.

In Fig.~\ref{fig:free_energy_fixed_eps}B and C, we show the free energy comparison of these same competing phases for elastically asymmetry diblocks:  $\epsilon = 0.3$, relatively stiffer tubular A block; and $\epsilon = 3.0$ relatively stiffer matrix B block. 
We observe windows of mDG stability over Lam and Hex phases for both limits of elastic asymmetry, consistent with our previously reported results \cite{Reddy2022}.
However, while both mDD and mDP structures are both far from stable for $\epsilon = 0.3$ (stiffer A block), mDD is in close competition with Lam and Hex phases for $\epsilon = 3.0$ (stiffer B block) in the window where DG is stable.
This indicates an asymmetry how medial packing relaxes entropic penalties in the two blocks: the cost of stretching for mDD and mDP is much larger than mDG when the A block is stiffer, whereas the three phases are have comparable stretching costs when the B block is stiffer.

To analyze how the relative thermodynamic stability of cubic network phases varies with elastic asymmetry, we compare their free energies at the compositions where Lam and Hex have equal free energy in Fig.~\ref{fig:fe_compare}.  
These points fall on the dashed line in Fig.~\ref{fig:fe_compare}A in the $f$-$\epsilon$ plane, falling within the DG stability windows at high and low $\epsilon$.  
We compare both the mSST predictions which model the asymptotic $\chi N \to \infty$ limit as well as finite-$\chi N$ SCFT calculations for increasingly large values of segregation ($\chi N = 40, 50, 60$ and 75), shown in Fig.~\ref{fig:fe_compare}B.  
As previously reported, both mSST and SCFT show that the free energy of DG relative to Lam and Hex competitors varies non-monotonically with $\epsilon$, decreasing both as $\epsilon$ gets larger and smaller than $\epsilon \approx 1$.  
This non-monotonic behavior suggests that the nature of sub-domain chain packing in DG adjusts to accommodate regimes of both relatively stiffer matrix and minority block regimes, while maintaining thermodynamically favorable aspects of the morphology, which we discuss in more detail below.  
In comparison, SCFT at these finite $\chi N$ predicts a lower free energy for DG than Hex or Lam for the full range of $\epsilon$, indicating that it retains an equilibrium stability window up through $\chi N = 75$.  
In comparison, mSST predicts that the free energy of DG exceeds that of Lam and Hex at intermediate $1.0 \lesssim \epsilon \lesssim 2.0$, corresponding to the vanishing equilibrium stability window, as previously reported \cite{Reddy2022}.
While we note some basic consistency in the non-monotonic variation of the DG to Lam free energy with $\epsilon$, mSST shows greater range in relative free energy variation.  
In part, we can attribute some of this difference to finite-$\chi N$ corrections to the free energy that are absent from the mSST calculation.  
Notably, the magnitudes of relative SCFT free energies increase considerably with $\chi N$ over this fairly modest range, suggesting that extending SCFT to the asymptotic $\chi N \to \infty$ would likely result in relative free energies more comparable to the mSST predictions. 
Prior studies \cite{Matsen2001,Matsen2010} suggest such finite-$\chi N$ corrections may be significant (on the scale of the few $\%$ differences) for $\chi N$ large as $10^3-10^4$.

Fig.~\ref{fig:fe_compare}B shows that the mSST predicts that the relative free energies of DD and DP decrease monotonically with $\epsilon$.  
This behavior is somewhat more intuitive than the non-monotonic variation of DG, as it is consistent with the interpretation that the tubular A blocks are the most costly regions in the morphology for chain stretching so that decreasing the entropic penalty for stretching in those blocks, by increasing $\epsilon$, diminishes the relative free energy gap of those phases relative to competitors.  
Furthermore, unlike the case of DG, which decreases for small $\epsilon$, mSST predictions imply that the medial packing of DD and DP structures is unable to reorganize sufficiently in the regime of relatively stiffer A blocks to mitigate the presumably large entropic costs of tubular node packing.  
Somewhat surprising in this context is the fact that SCFT for DD and DP show a non-monotonic variation of relative free energy, decreasing from a maximal value at intermediate elastic asymmetry for both $\epsilon \gtrsim 1$ {\it and} $\epsilon \lesssim 1$. 
As we describe and analyze in the next sections, this latter trend suggests that additional packing motifs, outside of the scope of strictly medial packing, are likely important for DD and DP, at least in this low-$\epsilon$ regime, whereas the overall consistency between mSST and SCFT predictions for DG over the full $\epsilon$ regime suggests that this morphology more closely follows medial packing in the strong-segregation limit.


\section{Subdomain packing geometry}

The thermodynamic comparisons of the previous section imply that medial packing accounts for a significant relaxation of the free energy of all networks relative to the skeletal packing ansatz, although that relaxation is less prominent for DD and DP than for DG, which gains thermodynamic stability for modest values of elastic asymmetry\footnote{Prior SST models did not predict stability window for neat diblocks for any network phase for $\epsilon \lesssim 9$.}.  
Moreover, large-$\chi N$ SCFT predictions for the free energy dependence on elastic asymmetry depart significantly from mSST predictions in the low-$\epsilon$ for DD and DP, suggesting that optimal modes of packing are likely to deviate from medial motifs.  
In this section, we analyze geometric features of the competing packing as predicted by both mSST and large-$\chi N$ SCFT in order to quantify the morphological signatures of packing frustration and understand the role of medial packing in the formation of different networks.

\begin{figure}[h!]
\centering
\includegraphics[width=3in]{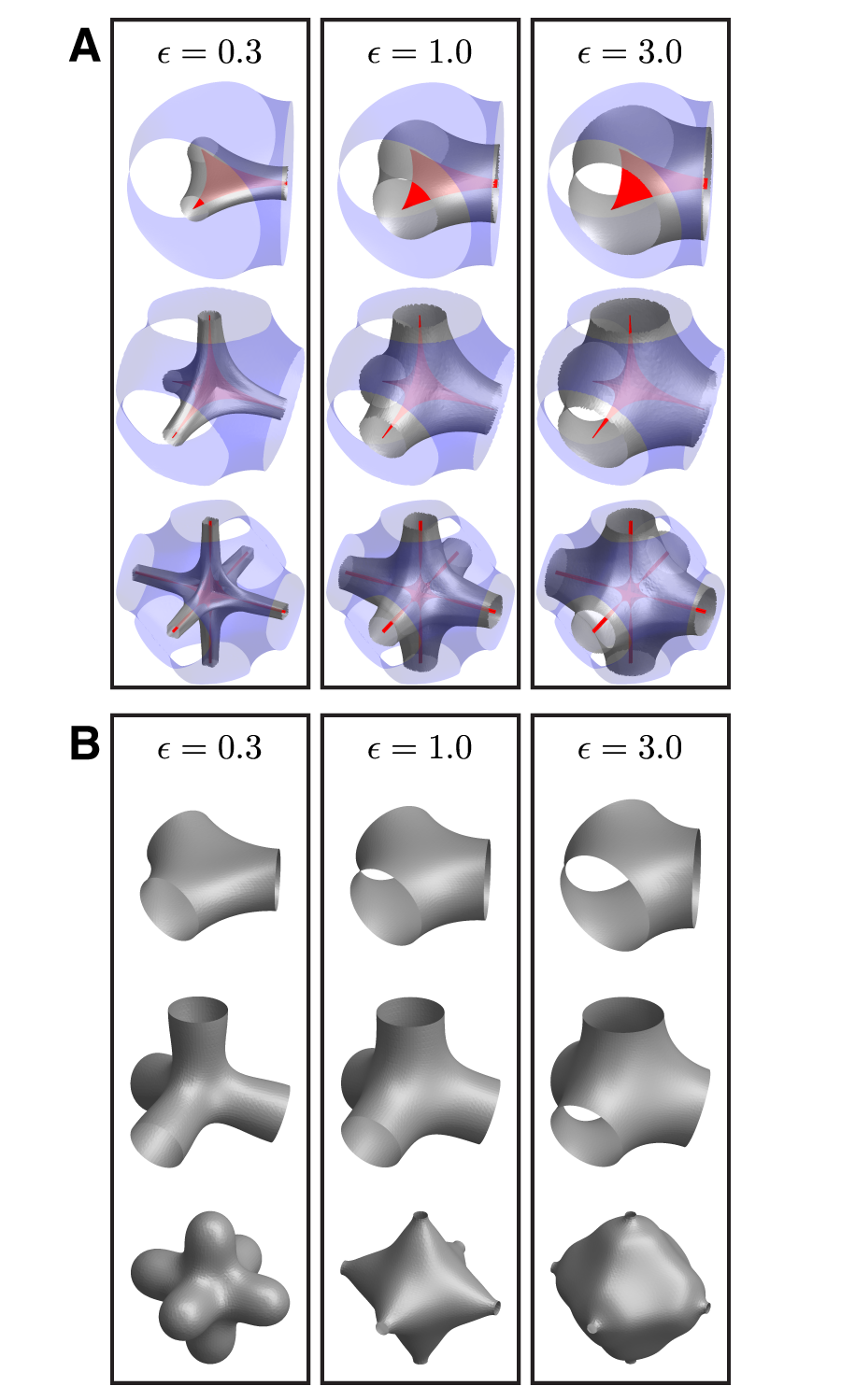}
\caption{\label{fig:imds_comparison} Morphologies predicted by (A) mSST and (B) SCFT at $\chi N = 75$ for various levels of elastic asymmetry $\epsilon \in \{ 0.3,\, 1.0,\, 3.0 \}$ at points were Lam and Hex have equal free energy. Note: for $\epsilon = 3.0$, due to numerical convergence issues, the $\chi N = 72.5$ DP structure is shown. The matrix medial surface is rendered translucent blue, the volume-balanced IMDS is rendered in a semi-opaque gray, and the tubular medial surface is red. Top row: DG mesoatoms; Middle row: DD mesoatoms; Bottom row: DP mesoatoms.  Note that mSST and SCFT predictions are carried out at different compositions, corresponding to differences in where free energies of Lam and Hex are equal, as indicated in Fig.~\ref{fig:free_energy_fixed_eps}.  This leads to visually obvious difference in minority domain thickness at low $\epsilon$.}
\end{figure}

\subsection{IMDS shapes}

We first consider the geometry of network morphologies by comparing the IDMS shapes in regimes where networks compete for stability (i.e.~at compositions where Lam and Hex are degenerate) for three different values of elastic asymmetry, spanning from stiffer tubular A-block to stiffer matrix B-block.  
In Fig.~\ref{fig:imds_comparison}A, we show the IMDS shapes for DG, DD and DP predicted by mSST, superposed with the respective terminal boundaries for A- and B-blocks. 
Foremost, these show a clear effect of the variable tubular domain fraction, with morphologies varying from relatively ``slender'' tubular domains of $f\approx 0.07$ for $\epsilon = 0.3$ to ``majority tubular'' networks at $f\approx 0.55$ for $\epsilon = 3.0$.  
In these examples, and most obviously for small $f$, it can be observed that the A domain sheaths the A-block terminal webs, leading to an IMDS that maintains roughly constant distance from the closest span of the web.  
For the cases of DD and DP, which have ``interior corners'' in their terminal webs, this sheathing leads then to inward puckering of the IMDS at positions where three leaves of the terminal web meet (i.e.~above nodal centers along $\langle 111\rangle$ directions).  
The smooth, corner-free terminal web of DG requires no inward curvature of the IMDS above the node centers.  

For comparison, we show also the IMDS shapes from $\chi N =75$ SCFT predictions in  Fig.~\ref{fig:imds_comparison}B (for other values of $\chi N$, see \note{SI Figs.~S2-S4}).  
These show that the gross features of the IMDS shape, and its variation with elastic asymmetry and composition, are well-captured by mSST for DG, as well as DD at least for cases of stiffer matrix blocks (i.e.~$\epsilon > 1$).  
SCFT calculations of IMDS shapes of DP show the greatest departure from mSST predictions.  
On one hand, the degree of inward pucker of the IMDS above the node centers is much less, if not outwardly curved, in SCFT solutions than is predicted by mSST.
Additionally, SCFT solutions show a greater variation of the local thickness of A-block domains {\it along the struts}, with distance from strut to IMDS dropping substantially more in SCFT solutions than mSST predictions.  
As discussed below, we attribute this difference to the extreme thermodynamic costs of packing the interior corners of the DP node, which are likely sub-optimally resolved by a strictly medial arrangement and exhibit additional degrees of freedom outside of the variational class of mSST morphologies studied here.  
Indeed we note that the bicontinuous network topology of DP actually becomes unstable for sufficiently low elastic asymmetry. 
The interconnected 6-valent network A-domains pinch off into disjoint and highly non-convex closed shapes (with a BCC symmetry) for $\epsilon \lesssim 0.8$ (see \note{SI Fig.~S4}).  

\subsection{Terminal boundary geometry}

\begin{figure}[h!]
\centering
\includegraphics[width=\textwidth]{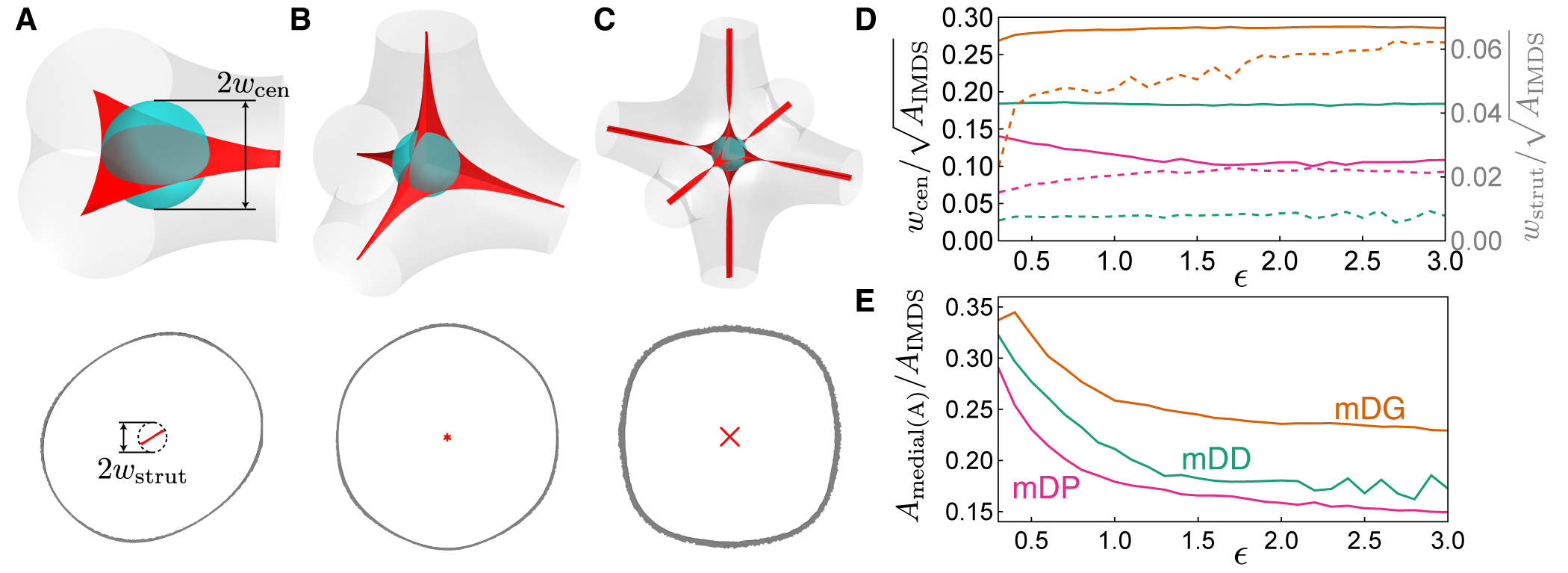}
\caption{\label{fig:medial_geometry} Geometric measures of the tubular medial surface. We define a center width $w_{\rm cen}$ and a strut width $w_{\rm strut}$ for the medial surfaces of (A) DG, (B) DD, and (C) DP. (D) shows plots of widths measured relative to the linear dimension of the IMDS, with $w_{\rm cen}/\sqrt{A_{\rm IMDS}}$ shown with solid curves (left axis) and $w_{\rm strut}/\sqrt{A_{\rm IMDS}}$ shown with dashed curves (right axis) along the Lam/Hex boundary. (E) shows a plots of the total area of the tubular medial surface $A_{\rm medial(A)}$ relative to $A_{\rm IMDS}$.}
\end{figure}

We next analyze the shapes of terminal boundaries predicted by mSST.  
As shown in Fig.~\ref{fig:imds_comparison}A above, the terminal boundaries of the B-block matrix domain maintain shapes that largely conform the corresponding triply-periodic minimal surface shapes, changing little over the large range of composition and elastic asymmetry explored.  
Hence, in Fig.~\ref{fig:medial_geometry}, we focus our attention on the previously unexplored shape aspects of the web-like, tubular A-block domain terminal surfaces, and their more obvious variations with BCP structure.  
The DG terminal sets are single sheets that twist as they connect between neighboring nodes, whereas the DD and DP medal sets have fins that intersect along the skeletal graph.
Since the skeletal graph lies within each terminal set, the distribution of terminal set widths can provide a measure of departure from skeletal packing.
We focus on the terminal set width measured at two locations of interest: about the node center and at the midpoint of a strut.
Here, the width is defined as the minimum radius of a sphere, centered at either the node or mid-strut ($w_{\rm cen}$ and $w_{\rm cen}$, respectively), that intersects with the boundary of the terminal set, as shown in Fig.~\ref{fig:medial_geometry}A-C.
In Fig.~\ref{fig:medial_geometry}D, we plot these two widths in comparison with a characteristic length of the IMDS (given by $\sqrt{A_{\rm IMDS}}$) for each mSST equilibrium structure along the Lam-Hex boundary.  

First, we note that DG has the largest terminal web in terms of its relative area compared to the IMDS.  
This is consistent with the observation from Fig.~\ref{fig:free_energy_fixed_eps}A above that the medial packing benefits DG relative to skeletal packing more than for DD and DP.  
That is, spreading of terminal ends away from the 1D skeletal graphs lowers the entropic costs of filling space, and crudely stated, the larger the terminal web, the greater the free energy relaxation from the skeletal ansatz.  

The larger terminal web for DG is somewhat counter intuitive, since the additional fins of the DD and DP webs would presumably raise their total area.  
The reduced area of the terminal webs of DD and DP relative to DG can therefore be attributed to a much narrower widths.  
The width $w_{\rm cen}$ characterizes the width of the terminal webs that meet at the center of the A-block terminal surface. 
We see that the DG terminal surface, which consists of a single contiguous flat piece, is the widest of the three, whereas the DP terminal surface, which has eight planar fins per node, has the smallest width (the inner terminal web for DD has four fins per node).  
As we discuss below, the corner-free surface of DG allows for quasi-planar packing of chains on either side of the 3-fold junction, whereas the interior corners of the DD and DP terminal boundaries require chain trajectories to tilt relative to the boundaries in order to fill those inner central points.  
Tilted chains in these ``multi-finned'' webs are relatively extended and the degree of tilt at the center can be expected to increase as the interior angle between fins {\it decreases}.  
Hence, we expect an entropic penalty for widening the central portion of the terminal web that progressively grows with the number fins meeting at the node, and substantially suppresses the relative size of $w_{\rm cen}$ for DD and DP relative to DG.
The strut width $w_{\rm strut}$ decreases as the minority domain is made stiffer for both DG and DP, which is likely due to imprinting of rotational symmetry of the web on the sheathed IMDS.   
At low $\epsilon$, when the domain thickness is small, this results in faceting of IDMS shapes, apparent in the mis-strut cross-sections, as shown in Fig.~\ref{fig:medial_geometry}A-C, reflecting the $n$-fold symmetry of terminal webs.  
Rapid undulations in the IMDS curvature caused by these facets introduce unfavorable interfacial cost, which in turn favors narrowing of the terminal web mid-strut. 
Notably, the DD phase always has the narrowest medial surface along the struts. 
This is due to the inversion symmetry about its strut center, which results in the disappearance of the dominant 3-fold rotational symmetry of the generating surface, leaving a generating surface that has a lower-order 6-fold rotational symmetry that is seen in the 6-fold symmetry of the medial surface.

As previously noted, for DG, $w_{\rm strut}$ drops substantially (a nearly 2-3 fold reduction), for low $\epsilon$, while $w_{\rm cen}$ retains a large width relative to the IMDS size.
This transition suggests that the smooth, single-leaf geometries of the inner web of DG permits a simple adaptation to the prerogatives of both stiffer matrix {\it and} stiffer tubular block chains, which is not available for the more complex, multi-leaf webs of DD and DP.
Furthermore, we attribute the re-entrant stability of DG relative to Lam and Hex for low $\epsilon$ to the mitigation of the otherwise destabilizing effects of variable stretching of tubular blocks, facilitated by the narrowing of $w_{\rm strut}$ when the A block becomes shorter and stiffer.

\subsection{Variable IMDS curvature}

We next consider packing frustration as measured through the mean curvature $H$ of the IMDS.  
A long-standing heuristic for understanding self-assembled amphiphilic phases focus on area-minimizing interfaces, which are favored in situations dominated by interfacial free energy.  
Ignoring, to a first approximation, the additional costs of chain packing and simply considering shapes that optimize IMDS area subject to a global volume constraint on adjoining subdomains would suggest that the optimal IMDS shapes are in the family of constant-mean curvature (CMC) surfaces \cite{Thomas1988}.  
However, it has been recognized that for BCP melts in general and for complex structures like bicontinuous networks in particular, variations in the local packing environments involve variations in the balance between chain entropy and interfacial enthalpy, which result in local departures of the IMDS from strictly area-minimizing (i.e.~CMC) shapes.
Thus, the variance in mean curvature, $\langle \Delta H^2 \rangle \equiv \langle H^2 \rangle - \langle H \rangle^2$, can be regarded as a measure of packing frustration.  
This variance of mean-curvature was first quantified by Matsen and Bates based on intermediate-$\chi N$ SCFT predictions \cite{Matsen1996}.  
More recently, both experimental tomographic reconstructions of DG assemblies, as well as SCFT calculations over a larger range of composition, segregation and conformational asymmetry, suggest the deviation from or agreement with CMC-like curvature itself varies considerably with interaction and structural parameters of the chains \cite{Feng2019}.

\begin{figure}[h]
\centering
\includegraphics[width=3in]{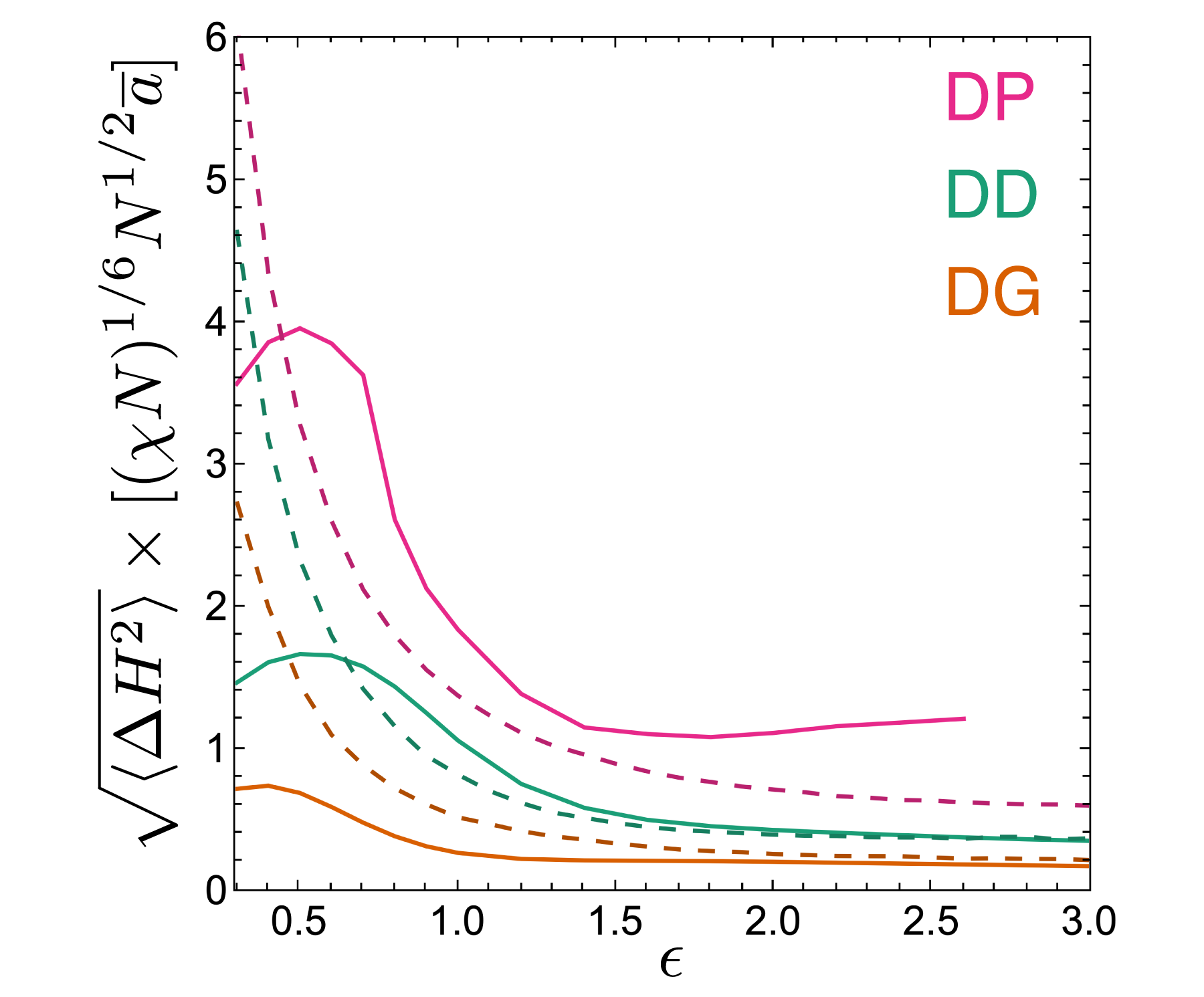}
\caption{\label{fig:curvature} Standard deviation of the mean curvature $H$ for structures computed by SCFT at $\chi N = 75$ (solid curves) and mSST (dashed curves) along the Lam/Hex phase boundary.}
\end{figure}

As shown in Fig.~\ref{fig:curvature}, the mean curvature variance $\langle \Delta H^2 \rangle$ of the DG phase is consistently smaller than that of the DD and DP phases along the Lam-Hex phase boundary.
Here, we compare the results of mSST calculations to SCFT calculations at finite segregation $\chi N$ and find good agreement, particularly for large values of $\epsilon$.
The most significant differences appear in the DP phase, for which the IMDS calculated using mSST is qualitatively different from that calculated using SCFT, which exhibits a topological transition for $\epsilon \lesssim 0.8$.
Nevertheless, since mSST involves a direct construction of the variable packing environments, it provides a clear picture of the sources of $\langle \Delta H^2 \rangle$.
These variations arise from the major geometric differences between the smooth matrix medial surface and tubular medial surface, which has edges and corners.
The elastic asymmetry parameter $\epsilon$ adjusts which medial surface plays the dominant role in determining the geometry of the IMDS.
When the tubular domain is stiffer ($\epsilon < 1$), the chains conform to the tubular medial surface, forming a layer with minimal thickness variations, as shown in Fig.~\ref{fig:imds_comparison}A, and thickness variations are relegated to the matrix domain.
Conversely, when the matrix domain is stiffer ($\epsilon > 1$), chains conform to the matrix medial surface.
Since the matrix medial surface is a CMC surface (the $H = 0$ surface), the IMDS is closer to CMC when the matrix phase is stiffer, but develops larger curvature variations due to the singular features of the tubular medial surface as the tubular domain grows stiffer (i.e.~decreasing $\epsilon$); this is clearly seen in the mean and Gaussian curvature distributions in \note{SI Figs.~S5 and S6}.
This suggests a simple heuristic for the stability of DG over DD and DP based on how singular the tubular medial surfaces are.
The DG tubular medial surface is a twisting ribbon that promotes the development of lamellar-like packing environments and reduced curvature of the IMDS, with edges that lead to saddle-like curvature of the IMDS (as illustrated in Fig.~\ref{fig:network_geometry}B).
Meanwhile, the DD and DP tubular medial surfaces have multiple flat, planar patches that intersect in creases and corners, which are regions of singular curvature that lead to strong curvature variations on the IMDS.
Due to the tetrahedral coordination of the DD nodal region, the planar patches intersect at a larger angle than those of the DP, and thus the resulting curvature variation of the IMDS is reduced in comparison with the DP.

Note that the SCFT calculations show a non-monotonic decrease in curvature variations for decreasing $\epsilon$ that is not captured by mSST.
This highlights a regime in which the medial map is not an accurate representation of chain trajectories, which we discuss later.

\subsection{Chain tilt}

Here, we analyze distributions of chain tilt with respect to the IMDS.  
On one hand, non-zero local tilt implies that chain stretching and IMDS area per chain are locally enhanced relative to strictly normal packing; hence, variable tilt in a complex packing is an elementary signature of packing frustration.  
Beyond this, local tilt is a basic consequence of local volume balance in the medial construction underlying mSST for diblock melts.  
Notably, variable chain tilt has been well-appreciated as a mechanism to relax frustration in bilayer models of lyotropic surfactant assemblies, as a means of optimizing the compromise between favorable curvature and thickness subject to constraints of filling space in regions occupied by solvophobic chains.\cite{Hamm1998,Chen2017}  

\begin{figure}[h!]
\centering
\includegraphics[width=\textwidth]{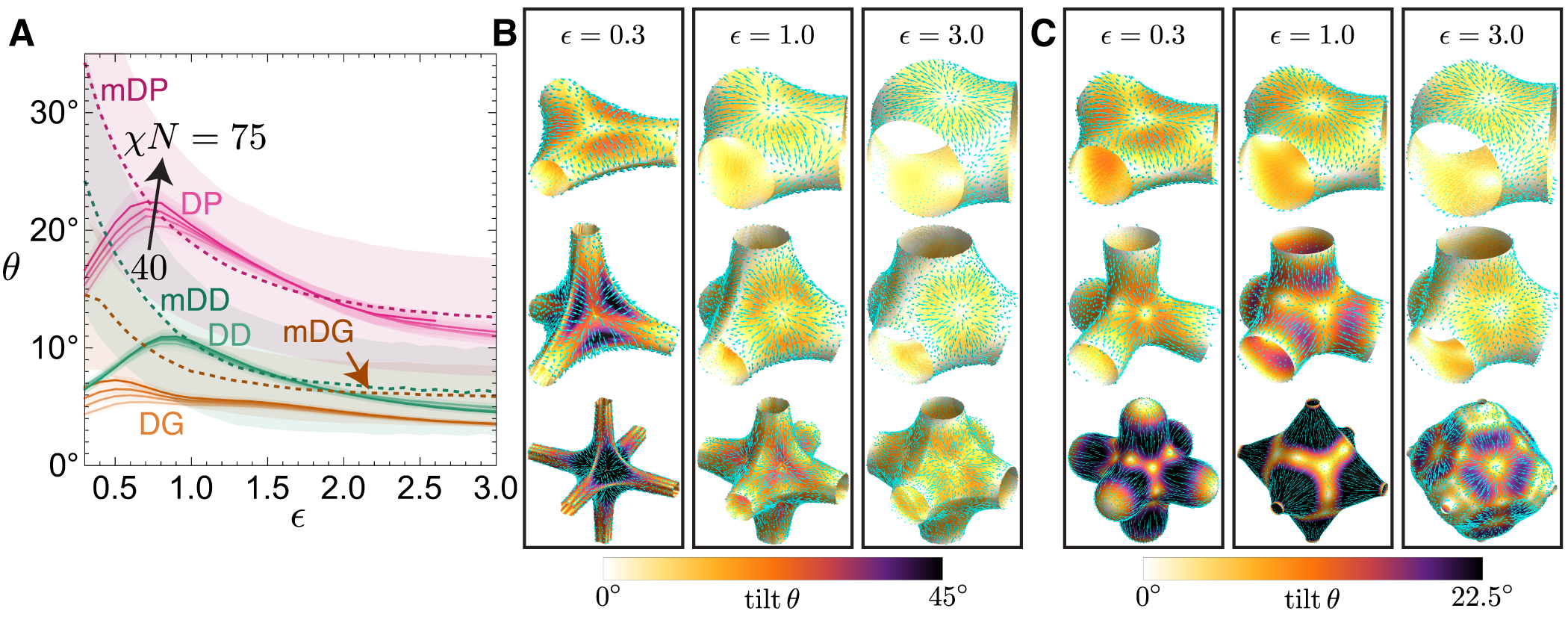}
\caption{\label{fig:tilt} 
(A) Average tilt angle as a function of elastic asymmetry $\epsilon$ along the Lam/Hex boundary; shaded region has a width of one standard deviation on either side of the mean. Dashed lines correspond to mSST results whereas solid lines correspond to results of SCFT calculations at $\chi N = 75$. (B,C) Spatial maps of chain tilt, computed from mSST (B) and SCFT (C) at $\chi N = 75$ (note: due to convergence problems, the $\epsilon = 3.0$ DP structure is actually $\chi N = 72.5$). The color gradient shows the magnitude of the local tilt and the arrows show the direction of the tilt, projected onto the the IMDS. Note that the arrow lengths are scaled independently for each surface for ease of visualization.
}
\end{figure}

The mSST approximates chain trajectories as lying along a collection of straight-line paths joining the two terminal surfaces, with those terminal surfaces deriving from the medial map of the generating surface, which ultimately differs, at least slightly, from the IDMS. 
The geometry of chain trajectories is modeled by the tessellation of wedge-like volumes spanning between the A and B terminal boundaries. 
As depicted in Fig.~\ref{fig:wedges_schematic}, each wedge is a thin bundle of these straight-line paths with orientations that vary gently about a local mean orientation, supplied by centroidal vector $\mathbf{C}$ joining the centroid of the wedge's triangular facet on the tubular terminal surface to that of the matrix terminal surface, with the mean chain trajectories in the wedge extending along $\hat{\mathbf{C}}$. 
As described above, due to the local volume balance constraint requiring the relative volume of A to B portions to satisfy $f:(1-f)$ at all points, the ultimate IMDS patch within each wedge is typically tilted relative to the average chain trajectory.  
This tilt, quantified in Fig.~\ref{fig:tilt}A is characterized by an angle $\theta$ between the local IMDS normal unit vector $\hat{\mathbf{N}}$ and the corresponding wedge centroidal unit vector $\hat{\mathbf{C}}$, such that $\cos \theta = \hat{\mathbf{N}} \cdot \hat{\mathbf{C}}$.
Since generating surfaces that yield a specific pair of medial surfaces have normal vectors that lie along the collection of centroidal vectors $\mathbf{C}$, the terminal surfaces of a tilted IMDS are generally not medial surfaces of that IMDS.
Therefore, while the medial map leads to polymer trajectories that minimize the cost of stretching from any given generating surface, these are not trajectories that minimize the cost of stretching to the IMDS.
In this sense, enforcing the local volume balance constraint \emph{frustrates} the ability of chains to lie along trajectories that optimize the cost of stretching, and tilt is a measure of this form of packing frustration (i.e.~deviation from strictly medial packing).  

A similar notion of tilt can be computed at finite-$\chi N$ using results of SCFT. 
Here, tilt is given by the the average chain orientation near the IMDS, which is supplied by the polar order parameter $\mathbf{p}$; this is computed using the method developed by Prasad, et al\cite{Prasad2017}.
The tilt angle $\theta$ is then the angle between $\hat{\mathbf{p}}$ evaluated at the IMDS and the local surface normal $\hat{\mathbf{N}}$ via $\tan \theta = \hat{\mathbf{N}} \cdot \hat{\mathbf{p}}$.

We show the mean tilt at the IMDS (relative to the local normal) for all three competitor networks in Fig.~\ref{fig:tilt}A for both mSST and $\chi N =75$ SCFT.  
The mean tilt angle $\theta$ decreases monotonically with increasing $\epsilon$ along the Lam-Hex degeneracy line, which indicates a larger effect of packing frustration for lower A-block fraction $f$.  
Chain packing, on average, tends more towards being strictly medial in the regime when matrix domains have relatively stiffer segments and larger domain fractions.  
Out of the three network morphologies, DG is least tilted ($\lesssim 10^\circ$) and DP is most tilted ($\approx 15-25^\circ$), particularly for low $\epsilon$, suggesting that this packing frustration effect is particularly sensitive to the geometry of the tubular medial surface for low $f$.   
Notably, the degree of tilt measured from SCFT is always at least slightly less than mSST predictions, suggesting a measure of relaxation relative to the mSST packing, particularly for the low-$\epsilon$ regime.  
As illustrated in Fig.~\ref{fig:imds_comparison}A above, in this narrow-tube regime, mSST predicts that the narrow A-block domains tightly wrap the terminal webs.  
We note that the magnitudes of variable tilt, as well as the much larger value for DP, for these diblock melt predictions are generally consistent with elastic bilayer model calculations of so-called ``normal'' lytropic cubic phases, where the hydrocarbons occupy the tubular regions (predicted tilts are considerably lower for the inverse phases\cite{Chen2017}).


Local distributions of tilt, as shown in Fig.~\ref{fig:tilt}B, show the tilt texture on the IDMS, highlighting where this frustration of medial packing is the largest.  
Notably we find basic agreement in the spatial patterns between mSST and SCFT (shown in Fig.~\ref{fig:tilt}C), as least for DG and DD structures, particularly for larger values of $\epsilon$.  
For each structure, the tilt is maximal for chains that terminate on the flat portions of the tubular medial surface surrounding the nodal center, but reaches local minima near the rotational symmetry axes along $\langle 111\rangle$, including the regions on the IMDS where chains stretch to the nodal center of the tubular medial surface.  
These high symmetry points are sources and sinks of the vector field $\mathbf{C}_{\perp}$, obtained by projecting the molecular orientation onto the surface via $\mathbf{C}_{\perp} \equiv \mathbf{C} - (\hat{\mathbf{N}}\cdot \mathbf{C})\hat{\mathbf{N}}$.

The qualitative features of the tilt patterns and the differences between tilt magnitudes between competing cubic networks derive from features of the underlying medial map that serves as template for chain packing.  
In particular, chain trajectories are, in general, inclined with respect to the inner (A block) terminal webs that derive from the medial surfaces of the tubular network generating surfaces.  
Since the IMDS tends to envelop these medial webs, the tilt pattern relative to the webs tends to imprint onto tilt at the IMDS, particularly as the A-block composition (and thus the sub-domain thickness) decreases (see e.g.~\note{SI Fig.~S7}).  
The respective magnitudes of tilt can then be understood in terms of the inclination of the trajectories with respect to the medial webs, which is minimal along the ``monkey saddle’’ directions, i.e.~the $\langle 111 \rangle$ axes that reach the node centers.  
While the medial web of DG is normal to this axis, the fact that webs of DD and DP are composed of multiple leaves implies non-zero inclination between the $\langle 111 \rangle$ and the web normals where those leaves join at the node center.  
Those inclination angles are $\arccos (\sqrt{2/3}) \simeq 35.3^\circ$ and $\arccos (\sqrt{1/3}) \simeq 54.7^\circ$ for DD and DP, respectively.   
Correspondingly, extent of chain tilt imprinted on the IMDS in mSST models, particularly near to the boundary between Lam and Hex at low $\epsilon$, is increasingly larger for DD, and then DP, relative to DG.  
This observation, combined with the fact that SCFT predictions show significant deviations for DD and DP morphologies from medial packing in this regime, confirm the connection between the optimal thermodynamics of DG and the smooth geometry of its inner (tubular) medial webs.

\subsection{Chain bending}

While the medial packing ansatz represents a significantly lower {\it upper bound} on the SST free energy for cubic network phases relative to the prior skeletal ansatz, several aspects of the comparisons between mSST and large-$\chi N$ SCFT predictions suggest that the straight-line trajectories are, at times, far from optimal.  
This is most notable for the cases of stiffer tubular domains for DD and DP which have more complex inner terminal webs and show departures from even qualitative dependence of free energies and IMDS curvature  on $\epsilon$ in these regimes.  
One alternative motif that has been proposed and previously explored is the possibility of chain ``kinking,'' in which chain trajectories bend sharply at the IMDS, which provides additional degrees of freedom for the morphology to relax its free energy at uniform filling.

\begin{figure}[h!]
\centering
\includegraphics[width=3in]{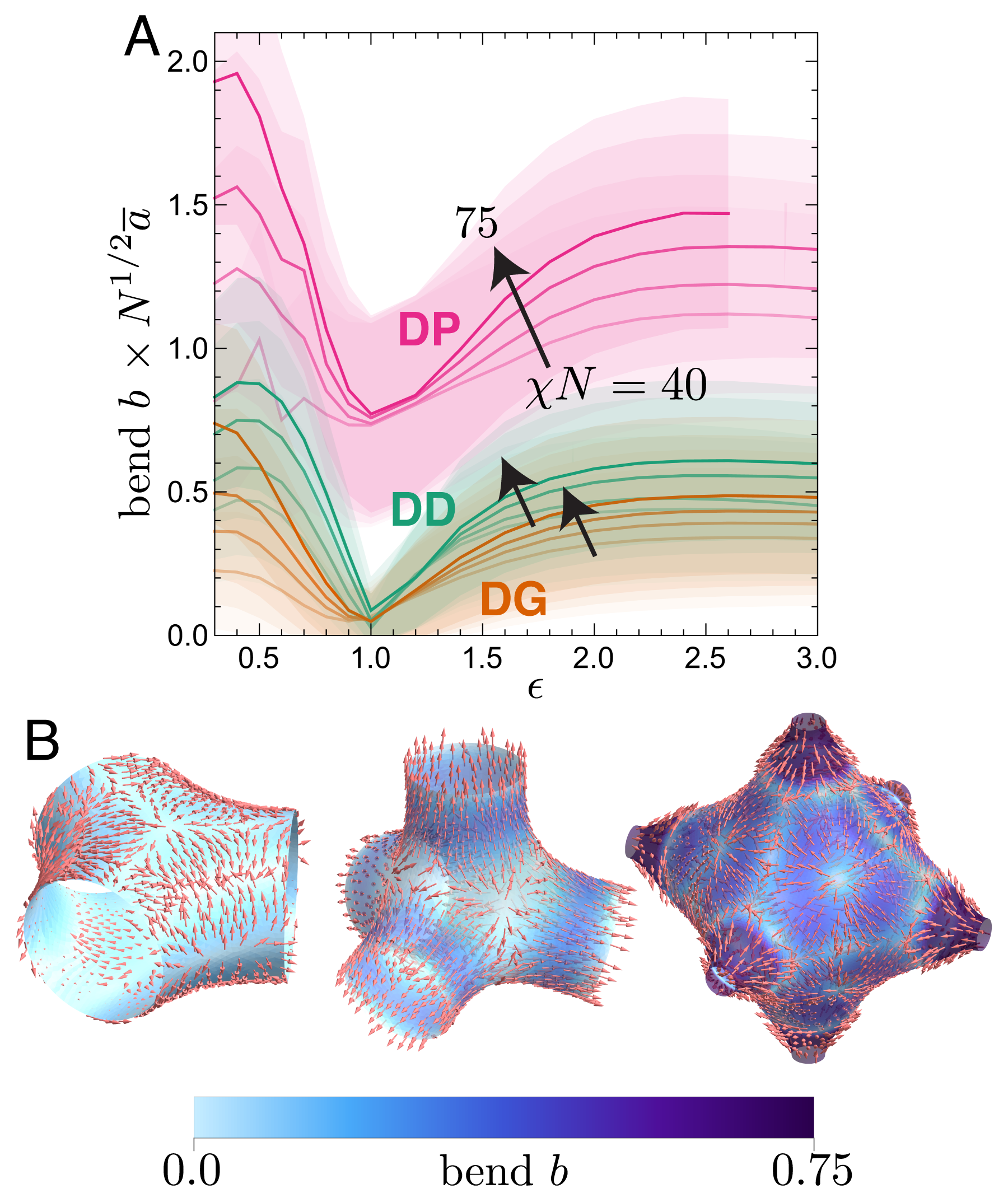}
\caption{\label{fig:bend} (A) Average bend $b$ measured at the IMDS for each phase, as calculated by SCFT for $\chi N = 40, 50, 60$ and 75. The bars represent a single standard deviation above and below the mean. (B) Distribution of bend, plotted on the nodal IMDSs at $\chi N = 75$, $\epsilon = 1$, $f \approx 0.32$. Arrows represent the direction in which chains deflect from a straight trajectory and the color gradient represents the magnitude of $b$.}
\end{figure}

One way of directly quantifying the expected departure of chain trajectories from straight (i.e.~medial) paths is by computing the average bend $b(\mathbf{r}) \equiv |(\hat{\mathbf{p}}\cdot\bm{\nabla}) \hat{\mathbf{p}}|$ of the trajectories predicted from SCFT, where $\mathbf{p}$ is polar order parameter.  
Notably, strictly straight-line trajectories correspond to $b(\mathbf{r})=0$.  
As a measure of potential kinking, we analyze the distributions of $b(\mathbf{r})$ at the IMDS, as the integrated bend through the IMDS (along chain trajectories) gives the bend angle of polar order parameter from A- to B-side of the subdomain.
The average molecular bend at the IMDS is shown in Fig.~\ref{fig:bend}A.
We observe that the average bend of the DG is always less than that of the DD and DP phases, suggesting that among competing networks it most closely follows the straight path ansatz of mSST.
Furthermore, the bend quickly increases as the tubular domain becomes stiffer than the matrix domain ($\epsilon < 1$) particularly for DD and DP, confirming our expectation that the IMDS warping due to the medial packing \emph{ansatz} is partially relieved by highly curved chain trajectories.

To further explore how chains bend at the IMDS, we map the magnitude $b$ and direction $\hat{\mathbf{b}}$ of the bend on the IMDS for $\epsilon = 1$ in Fig.~\ref{fig:bend}B.
These bend maps reveal distinct patterns of minimal and maximal bend, with bend reaching a minimum along the $\langle 111 \rangle$ directions and chains generally curving away as they pass from the A-block subdomain to the B-block subdomain.
The enormous bend near the mid-strut region of the IMDS for DP, along with the bend direction, suggests that the A-block chain ends are drawn towards the node even when the junctions are closer to the strut regions of the IMDS.
This is further supported by the substantial difference in A-block domain thickness near the node as compared with the thickness of the mid-strut and may provide a rationale for the change in topology of the DP phase when the A-block chains are increased in stiffness.
Indeed, we see a dramatically different distribution of bend for $\epsilon = 0.3$, as shown in \note{SI Fig.~S8}.

Incorporating chain bending into SST remains challenging and has been studied in most detail for 2D phases, using the so-called ``kinked path'' ansatz for computing the free energy of cylinder phases \cite{Olmsted1998, Grason2004}.  
These results show that packing frustration introduced by the corners of Voronoi cells in the cylinder lattice can be accommodated by tilting the chain trajectories in the matrix domain while maintaining purely radial paths in the cylinder domain.  
An effectively kinked path construction was used by Likhtman and Semenov for network phases, but with A trajectories extending skeletal terminal boundary \cite{Likhtman1994}.  
Comparing these results for $\epsilon=1$ it would appear that this ansatz led to free energies that are almost identical to the straight-path, skeletal ansatz of Omsted and Milner, and still $~2-3\%$ larger than mSST, which suggests that relaxation of the A block terminal boundary may be much more important for at least the DD morphology under the conditions reported in ref.~\cite{Likhtman1994}. 

The large values of bend for cases of elastic asymmetry, particularly for DP, suggest that incorporation of kinked paths into mSST would allow for significant structural relaxation.
From the bend, we can estimate an effective ``kinking angle'' at the IMDS, approximated by the bend $b$ times the thickness $w$ of the IMDS, which is approximated by the gradient of the A-block density field, evaluated at the IMDS, i.e.~$w\sim |\nabla \phi|^{-1}_{\rm IMDS}$.
Averaging the product $b/|\nabla \phi|_{\rm IMDS}$, we find a lower range estimate of $\sim 0.3^\circ$ for DG at $\epsilon = 1.0$ and an upper range estimate of $\sim 20^\circ$ for DP at $\epsilon = 0.3$ (both at $\chi N = 75$).

In general, a kinked path can relax the cost of tilting chains at the IMDS, allowing a single block to be less tilted at the cost of making the other block more tilted.
In the case of elastic asymmetry, it is more favorable for IMDS to be oriented such that the tilt of the stiffer block is minimized, leading to increased tilt of the softer block which effectively ``absorbs" the more of consequences of packing frustration, resulting in greater kinking of the chain path.
This is supported by an observed reversal in the bend direction as the B-block becomes stiffer than the A-block (see \note{SI Fig.~S8}).
The scale of this effect depends on the ratio of the entropic stiffnesses of the two blocks and is somewhat analogous to the refraction of light, arising from differences in the phase velocity of electromagnetic waves passing through different media. 
Indeed, we find that the bend is minimal in the case of elastic symmetry ($\epsilon = 1$), but the persistence of non-zero bend points towards further dependence on chain length constraints as well as non-local space filling requirements.

\section{Subdomain chain thermodynamics}

We now turn to the distribution of thermodynamic costs of BCP chain filling based on mSST models of cubic networks.  
Here, the goal is to analyze the optimal free energy compromise between (interfacial) enthapic and (stretching) entropic driving forces in each morphology, how those distinct thermodynamic costs are distributed spatially in the structure, correlating with geometrical features of packings, and how to assess the role of variance in these quantities in their thermodynamic stability relative to other morphologies.

\subsection{Co-variation of local chain enthalpy and entropy}

Here, we analyze the distribution of free energies per chain, to understand the relative variations of enthalpy and entropy in each structure.  
In essence, we consider the BCPs that are associated to particular point on the IMDS, i.e.~the chains whose junctions lie at a given point.  
Since each area element of the IMDS corresponds to a local density of chains, the enthalpic costs are proportional the IMDS \textit{area per chain}.  
Additionally, the A and B blocks departing from a given area element contribute to the strongly-stretched brush regions that extend up to the terminal boundaries of the associated subdomain, and are characterized by entropic costs that are proportional to the {\it stretching per chain}.  
While notions of packing frustration are often discussed and diagnosed in terms of particularly large costs for \textit{either} areal \textit{or} entropic (stretching) free energy, these are inextricably linked for any complex morphology.  
Regions that are locally extended, and therefore have pronounced stretching cost, also tend to have a correspondingly low area per chain. 
This is most obvious when considering locally flat (i.e.~lamellar) geometries with a given subdomain thickness, $h$.  
Stretching costs per chain grow as $h^2$, while area per chain falls off with extension as $1/h$.  
While this special relationship is altered somewhat in non-flat geometries (e.g.~cylinder, spherical or saddle-like volume elements), it generally holds that, locally, the area per chain and stretching cost will be anti-correlated in a given structure.  
However, at the same time, as we illustrate below, thermodynamic equilibrium in the strong-segregation limit \cite{MatsenBates1997} requires that on {\it average} (over any an entire morphology) the mean costs of stretching and interfacial enthalpy maintain exact proportion.  
Taken together, as we shall show, these two features require that the local entropic and enthalpic costs exhibit a complex co-variation within a given morphology, leading to a nuanced picture of inhomogeneous packing thermodynamics and its interpretation in the context of packing frustration.

Our mSST construction decomposes space into narrow wedge-like packing environments, the $\mu$\textsuperscript{th} wedge corresponding to a chains associated to a specific IMDS point.   
The (per chain) free energy $\hat{F}$ in a given morphology is a sum over free energy contributions from each wedge, i.e.~$\hat{F} = n^{-1}_{\rm ch}\sum_{\mu} n_{{\rm ch},\mu}\hat{F}_\mu$, where $\hat{F}_\mu$ is the free energy per chain (addressed to IMDS location $\mu$) and $n_{{\rm ch},\mu}$ is the mean number of chains in the wedge (i.e.~its volume times $\rho_0/N$) and $n_{\rm ch}=\sum_{\mu} n_{{\rm ch},\mu} $ is the total chain number.
The IMDS-addressed free energy per chain is $\hat{F}_\mu = \hat{S}_\mu + \hat{H}_\mu$ and has two parts: a stretching free energy per chain,
\begin{equation}
\hat{S}_{\mu} = \frac{\pi^{2/3}}{4}\overline{\lambda}^2 \frac{\overline{\kappa}_{\rm A} \tilde{I}_{{\rm A},\mu} + \overline{\kappa}_{\rm B} \tilde{I}_{{\rm B},\mu}}{\tilde{V}_\mu} \left[(\chi N)^{1/3}k_{\rm B} T\right] \, ,
\end{equation}
and an enthalpy per chain,
\begin{equation}
    \hat{H}_\mu = \frac{\pi^{2/3}}{2\overline{\lambda}}\frac{\overline{\gamma}\tilde{A}_\mu}{\tilde{V}_\mu}\left[(\chi N)^{1/3}k_{\rm B} T\right] \, ,
\end{equation}
where $\tilde{A}_{\mu}$ is the area of the IMDS surface patch, $\tilde{V}_\mu$ is the corresponding wedge's total volume, and $\tilde{I}_{{\rm A},\mu}$ and $\tilde{I}_{{\rm B},\mu}$ are the second moments of volume of each subdomain of the wedge.
While the IMDS-addressed free energy involves the local geometry of each wedge, information about the full structure is encoded in the equilibrated value of the scale factor $\overline{\lambda}$, given by Eq.~\ref{eq:scaling_factor2}.  
Note that we may consider free energy per chain as a function of size, by rescaling its dimensions by a factor $\lambda/\overline{\lambda}$, which leads to 
\begin{equation}
\label{eq: flam}
  \hat{F}(\lambda/\overline{\lambda}) = \left(\overline{\lambda}/\lambda\right)\hat{H} + \left(\lambda/\overline{\lambda}\right)^2\hat{S} , 
\end{equation}
where $\hat{S} = n^{-1}_{\rm ch}\sum_{\mu} n_{{\rm ch},\mu} \hat{S}_\mu$ is  \emph{average} entropy per chain and $\hat{H} = n^{-1}_{\rm ch}\sum_{\mu} n_{{\rm ch},\mu} \hat{H}_\mu$ \emph{average} enthalpy per chain.  
Since equilibrium requires optimality with respect to size, in this case occurring for $\lambda=\overline{\lambda}$ by construction, it is straightforward to see from the minimization of eq.~\ref{eq: flam} that average enthalpy, entropic cost and total free energy per chain maintain the relationship
\begin{equation}
  \hat{H} = 2\hat{S} = \frac{2}{3}\hat{F} .
\end{equation}
Notice that this strict proportionality on the averages holds notwithstanding the spread of local entropic and enthalpic costs throughout the assembly, facilitated by the adjustment of the equilibrium dimensions.  
According to $\overline{\lambda} \propto (\hat{H} /\hat{S})^{1/3}$, structures with particularly pronounced stretching or interfacial costs adjust equilibrium dimensions accordingly to maintain the thermodynamically optimal balance between enthalpy and entropy in each morphology. 

\begin{figure}[h!]
\centering
\includegraphics[width=\textwidth]{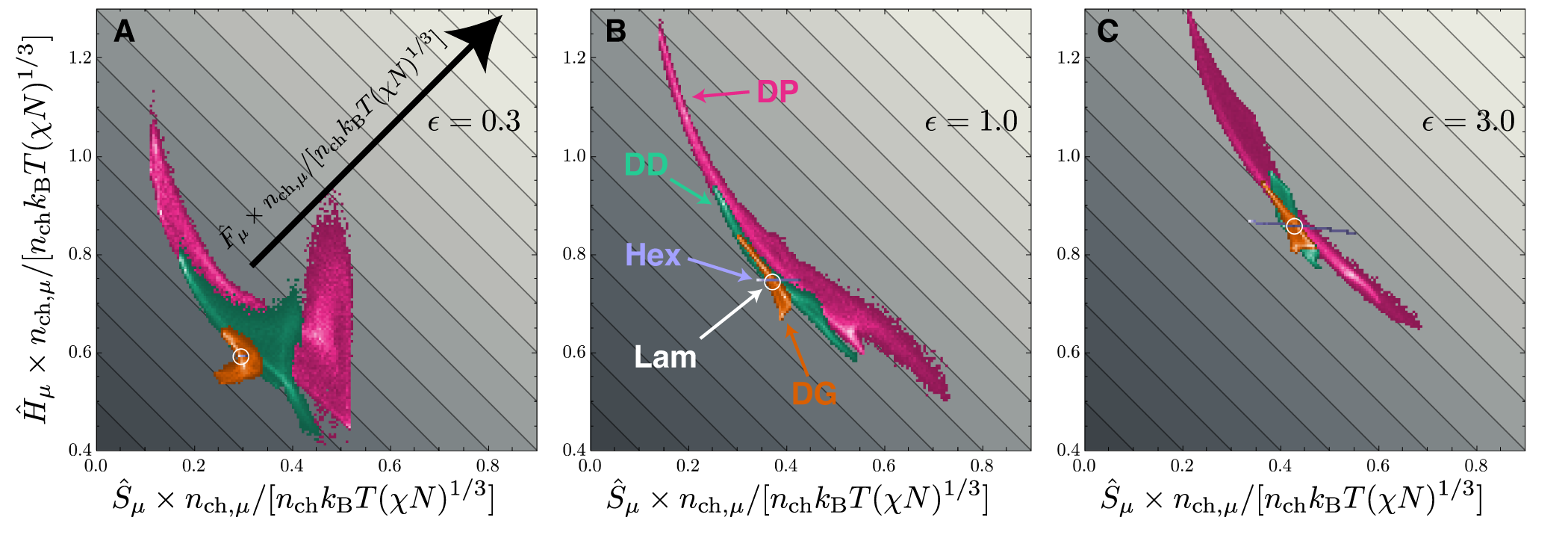}
\caption{\label{fig:correlation_figure} Distributions of IMDS-addressed ($\mu$ indexes specific regions of the IMDS) interfacial enthalpy $\hat{H}_\mu$ and chain stretching $\hat{S}_\mu$ contributions to the free energy per chain $\hat{F}_\mu$ from local packing environments for (A) $\epsilon = 0.3$, (B) $\epsilon = 1.0$, and (C) $\epsilon = 3.0$ along the Lam-Hex boundary. Gray-shaded contours are level sets of $\hat{F}_\mu$. The white circle is centered on the free energy contributions for the single lamellar environment for each value of $\epsilon$. Each distribution is a histogram, where lighter bins represent a larger population of packing environments with that bin's range of free energy components.}
\end{figure}

The full distribution of interfacial and stretching costs for each network phase, along with the lamellar and hexagonal cylinder phases are shown in Fig.~\ref{fig:correlation_figure} for three values of elastic asymmetry along the points of Lam-Hex degeneracy.  
These are 2D histograms in the {\it stretching}-{\it enthalpy} plane, with diagonal grey contours highlighting value the total free energy of each point (i.e. $\hat{F}_\mu = \hat{S}_\mu + \hat{H}_\mu$).

Since the Lam phase consists of a single repeated packing environment, its distribution consists of a single point, whereas the Hex phase occupies a thin band with nearly constant interfacial free energy (for $\epsilon = 1$ and $\epsilon = 0.3$), due to a nearly circular shape of the IMDS, and a variable stretching free energy due primarily to the variable stretching of the matrix domain.  
Note that for case of stiffer matrices ($\epsilon = 0.3$) the IMDS for Hex becomes fairly faceted, leading to a more obvious tilt of the distribution in the {\it stretching}-{\it enthalpy} plane, as well as larger horizontal spread that is due to the amplified effect of B-block chains stretching from the curved IMDS to the polygonal terminal boundary, which lies on the Voronoi cells generated by the cylindrical domains.

For the three network phases, we see a more prominent anti-correlation between local enthalpic and entropic free energy costs, with regions of relatively low stretching cost corresponding to relatively large enthalpy (i.e.~area per chain) and \emph{vice versa}.  
The DG phase has the narrowest distribution of free energy per chain among the network phases, and thus the least variability in thermodynamic costs of different packing environments, whereas the DP phase has the broadest.  
In \note{SI Fig.~S9}, we also compare the individual histograms of A- and B-block stretching free energy.  
Moreover, despite the variation in proportion of stretching to interfacial free energy costs, the free energy per chain of DG closely conforms to the $\hat{F}_\mu =  {\rm const.}$ contours for $\epsilon \gtrsim 1$, indicating a general uniformity of the {\it total} free energy per chain of each packing environment. 
Both DD and DP show notable tails of particularly low stretching and high enthalpy that lead to a much larger spread in the total free energy per chain ($\hat{F}_\mu$) in these networks than in DG.  
In fact, the DG phase is even more uniform than the hexagonal cylinder phase in this regime of elastic asymmetry, which has a broad distribution of free energy per chain.
By comparison, the DD and DP phases have regions of large enthalpic cost where chain stretching is minimal.
These extreme enthalpy regions cross multiple $\hat{F}_\mu =  {\rm const.}$ lines and keep the DD and DP phases from competing with the DG phase.
For $\epsilon = 0.3$, the DG phase has larger variation in free energy per chain, but importantly has an excess of regions where the entropic and enthalpic costs fall below lamellar, promoting its stability for low $\epsilon$.
Interestingly, in this same regime, the DD and DP phases develop regions of simultaneous large entropic and enthalpic cost.

These free energy distributions show that \emph{enthalpy} plays a significant role the variability of chain free energy in networks.
While there are broad variations in chain stretching costs, the largest contributions to the average free energy tend to correspond to smallest stretching (i.e.~upper left regions of the $\hat{S}_\mu - \hat{H}_\mu$ plane), with exceptions when the tubular domain is stiffer than the matrix domain, which shows large enthalpy contributions for DD and DP at both large and small $\hat{S}_\mu$.  
Additionally, we note that DG is unique among the networks in that its lowest enthalpy chains are not the maximally stretched chains.  
Next, we consider the spatial distributions of these thermodynamic packing environments for DG, DD and DP structures.

\subsection{Spatial maps of chain free energy}

\begin{figure}[h!]
\centering
\includegraphics[width=3in]{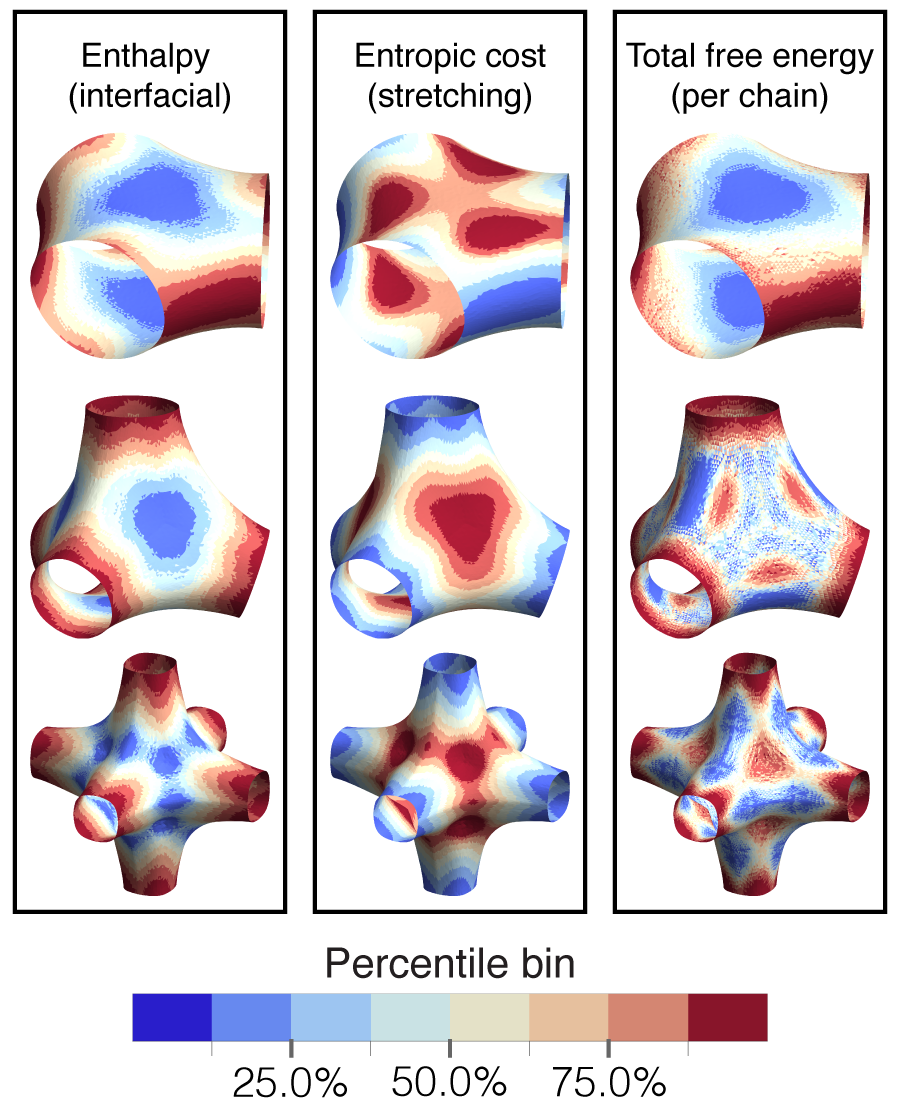}
\caption{\label{fig:free_energy_maps} Local free energy per chain contributions, mapped onto the IMDSs of the network phases for $\epsilon = 1.0$, $f \approx 0.29$. (A) shows the local interfacial enthalpy, (B) shows the local stretching free energy, (C) shows the total free energy per chain. Regions are binned according to the percentile rank of the corresponding free energy per chain component costs, such that the dark red corresponds to the regions containing the 12.5\% chains with the highest cost and the dark blue corresponds to the 12.5\% lowest cost.}
\end{figure}

Fig.~\ref{fig:free_energy_maps} shows a complementary view of the free energy per chain distribution and its components, mapped onto corresponding regions of the IMDS for each network structure, focusing on the case of $\epsilon = 1$ (see \note{SI Figs.~S10 and S11} for elastically asymmetric cases).
We find that the regions of maximal enthalpic cost are typically separated from regions of maximal entropic cost, which is consistent with their reciprocal relationship described above.

For DG, the regions of maximal enthalpic cost are located in the saddle-like ``elbow'' regions, whereas the regions of maximal stretch are along the flatter region, distributed around the 3-fold rotational symmetry axis, but notably away from the minimal enthalpy region centered above the 3-fold node.  
Hence, in contrast to the heuristic view that chains reaching to the node center suffer the largest stretching, we observe that maximal stretch for DG is associated with regions of high tilt in the quasi-lamellar packing in the 3-fold plane of the node.  
The distribution of free energy per chain shows that the enthalpic component dominates and the largest free energy per chain is in the elbow regions of the node.

By contrast, the DD and DP nodes have maximal enthalpy per chain halfway along the strut joining two nodal regions and maximal stretching entropy cost per chain on the monkey saddles of the IMDS, corresponding to chain trajectories that extend into the inner corners of the terminal webs at the node center (DP also has relatively large stretching costs in its elbow regions).
Due to the narrowing of the tubular portions of the IMDS, chains halfway along the strut are least stretched and thus, due to the incompressibility constraint, they occupy the largest area on the IMDS, leading to larger enthalpic costs.   
Quite unlike DG, the net free energy for DD and DP is smallest in the elbow regions.  
Also, the local maxima of total free energy is more complex, with large packing costs originating from both high entropic cost (on monkey saddles) as well as high enthalpic cost (on tubular struts).

While the normalized distributions shown in these maps depict high free energy costs of packing in the elbow regions of DG relative to other regions of the same morphology, it should be noted on absolute terms, packing is this region is comparable to the {\it median} value from chains in DD and the {\it minimal} value for chains in DP.

\section{Discussion}

\subsection{Medial strong-segregation picture of frustration in networks}

We have explored packing frustration in network phases through the medial SST model of diblock melt assembly, in comparison to SCFT predictions at finite, but large $\chi N$. 
The mSST construction assumes chain trajectories to be straight-line paths based on a prescribed association map between two terminal boundaries.  
Distinct from prior SST approaches that assumed the inner terminal boundaries to be 1D skeletal graphs of the networks, we explored the ansatz that the terminal surfaces are well-approximated by the medial surfaces of a class of generating surfaces whose symmetries are consistent with the network phase.  
Essentially, this construction is based on the premise that most efficient mode of chain packing in tubular network domains is to spread out termini in a geometrically optimal way, while maintaining co-linear trajectories in both blocks.  
This construction of the terminal surfaces is motivated by the fact that the medial map provides a distance-minimizing way of mapping all points the volumes on the generating surface.  
Assuming then that the generating surfaces roughly approximate the ultimate IMDS shapes, these maps would then provide minimal chain stretching with each subdomain.  
However, as we mentioned in our discussion of tilt, the condition of local volume balance to either side of the IMDS general requires some adjustment of this interface relative to the generating surface, so that the ultimate packing evaluated by SST is not strictly medial (i.e.~chain paths, in general, are not normal to the IMDS).  

Despite the ``frustration" of medial packing by local volume balance, we find in general that the upper bound on the free energy provided by mSST is substantially smaller (by several \%) than the SST calculations for all three cubic network phases based on the skeletal packing ansatz.  
Additionally, we find free energy predictions that are overall consistent with SCFT predictions (albeit at fairly modest range of finite $\chi N \leq 75$).  
Namely, the relative ratios of DG, DD and DP free energies predicted by mSST are reasonably comparable to SCFT.
Moreover, we find that DG enjoys thermodynamic stability between Hex and Lam for values of elastic asymmetry close to 1 and mSST captures the non-trivial, non-monotonic dependence of the free energy gap between DG and its competitor phases with elastic asymmetry.  

More careful comparison of both the free energy trends and the detailed geometric features of the morphologies predicted by mSST and SCFT suggest a level of disagreement that is both dependent on the network phase as well as the chain parameters.  
Notably, mSST appears to significantly depart from SCFT predictions for both DD and DP phases in the regime of $\epsilon \lesssim 1$ when tubular (minority) blocks are relatively stiff.  
Moreover, DD (for $\epsilon < 1$) and particularly DP show obvious differences in IMDS shapes between mSST and SCFT.  
These departures suggest that more complex packing motifs, such as kinked or bent chain trajectories, are needed to capture the full details of packing frustration and thermodynamics, at least for certain networks in certain parameter regimes.  
Indeed, SCFT shows that at least some measure of path bending for large $\chi N$ for all networks, but this is most significant for DD and DP and most prominent for the stiff minority regimes.  
All together, we understand this comparison to suggest that chain packing in DG is ``most medial'' among the cubic competitors.  
As such, the mSST model provides currently the most accurate and detailed picture of the underlying coupling between domain shapes, chain packing and thermodynamics that governs the formation of DG and its stability in diblocks, and to a large extent, its competition with sub-optimal DD and DP phases.  

In this context we largely attribute the stability of DG among cubic networks to the basic geometry of its terminal boundaries.  
In particular, the DG tubular medial surface, which approximates the terminal surface in the tubular subdomain, is a single twisting sheet without the singular features that plague the DD and DP tubular medial surfaces, namely corners.  
These singular ``interior corners'' result in pronounced variations in the mean curvature of the IMDS of DD and DP, which increases the interfacial contribution to the free energy.
Moreover, the local volume balance constraint results in an IMDS that is generally tilted relative to the chains, the consequence of which is that the medial map does not, in fact, minimize the distance from the IMDS to the terminal surfaces.  
The degree of tilt is larger for medial surfaces with singular features and is consequently largest for the DP phase.  
This frustration of medial packing consequently leads to substantially larger stretching free energy.  
In fact, we find that at finite segregation, chains will curve from straight paths, better optimizing the compromise between thermodynamics and the space filling constraint in the highly frustrated DP.  

In this light, mSST suggests that chain packing in the interior nodal volumes of DD and DP are indeed problematic, tending to destabilize these structures in neat diblock systems.  
This is consistent with predictions that DD and DP only become stable (if ever) in blend systems, presumably tailored to mitigate the costs of filling ``hot spots,'' which we discuss below.  
On the other hand, the smooth terminal geometry of DG, as well as its stability in the $\epsilon <1$ regime, actually suggests that chain packing in the nodal centers is not particularly frustrated, or at least not in qualitatively different ways than its close, non-network competitors.  
That is, in contrast to some widely held notions, the geometry of chain packing in DG on {\it both sides of the IMDS} would appear to be favorable, facilitating a generically optimal compromise intermediate to columnar and lamellar packing.

The thermodynamic cost of chain packing is far from homogeneous, even in DG, as revealed by spatial distributions of the free energy per chain predicted by mSST.  
We found that while each of the canonical network phases has regions where the entropic cost of stretching dominates over interfacial enthalpy, and \emph{vice versa}, there are regions of pronounced free energy cost.
The lamellar-like regions at the three-fold symmetry axes of the DG phase are regions of particularly low free energy (see Fig.~\ref{fig:free_energy_maps}), possessing simultaneously low enthalpic and entropic contributions to the free energy.
The DD and DP phases, by comparison, have regions of pronounced enthalpic costs in regions of relatively low stretching free energy costs, elevating the average free energy and keeping either structure from competing with the DG.  
In contrast, these regions of elevated enthalpic cost occur in the cylinder-like packing regions along the struts of the mSST models of DD and DP phases.

\subsection{Localized hot spots of stretching}

\begin{figure}[h!]
\centering
\includegraphics[width=\textwidth]{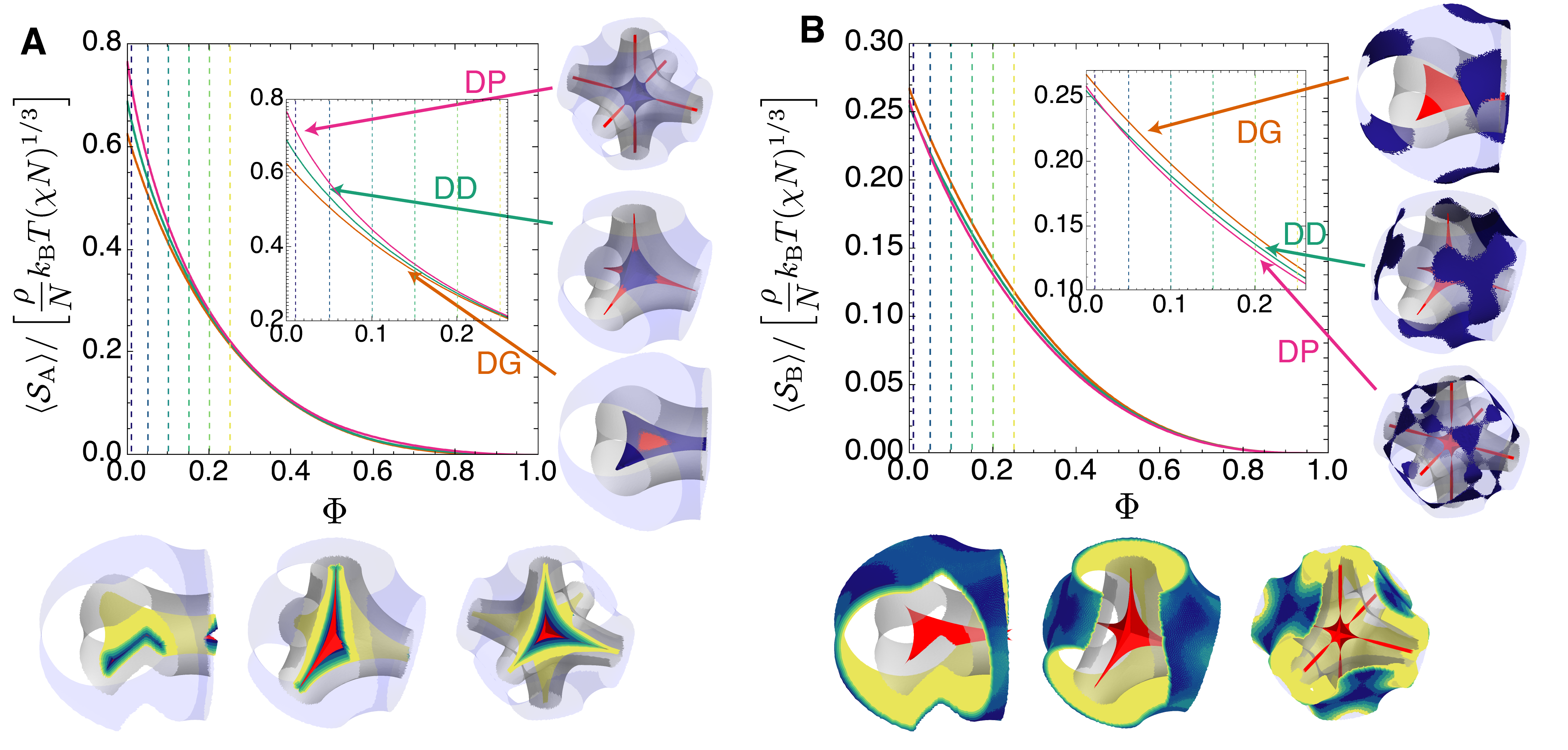}
\caption{\label{fig:hotspots} Average stretching free energy density $\langle \mathcal{S}_{{\rm A,B}} \rangle$ as a function of filling fraction $\Phi$ of the (A) A-block and (B) B-block domains. The insets focus on the behavior at low filling, with graphics showing the regions occupied at $\Phi = 1\%$, where the chains with the largest stretching free energy density are located. Additional graphics (below) show the regions representing progressively higher levels of filling, up to 25\% (yellow).}
\end{figure}


We conclude our discussion by exploiting the mSST predictions for networks to explore ``hot spots'' in the chain packing, which are often conceptually linked with notions of packing frustration in BCP and other surfactant systems\cite{Matsen1996,MatsenBates1997,Reddy2021}.  
These are regions of particularly high free-energy density, representing the regions in a given morphology where monomers experience the highest entropic cost.  
Heuristically, ``hot spots'' are regions that are expected to relax upon blending with some other ``guest'' molecules, for example nanoparticles\cite{Li2014} or homopolymers of the same chain chemistry as block where the hot spot is located\cite{Winey1991Macro}, which have been known to promote the stability of typically unstable ordered phases in neat diblock melts in both experiments\cite{Winey1992,Winey1992b,Takagi2017,Mueller2020,Takagi2021} and simulation studies\cite{Matsen1995PRL,Martinez-Veracoechea2009,Martinez-Veracoechea2009b,Cheong2020,Lai2021,Xie2021}.  
While we do not consider an extension of the mSST framework to account for the thermodynamics of such blended systems here, we nevertheless might consider the distributions of these hot spots as useful proxies for understanding the susceptibility of neat morphologies to the uptake of guests in blends.  
Specifically, we analyze the volume fractions of A and B subdomains that account for the largest portions of the stretching energy in those brush-like regions of each network. 

This information is encoded in the stretching contributions to the free energy: the contribution to the stretching free energy $\delta S_{\alpha}(z)$ (for $\alpha \in \{{\rm A},\, {\rm B}\}$) from chains in a small volume $\delta V$ a distance $z$ from the IMDS is given by $\delta S_{\alpha}(z) = (\kappa_\alpha/2)z^2\delta V$.
Therefore, the stretching free energy per volume is 
\begin{equation}
\mathcal{S}_\alpha(z) \equiv  (\kappa_\alpha/2)z^2 = \frac{\pi^{2/3}}{4}\overline{\lambda}^2\overline{\kappa}_\alpha \tilde{z}^2 \left[\frac{\rho}{N}(\chi N)^{1/3}k_{\rm B}T\right] \, ,
\end{equation}
where $\tilde{z} = z/D$ is the distance from the IMDS relative to the unit cell size.
Subdividing each wedge into small sub-wedge voxels, we determine a distribution of $\mathcal{S}_\alpha$ over the entire nodal region of each network morphology, what has been dubbed the ``mesoatomic'' unit in \cite{GrasonThomas2023}.
We then construct a cumulative distribution $\Omega_\alpha(\mathcal{S})$, consisting of the fraction $\Phi$ of the volume of subdomain $\alpha$ that has a stretching free energy cost greater than $\mathcal{S}$; since $\mathcal{S}_\alpha(z) \to 0$ as $z$ approaches the IMDS and every other point in the subdomain has a positive stretching free energy density, $\Omega$ is normalized such that $\Omega(0) = 1$.
Using the cumulative distribution, we then compute residual stretching energy $\langle \mathcal{S}_\alpha \rangle$ of the $\alpha$ subdomain associated with the lowest $1-\Phi$ fraction of that subdomain.  
In other words, $\langle \mathcal{S}_\alpha \rangle$ is the remaining stretching energy associated with ``filling in" the upper $\Phi$ proportional of the $\alpha$ region and removing its stretching costing.  
This residual stretching free energy $\langle \mathcal{S}_\alpha\rangle$, as a function of $\Phi$, is given by
\begin{equation}
\langle \mathcal{S}_\alpha \rangle(\Phi) \equiv \int_\Phi^1 {\rm d}\Omega_{\alpha}\, \mathcal{S}_{\alpha} \, ,
\end{equation}
which is a function that reaches its maximum at $\Phi = 0$, when the average is taken over the entire subdomain (i.e.~the pure melt), and decays to $0$ as $\Phi \to 1$, corresponding to the case where the entire region is filled in.

In Fig.~\ref{fig:hotspots}A and B, we show this residual stretching free energy of each of the network structures at $\epsilon = 1$, $f \approx 0.29$ as a function of the ``fill fraction'' $\Phi$ for A and B subdomains, respectively.
The convergence of the residual stretching free energy for larger $\Phi$ reflects a uniformity of the packing environments of the three structures near the IMDS, whereas the differences as $\Phi \to 0$ is due to the differences in the stretching thermodynamics near to the terminal boundaries, where the chains are have the largest stretch within the subdomain.
Inset graphics (right) show the regions occupied at 1\% infill of the subdomain ($\Phi = 0.01$), highlighting the maximal stretching free energy density regions of both A and B blocks, with the figures below illustrating isocontours of increasingly large $\Phi$.

Notably, whereas the largest A-block stretching free energy density lies at the node of the DD and DP terminal surfaces and the corners around the node, the most expensive parts of the DG subdomain volume lie along the edges of the terminal surface, away from the central node.  
This re-enforces the conclusion that the medial geometry of the DG morphology is uniquely optimal amongst the cubic network phases, as it supports an extended lamellar-like packing environment without singular folds or corners.  
Additionally, we note that while the A-block stretching free energy is lowest for DG at $\Phi=0$, the slopes of the stretching free energy with fill fraction are larger from DD and DP, suggesting that these structures are plausibly \emph{more susceptible} to in-filling by guest molecules that might relax large fractions of their entropic cost.  
Indeed, the curves for $\langle \mathcal{S}_A\rangle$ tend to converge around $\Phi \approx 0.2$, consistent with the range of homopolymer blending at which DD (or DP) have been reported to gain equilibrium stability in diblock/homopolymer blend systems\cite{Takagi2017,Takagi2021}.

Comparing this to Fig.~\ref{fig:hotspots}B we note that the overall scale of stretching free energy density in B blocks is roughly a factor of 3 smaller, suggesting that ``hot spots" of the matrix domains are somewhat ``cooler'' than in the A domains.  
As might be anticipated, these are distributed around CMC-like surfaces that constitute the matrix terminal boundaries; however, for sufficiently small $\Phi$, these maximal stretching regions localized to patchy regions.  
For DD and DP, maximal stretching hot spots concentrate over the monkey saddle positions, whereas for DG, maximal matrix stretching occurs away from this directions in bands that flank the elbow portions of the trihedral mesoatom.  

Taken together, these hot spot distributions present a possible way to interpret the stability of distinct morphologies in blended systems.  
Based on distinct features of the terminal boundaries, particularly the inner tubular subdomains, it is perhaps intuitive to imagine that the total volumes of hot spots which might be filled by guest molecules could increase with the total cost of packing frustration.  
What is less clear is what controls the ability of distinct morphologies to ``host'' these guest molecules.  
That is, in order to understand why DD can overtake DG when blended with A-block homopolymer, it is not only necessary to understand why the free energy of DD is lowered by hot spot filling, but it must also be understood why is not comparably lowered in DG.  
In other words, what makes DG (not to mention competitor Lam and Hex) a relatively bad host in those blends where DD or DP become the equilibrium phase?  
Answering this question requires a model of how chain packing adapts progressively as hot spots are filled, collectively relaxing the both enthalpic and entropic contributions to the free energy over the entire morphology\cite{Winey1991JCP}.

Finally, we note from Fig.~\ref{fig:hotspots} that hot spots cover and conform to these quasi-2D regions for $\Phi$ of a few percent. 
Provided that guest molecules are highly localized to hot spots, say for the case of weakly-interaction high-molecular weight homopolymers, imaging their 3D distributions (say be selective labeling of high-contrast guests\cite{Mayes1992}) at low-blend fractions may provide an indirect means to observe the shapes of the terminal boundaries of the host morphologies.

\section{Conclusions and Outlook}

Using the medial strong segregation theory (mSST) formalism, we have explored the multiple facets of packing frustration, linking geometric features of packing to their associated thermodynamic costs.
In doing so, we have demonstrated that packing frustration presents in a combination of heightened chain stretching costs, but also high interfacial costs, which are typically correlated with regions of low stretching.
In particular, we have shown that the terminal surface geometry of the DG morphology, as modeled by a set of medial surfaces, promotes stability over the competitor DD and DP network phases, uniquely taking advantage of packing environments that combine aspects of lamellar and cylindrical morphologies.

Nevertheless, a number of open questions remain regarding optimal packing in network phases.
We have provided evidence that chain bending or kinking can play a significant role in relaxing the thermodynamic costs of network structures.
The differences between IMDS morphologies predicted with SCFT calculations and our mSST calculations can be attributed to our straight-path \emph{ansatz} for chain trajectories.
Indeed, SST calculations that appropriately capture IMDS faceting for cylindrical phases rely on a kinked-path \emph{ansatz}; developing a similar kinked-path formalism that works \emph{in tandem} with medial surfaces remains an open challenge.
Moreover, while we have provided a plausible prediction of the susceptibility of each network phase to the uptake of guest molecules, the incorporation of guests into mSST (and the resulting structural relaxations) remains a challenge.

\begin{acknowledgement}

The authors thank to E. Thomas and B. Greenvall for stimulating discussions and valuable comments on this work. We also thank P. Olmsted for useful discussions about prior SST calculations.  This research was
supported by the US Department of Energy (DOE), Office of Basic Energy Sciences, Division
of Materials Sciences and Engineering under award DE-SC0022229.
Medial SST and SCF calculations were performed on the UMass Cluster at the Massachusetts
Green High Performance Computing Center.

\end{acknowledgement}

\begin{suppinfo}
Additional details for the mSST calculation methods and analysis results for the variable morphology predictions from both mSST and SCFT models. 

\end{suppinfo}

\bibliography{refs}

\providecommand{\latin}[1]{#1}
\makeatletter
\providecommand{\doi}
  {\begingroup\let\do\@makeother\dospecials
  \catcode`\{=1 \catcode`\}=2 \doi@aux}
\providecommand{\doi@aux}[1]{\endgroup\texttt{#1}}
\makeatother
\providecommand*\mcitethebibliography{\thebibliography}
\csname @ifundefined\endcsname{endmcitethebibliography}
  {\let\endmcitethebibliography\endthebibliography}{}
\begin{mcitethebibliography}{71}
\providecommand*\natexlab[1]{#1}
\providecommand*\mciteSetBstSublistMode[1]{}
\providecommand*\mciteSetBstMaxWidthForm[2]{}
\providecommand*\mciteBstWouldAddEndPuncttrue
  {\def\EndOfBibitem{\unskip.}}
\providecommand*\mciteBstWouldAddEndPunctfalse
  {\let\EndOfBibitem\relax}
\providecommand*\mciteSetBstMidEndSepPunct[3]{}
\providecommand*\mciteSetBstSublistLabelBeginEnd[3]{}
\providecommand*\EndOfBibitem{}
\mciteSetBstSublistMode{f}
\mciteSetBstMaxWidthForm{subitem}{(\alph{mcitesubitemcount})}
\mciteSetBstSublistLabelBeginEnd
  {\mcitemaxwidthsubitemform\space}
  {\relax}
  {\relax}

\bibitem[Israelachvili(2011)]{Isrealachvili2011}
Israelachvili,~J.~N. In \emph{Intermolecular and Surface Forces (Third
  Edition)}, third edition ed.; Israelachvili,~J.~N., Ed.; Academic Press: San
  Diego, 2011; pp 535--576\relax
\mciteBstWouldAddEndPuncttrue
\mciteSetBstMidEndSepPunct{\mcitedefaultmidpunct}
{\mcitedefaultendpunct}{\mcitedefaultseppunct}\relax
\EndOfBibitem
\bibitem[Hyde \latin{et~al.}(1997)Hyde, Ninham, Andersson, Larsson, Landh,
  Blum, and Lidin]{Hyde1997_ch4}
Hyde,~S.; Ninham,~B.~W.; Andersson,~S.; Larsson,~K.; Landh,~T.; Blum,~Z.;
  Lidin,~S. In \emph{The Language of Shape}; Hyde,~S., Ninham,~B.~W.,
  Andersson,~S., Larsson,~K., Landh,~T., Blum,~Z., Lidin,~S., Eds.; Elsevier
  Science B.V.: Amsterdam, 1997; pp 141--197\relax
\mciteBstWouldAddEndPuncttrue
\mciteSetBstMidEndSepPunct{\mcitedefaultmidpunct}
{\mcitedefaultendpunct}{\mcitedefaultseppunct}\relax
\EndOfBibitem
\bibitem[Scriven(1976)]{Scriven1976}
Scriven,~L.~E. {Equilibrium bicontinuous structure}. \emph{Nature}
  \textbf{1976}, \emph{263}, 123--125\relax
\mciteBstWouldAddEndPuncttrue
\mciteSetBstMidEndSepPunct{\mcitedefaultmidpunct}
{\mcitedefaultendpunct}{\mcitedefaultseppunct}\relax
\EndOfBibitem
\bibitem[Thomas \latin{et~al.}(1988)Thomas, Anderson, Henkee, and
  Hoffman]{Thomas1988}
Thomas,~E.~L.; Anderson,~D.~M.; Henkee,~C.~S.; Hoffman,~D. {Periodic
  area-minimizing surfaces in block copolymers}. \emph{Nature} \textbf{1988},
  \emph{334}, 598--601\relax
\mciteBstWouldAddEndPuncttrue
\mciteSetBstMidEndSepPunct{\mcitedefaultmidpunct}
{\mcitedefaultendpunct}{\mcitedefaultseppunct}\relax
\EndOfBibitem
\bibitem[Schoen(2012)]{Schoen2012}
Schoen,~A.~H. {Reflections concerning triply-periodic minimal surfaces}.
  \emph{Interface Focus} \textbf{2012}, \emph{2}, 658--668\relax
\mciteBstWouldAddEndPuncttrue
\mciteSetBstMidEndSepPunct{\mcitedefaultmidpunct}
{\mcitedefaultendpunct}{\mcitedefaultseppunct}\relax
\EndOfBibitem
\bibitem[Anderson \latin{et~al.}(1988)Anderson, Gruner, and
  Leibler]{Anderson1988}
Anderson,~D.~M.; Gruner,~S.~M.; Leibler,~S. {Geometrical aspects of the
  frustration in the cubic phases of lyotropic liquid crystals.}
  \emph{Proceedings of the National Academy of Sciences} \textbf{1988},
  \emph{85}, 5364--5368\relax
\mciteBstWouldAddEndPuncttrue
\mciteSetBstMidEndSepPunct{\mcitedefaultmidpunct}
{\mcitedefaultendpunct}{\mcitedefaultseppunct}\relax
\EndOfBibitem
\bibitem[Matsen and Bates(1996)Matsen, and Bates]{Matsen1996}
Matsen,~M.~W.; Bates,~F.~S. {Origins of Complex Self-Assembly in Block
  Copolymers}. \emph{Macromolecules} \textbf{1996}, \emph{29}, 7641--7644\relax
\mciteBstWouldAddEndPuncttrue
\mciteSetBstMidEndSepPunct{\mcitedefaultmidpunct}
{\mcitedefaultendpunct}{\mcitedefaultseppunct}\relax
\EndOfBibitem
\bibitem[Grason(2006)]{Grason2006}
Grason,~G.~M. {The packing of soft materials: Molecular asymmetry, geometric
  frustration and optimal lattices in block copolymer melts}. \emph{Physics
  Reports} \textbf{2006}, \emph{433}, 1--64\relax
\mciteBstWouldAddEndPuncttrue
\mciteSetBstMidEndSepPunct{\mcitedefaultmidpunct}
{\mcitedefaultendpunct}{\mcitedefaultseppunct}\relax
\EndOfBibitem
\bibitem[Shi(2021)]{ACShi2021}
Shi,~A.-C. Frustration in block copolymer assemblies. \emph{J. Phys.: Condens.
  Matter} \textbf{2021}, \emph{33}, 253001\relax
\mciteBstWouldAddEndPuncttrue
\mciteSetBstMidEndSepPunct{\mcitedefaultmidpunct}
{\mcitedefaultendpunct}{\mcitedefaultseppunct}\relax
\EndOfBibitem
\bibitem[Duesing \latin{et~al.}(1997)Duesing, Templer, and Seddon]{Duesing1997}
Duesing,~P.~M.; Templer,~R.~H.; Seddon,~J.~M. Quantifying Packing Frustration
  Energy in Inverse Lyotropic Mesophases. \emph{Langmuir} \textbf{1997},
  \emph{13}, 351--359\relax
\mciteBstWouldAddEndPuncttrue
\mciteSetBstMidEndSepPunct{\mcitedefaultmidpunct}
{\mcitedefaultendpunct}{\mcitedefaultseppunct}\relax
\EndOfBibitem
\bibitem[Matsen and Bates(1997)Matsen, and Bates]{MatsenBates1997}
Matsen,~M.~W.; Bates,~F.~S. {Block copolymer microstructures in the
  intermediate-segregation regime}. \emph{The Journal of Chemical Physics}
  \textbf{1997}, \emph{106}, 2436--2448\relax
\mciteBstWouldAddEndPuncttrue
\mciteSetBstMidEndSepPunct{\mcitedefaultmidpunct}
{\mcitedefaultendpunct}{\mcitedefaultseppunct}\relax
\EndOfBibitem
\bibitem[Schr\"{o}der-Turk \latin{et~al.}(2007)Schr\"{o}der-Turk, Fogden, and
  Hyde]{Schroder-Turk2007}
Schr\"{o}der-Turk,~G.~E.; Fogden,~A.; Hyde,~S.~T. Local v/a variations as a
  measure of structural packing frustration in bicontinuous mesophases, and
  geometric arguments for an alternating {${\rm Im}\overline{3}{\rm m}$} (I-WP)
  phase in block-copolymers with polydispersity. \emph{The European Physical
  Journal B} \textbf{2007}, \emph{59}, 115--126\relax
\mciteBstWouldAddEndPuncttrue
\mciteSetBstMidEndSepPunct{\mcitedefaultmidpunct}
{\mcitedefaultendpunct}{\mcitedefaultseppunct}\relax
\EndOfBibitem
\bibitem[Thomas \latin{et~al.}(1986)Thomas, Alward, Kinning, Martin, Handlin,
  and Fetters]{Thomas1986}
Thomas,~E.~L.; Alward,~D.~B.; Kinning,~D.~J.; Martin,~D.~C.; Handlin,~D.~L.;
  Fetters,~L.~J. {Ordered bicontinuous double-diamond structure of star block
  copolymers: a new equilibrium microdomain morphology}. \emph{Macromolecules}
  \textbf{1986}, \emph{19}, 2197--2202\relax
\mciteBstWouldAddEndPuncttrue
\mciteSetBstMidEndSepPunct{\mcitedefaultmidpunct}
{\mcitedefaultendpunct}{\mcitedefaultseppunct}\relax
\EndOfBibitem
\bibitem[Hajduk \latin{et~al.}(1994)Hajduk, Harper, Gruner, Honeker, Kim,
  Fetters, Kim, Thomas, Fetters, and Kim]{Hajduk1994}
Hajduk,~D.~A.; Harper,~P.~E.; Gruner,~S.~M.; Honeker,~C.~C.; Kim,~G.;
  Fetters,~L.~J.; Kim,~G.; Thomas,~E.~L.; Fetters,~L.~J.; Kim,~G. {The Gyroid:
  A New Equilibrium Morphology in Weakly Segregated Diblock Copolymers}.
  \emph{Macromolecules} \textbf{1994}, \emph{27}, 4063--4075\relax
\mciteBstWouldAddEndPuncttrue
\mciteSetBstMidEndSepPunct{\mcitedefaultmidpunct}
{\mcitedefaultendpunct}{\mcitedefaultseppunct}\relax
\EndOfBibitem
\bibitem[Foerster \latin{et~al.}(1994)Foerster, Khandpur, Zhao, Bates, Hamley,
  Ryan, and Bras]{Foerster1994}
Foerster,~S.; Khandpur,~A.~K.; Zhao,~J.; Bates,~F.~S.; Hamley,~I.~W.;
  Ryan,~A.~J.; Bras,~W. Complex Phase Behavior of Polyisoprene-Polystyrene
  Diblock Copolymers Near the Order-Disorder Transition. \emph{Macromolecules}
  \textbf{1994}, \emph{27}, 6922--6935\relax
\mciteBstWouldAddEndPuncttrue
\mciteSetBstMidEndSepPunct{\mcitedefaultmidpunct}
{\mcitedefaultendpunct}{\mcitedefaultseppunct}\relax
\EndOfBibitem
\bibitem[Khandpur \latin{et~al.}(1995)Khandpur, Foerster, Bates, Hamley, Ryan,
  Bras, Almdal, and Mortensen]{Khandpur1995}
Khandpur,~A.~K.; Foerster,~S.; Bates,~F.~S.; Hamley,~I.~W.; Ryan,~A.~J.;
  Bras,~W.; Almdal,~K.; Mortensen,~K. Polyisoprene-Polystyrene Diblock
  Copolymer Phase Diagram near the Order-Disorder Transition.
  \emph{Macromolecules} \textbf{1995}, \emph{28}, 8796--8806\relax
\mciteBstWouldAddEndPuncttrue
\mciteSetBstMidEndSepPunct{\mcitedefaultmidpunct}
{\mcitedefaultendpunct}{\mcitedefaultseppunct}\relax
\EndOfBibitem
\bibitem[Olmsted and Milner(1994)Olmsted, and Milner]{Olmsted1994}
Olmsted,~P.~D.; Milner,~S.~T. Strong-segregation theory of bicontinuous phases
  in block copolymers. \emph{Phys. Rev. Lett.} \textbf{1994}, \emph{72},
  936--939\relax
\mciteBstWouldAddEndPuncttrue
\mciteSetBstMidEndSepPunct{\mcitedefaultmidpunct}
{\mcitedefaultendpunct}{\mcitedefaultseppunct}\relax
\EndOfBibitem
\bibitem[Olmsted and Milner(1995)Olmsted, and Milner]{Olmsted1995}
Olmsted,~P.~D.; Milner,~S.~T. {Strong-Segregation Theory of Bicontinuous Phases
  in Block Copolymers}. \emph{Physical Review Letters} \textbf{1995},
  \emph{74}, 829--829\relax
\mciteBstWouldAddEndPuncttrue
\mciteSetBstMidEndSepPunct{\mcitedefaultmidpunct}
{\mcitedefaultendpunct}{\mcitedefaultseppunct}\relax
\EndOfBibitem
\bibitem[Olmsted and Milner(1998)Olmsted, and Milner]{Olmsted1998}
Olmsted,~P.~D.; Milner,~S.~T. {Strong Segregation Theory of Bicontinuous Phases
  in Block Copolymers}. \emph{Macromolecules} \textbf{1998}, \emph{31},
  4011--4022\relax
\mciteBstWouldAddEndPuncttrue
\mciteSetBstMidEndSepPunct{\mcitedefaultmidpunct}
{\mcitedefaultendpunct}{\mcitedefaultseppunct}\relax
\EndOfBibitem
\bibitem[Likhtman and Semenov(1994)Likhtman, and Semenov]{Likhtman1994}
Likhtman,~A.~E.; Semenov,~A.~N. {Stability of the OBDD Structure for Diblock
  Copolymer Melts in the Strong Segregation Limit}. \emph{Macromolecules}
  \textbf{1994}, \emph{27}, 3103--3106\relax
\mciteBstWouldAddEndPuncttrue
\mciteSetBstMidEndSepPunct{\mcitedefaultmidpunct}
{\mcitedefaultendpunct}{\mcitedefaultseppunct}\relax
\EndOfBibitem
\bibitem[Likhtman and Semenov(1997)Likhtman, and Semenov]{Likhtman1997}
Likhtman,~A.~E.; Semenov,~A.~N. {Theory of Microphase Separation in Block
  Copolymer/Homopolymer Mixtures}. \emph{Macromolecules} \textbf{1997},
  \emph{30}, 7273--7278\relax
\mciteBstWouldAddEndPuncttrue
\mciteSetBstMidEndSepPunct{\mcitedefaultmidpunct}
{\mcitedefaultendpunct}{\mcitedefaultseppunct}\relax
\EndOfBibitem
\bibitem[Matsen and Bates(1996)Matsen, and Bates]{Matsen1996b}
Matsen,~M.~W.; Bates,~F.~S. {Unifying Weak- and Strong-Segregation Block
  Copolymer Theories}. \emph{Macromolecules} \textbf{1996}, \emph{29},
  1091--1098\relax
\mciteBstWouldAddEndPuncttrue
\mciteSetBstMidEndSepPunct{\mcitedefaultmidpunct}
{\mcitedefaultendpunct}{\mcitedefaultseppunct}\relax
\EndOfBibitem
\bibitem[Cochran \latin{et~al.}(2006)Cochran, Garcia-Cervera, and
  Fredrickson]{Cochran2006}
Cochran,~E.~W.; Garcia-Cervera,~C.~J.; Fredrickson,~G.~H. Stability of the
  Gyroid Phase in Diblock Copolymers at Strong Segregation.
  \emph{Macromolecules} \textbf{2006}, \emph{39}, 2449--2451\relax
\mciteBstWouldAddEndPuncttrue
\mciteSetBstMidEndSepPunct{\mcitedefaultmidpunct}
{\mcitedefaultendpunct}{\mcitedefaultseppunct}\relax
\EndOfBibitem
\bibitem[Davidock \latin{et~al.}(2003)Davidock, Hillmyer, and
  Lodge]{Davidock2003}
Davidock,~D.~A.; Hillmyer,~M.~A.; Lodge,~T.~P. Persistence of the Gyroid
  Morphology at Strong Segregation in Diblock Copolymers. \emph{Macromolecules}
  \textbf{2003}, \emph{36}, 4682--4685\relax
\mciteBstWouldAddEndPuncttrue
\mciteSetBstMidEndSepPunct{\mcitedefaultmidpunct}
{\mcitedefaultendpunct}{\mcitedefaultseppunct}\relax
\EndOfBibitem
\bibitem[Reddy \latin{et~al.}(2021)Reddy, Feng, Thomas, and Grason]{Reddy2021}
Reddy,~A.; Feng,~X.; Thomas,~E.~L.; Grason,~G.~M. Block Copolymers beneath the
  Surface: Measuring and Modeling Complex Morphology at the Subdomain Scale.
  \emph{Macromolecules} \textbf{2021}, \emph{54}, 9223--9257\relax
\mciteBstWouldAddEndPuncttrue
\mciteSetBstMidEndSepPunct{\mcitedefaultmidpunct}
{\mcitedefaultendpunct}{\mcitedefaultseppunct}\relax
\EndOfBibitem
\bibitem[Prasad \latin{et~al.}(2018)Prasad, Jinnai, Ho, Thomas, and
  Grason]{Prasad2018}
Prasad,~I.; Jinnai,~H.; Ho,~R.-M.; Thomas,~E.~L.; Grason,~G.~M. Anatomy of
  triply-periodic network assemblies: characterizing skeletal and inter-domain
  surface geometry of block copolymer gyroids. \emph{Soft Matter}
  \textbf{2018}, \emph{14}, 3612--3623\relax
\mciteBstWouldAddEndPuncttrue
\mciteSetBstMidEndSepPunct{\mcitedefaultmidpunct}
{\mcitedefaultendpunct}{\mcitedefaultseppunct}\relax
\EndOfBibitem
\bibitem[Reddy \latin{et~al.}(2022)Reddy, Dimitriyev, and Grason]{Reddy2022}
Reddy,~A.; Dimitriyev,~M.; Grason,~G. Medial packing and elastic asymmetry
  stabilize the double-gyroid in block copolymers. \emph{Nature Communications}
  \textbf{2022}, \emph{13}\relax
\mciteBstWouldAddEndPuncttrue
\mciteSetBstMidEndSepPunct{\mcitedefaultmidpunct}
{\mcitedefaultendpunct}{\mcitedefaultseppunct}\relax
\EndOfBibitem
\bibitem[Schr{\"{o}}der \latin{et~al.}(2003)Schr{\"{o}}der, Ramsden, Christy,
  and Hyde]{Schroder2003}
Schr{\"{o}}der,~G.~E.; Ramsden,~S.~J.; Christy,~A.~G.; Hyde,~S.~T. {Medial
  surfaces of hyperbolic structures}. \emph{The European Physical Journal B -
  Condensed Matter} \textbf{2003}, \emph{35}, 551--564\relax
\mciteBstWouldAddEndPuncttrue
\mciteSetBstMidEndSepPunct{\mcitedefaultmidpunct}
{\mcitedefaultendpunct}{\mcitedefaultseppunct}\relax
\EndOfBibitem
\bibitem[Templer \latin{et~al.}(1998)Templer, Seddon, Duesing, Winter, and
  Erbes]{Templer1998}
Templer,~R.~H.; Seddon,~J.~M.; Duesing,~P.~M.; Winter,~R.; Erbes,~J. Modeling
  the Phase Behavior of the Inverse Hexagonal and Inverse Bicontinuous Cubic
  Phases in 2:1 Fatty Acid/Phosphatidylcholine Mixtures. \emph{The Journal of
  Physical Chemistry B} \textbf{1998}, \emph{102}, 7262--7271\relax
\mciteBstWouldAddEndPuncttrue
\mciteSetBstMidEndSepPunct{\mcitedefaultmidpunct}
{\mcitedefaultendpunct}{\mcitedefaultseppunct}\relax
\EndOfBibitem
\bibitem[Kulkarni \latin{et~al.}(2010)Kulkarni, Tang, Seddon, Seddon, Ces, and
  Templer]{Kulkarni2010}
Kulkarni,~C.~V.; Tang,~T.-Y.; Seddon,~A.~M.; Seddon,~J.~M.; Ces,~O.;
  Templer,~R.~H. Engineering bicontinuous cubic structures at the
  nanoscale—the role of chain splay. \emph{Soft Matter} \textbf{2010},
  \emph{6}, 3191--3194\relax
\mciteBstWouldAddEndPuncttrue
\mciteSetBstMidEndSepPunct{\mcitedefaultmidpunct}
{\mcitedefaultendpunct}{\mcitedefaultseppunct}\relax
\EndOfBibitem
\bibitem[Schr{\"{o}}der-Turk \latin{et~al.}(2006)Schr{\"{o}}der-Turk, Fogden,
  and Hyde]{Schroder-Turk2006}
Schr{\"{o}}der-Turk,~G.~E.; Fogden,~A.; Hyde,~S.~T. {Bicontinuous geometries
  and molecular self-assembly: comparison of local curvature and global packing
  variations in genus-three cubic, tetragonal and rhombohedral surfaces}.
  \emph{The European Physical Journal B} \textbf{2006}, \emph{54},
  509--524\relax
\mciteBstWouldAddEndPuncttrue
\mciteSetBstMidEndSepPunct{\mcitedefaultmidpunct}
{\mcitedefaultendpunct}{\mcitedefaultseppunct}\relax
\EndOfBibitem
\bibitem[Chen \latin{et~al.}(2022)Chen, Mahanthappa, and Dorfman]{Chen2022}
Chen,~P.; Mahanthappa,~M.~K.; Dorfman,~K.~D. Stability of cubic single network
  phases in diblock copolymer melts. \emph{Journal of Polymer Science}
  \textbf{2022}, \emph{60}, 2543--2552\relax
\mciteBstWouldAddEndPuncttrue
\mciteSetBstMidEndSepPunct{\mcitedefaultmidpunct}
{\mcitedefaultendpunct}{\mcitedefaultseppunct}\relax
\EndOfBibitem
\bibitem[Matsen(1995)]{Matsen1995}
Matsen,~M.~W. {Stabilizing New Morphologies by Blending Homopolymer with Block
  Copolymer}. \emph{Physical Review Letters} \textbf{1995}, \emph{74},
  4225--4228\relax
\mciteBstWouldAddEndPuncttrue
\mciteSetBstMidEndSepPunct{\mcitedefaultmidpunct}
{\mcitedefaultendpunct}{\mcitedefaultseppunct}\relax
\EndOfBibitem
\bibitem[Martinez-Veracoechea and Escobedo(2009)Martinez-Veracoechea, and
  Escobedo]{Martinez-Veracoechea2009}
Martinez-Veracoechea,~F.~J.; Escobedo,~F.~A. {The Plumber's Nightmare Phase in
  Diblock Copolymer/Homopolymer Blends. A Self-Consistent Field Theory Study.}
  \emph{Macromolecules} \textbf{2009}, \emph{42}, 9058--9062\relax
\mciteBstWouldAddEndPuncttrue
\mciteSetBstMidEndSepPunct{\mcitedefaultmidpunct}
{\mcitedefaultendpunct}{\mcitedefaultseppunct}\relax
\EndOfBibitem
\bibitem[Martínez-Veracoechea and Escobedo(2009)Martínez-Veracoechea, and
  Escobedo]{Martinez-Veracoechea2009b}
Martínez-Veracoechea,~F.~J.; Escobedo,~F.~A. Bicontinuous Phases in Diblock
  Copolymer/Homopolymer Blends: Simulation and Self-Consistent Field Theory.
  \emph{Macromolecules} \textbf{2009}, \emph{42}, 1775--1784\relax
\mciteBstWouldAddEndPuncttrue
\mciteSetBstMidEndSepPunct{\mcitedefaultmidpunct}
{\mcitedefaultendpunct}{\mcitedefaultseppunct}\relax
\EndOfBibitem
\bibitem[Cheong \latin{et~al.}(2020)Cheong, Bates, and Dorfman]{Cheong2020}
Cheong,~G.~K.; Bates,~F.~S.; Dorfman,~K.~D. {Symmetry breaking in
  particle-forming diblock polymer/homopolymer blends}. \emph{Proceedings of
  the National Academy of Sciences} \textbf{2020}, \emph{117},
  16764--16769\relax
\mciteBstWouldAddEndPuncttrue
\mciteSetBstMidEndSepPunct{\mcitedefaultmidpunct}
{\mcitedefaultendpunct}{\mcitedefaultseppunct}\relax
\EndOfBibitem
\bibitem[Xie and Shi(2021)Xie, and Shi]{Xie2021}
Xie,~J.; Shi,~A.-C. Formation of complex spherical packing phases in diblock
  copolymer/homopolymer blends. \emph{Giant} \textbf{2021}, \emph{5},
  100043\relax
\mciteBstWouldAddEndPuncttrue
\mciteSetBstMidEndSepPunct{\mcitedefaultmidpunct}
{\mcitedefaultendpunct}{\mcitedefaultseppunct}\relax
\EndOfBibitem
\bibitem[Lai and Shi(2021)Lai, and Shi]{Lai2021}
Lai,~C.~T.; Shi,~A.-C. Binary Blends of Diblock Copolymers: An Effective Route
  to Novel Bicontinuous Phases. \emph{Macromolecular Theory and Simulations}
  \textbf{2021}, \emph{30}, 2100019\relax
\mciteBstWouldAddEndPuncttrue
\mciteSetBstMidEndSepPunct{\mcitedefaultmidpunct}
{\mcitedefaultendpunct}{\mcitedefaultseppunct}\relax
\EndOfBibitem
\bibitem[Takagi and Yamamoto(2021)Takagi, and Yamamoto]{Takagi2021}
Takagi,~H.; Yamamoto,~K. Effect of Block Copolymer Composition and Homopolymer
  Molecular Weight on Ordered Bicontinuous Double-Diamond Structures in Binary
  Blends of Polystyrene–Polyisoprene Block Copolymer and Polyisoprene
  Homopolymer. \emph{Macromolecules} \textbf{2021}, \emph{54}, 5136--5143\relax
\mciteBstWouldAddEndPuncttrue
\mciteSetBstMidEndSepPunct{\mcitedefaultmidpunct}
{\mcitedefaultendpunct}{\mcitedefaultseppunct}\relax
\EndOfBibitem
\bibitem[Milner(1994)]{Milner1994}
Milner,~S.~T. {Chain Architecture and Asymmetry in Copolymer Microphases}.
  \emph{Macromolecules} \textbf{1994}, \emph{27}, 2333--2335\relax
\mciteBstWouldAddEndPuncttrue
\mciteSetBstMidEndSepPunct{\mcitedefaultmidpunct}
{\mcitedefaultendpunct}{\mcitedefaultseppunct}\relax
\EndOfBibitem
\bibitem[Grason and Kamien(2004)Grason, and Kamien]{Grason2004}
Grason,~G.~M.; Kamien,~R.~D. {Interfaces in diblocks: A study of miktoarm star
  copolymers}. \emph{Macromolecules} \textbf{2004}, \emph{37}, 7371--7380\relax
\mciteBstWouldAddEndPuncttrue
\mciteSetBstMidEndSepPunct{\mcitedefaultmidpunct}
{\mcitedefaultendpunct}{\mcitedefaultseppunct}\relax
\EndOfBibitem
\bibitem[Semenov(1985)]{Semenov1985}
Semenov,~A.~N. {Contribution to the theory of microphase layering in
  block-copolymer melts}. \emph{Journal of Experimental and Theoretical
  Physics} \textbf{1985}, \emph{88}, 1242\relax
\mciteBstWouldAddEndPuncttrue
\mciteSetBstMidEndSepPunct{\mcitedefaultmidpunct}
{\mcitedefaultendpunct}{\mcitedefaultseppunct}\relax
\EndOfBibitem
\bibitem[Goveas \latin{et~al.}(1997)Goveas, Milner, and Russel]{Goveas1997}
Goveas,~J.~L.; Milner,~S.~T.; Russel,~W.~B. Corrections to Strong-Stretching
  Theories. \emph{Macromolecules} \textbf{1997}, \emph{30}, 5541--5552\relax
\mciteBstWouldAddEndPuncttrue
\mciteSetBstMidEndSepPunct{\mcitedefaultmidpunct}
{\mcitedefaultendpunct}{\mcitedefaultseppunct}\relax
\EndOfBibitem
\bibitem[Helfand and Sapse(1975)Helfand, and Sapse]{Helfand1975}
Helfand,~E.; Sapse,~A.~M. {Theory of unsymmetric polymer–polymer interfaces}.
  \emph{The Journal of Chemical Physics} \textbf{1975}, \emph{62},
  1327--1331\relax
\mciteBstWouldAddEndPuncttrue
\mciteSetBstMidEndSepPunct{\mcitedefaultmidpunct}
{\mcitedefaultendpunct}{\mcitedefaultseppunct}\relax
\EndOfBibitem
\bibitem[Milner \latin{et~al.}(1988)Milner, Witten, and Cates]{Milner1988}
Milner,~S.~T.; Witten,~T.~A.; Cates,~M.~E. Theory of the grafted polymer brush.
  \emph{Macromolecules} \textbf{1988}, \emph{21}, 2610--2619\relax
\mciteBstWouldAddEndPuncttrue
\mciteSetBstMidEndSepPunct{\mcitedefaultmidpunct}
{\mcitedefaultendpunct}{\mcitedefaultseppunct}\relax
\EndOfBibitem
\bibitem[Ball \latin{et~al.}(1991)Ball, Marko, Milner, and Witten]{Ball1991}
Ball,~R.~C.; Marko,~J.~F.; Milner,~S.~T.; Witten,~T.~A. {Polymers grafted to a
  convex surface}. \emph{Macromolecules} \textbf{1991}, \emph{24},
  693--703\relax
\mciteBstWouldAddEndPuncttrue
\mciteSetBstMidEndSepPunct{\mcitedefaultmidpunct}
{\mcitedefaultendpunct}{\mcitedefaultseppunct}\relax
\EndOfBibitem
\bibitem[Belyi(2004)]{Belyi2004}
Belyi,~V.~A. {Exclusion zone of convex brushes in the strong-stretching limit}.
  \emph{The Journal of Chemical Physics} \textbf{2004}, \emph{121},
  6547--6554\relax
\mciteBstWouldAddEndPuncttrue
\mciteSetBstMidEndSepPunct{\mcitedefaultmidpunct}
{\mcitedefaultendpunct}{\mcitedefaultseppunct}\relax
\EndOfBibitem
\bibitem[Dimitriyev and Grason(2021)Dimitriyev, and Grason]{Dimitriyev2021}
Dimitriyev,~M.~S.; Grason,~G.~M. End exclusion zones in strongly stretched,
  molten polymer brushes of arbitrary shape. \emph{J. Chem. Phys.}
  \textbf{2021}, \emph{155}, 224901\relax
\mciteBstWouldAddEndPuncttrue
\mciteSetBstMidEndSepPunct{\mcitedefaultmidpunct}
{\mcitedefaultendpunct}{\mcitedefaultseppunct}\relax
\EndOfBibitem
\bibitem[Blum(1967)]{Blum1967}
Blum,~H. In \emph{Models for Perception of Speech and Visual Form};
  Wathen-Dunn,~W., Ed.; MIT Press: Cambridge, MA, 1967; pp 362--380\relax
\mciteBstWouldAddEndPuncttrue
\mciteSetBstMidEndSepPunct{\mcitedefaultmidpunct}
{\mcitedefaultendpunct}{\mcitedefaultseppunct}\relax
\EndOfBibitem
\bibitem[Nackman and Pizer(1985)Nackman, and Pizer]{Nackman1985}
Nackman,~L.~R.; Pizer,~S.~M. Three-Dimensional Shape Description Using the
  Symmetric Axis Transform I: Theory. \emph{IEEE Transactions on Pattern
  Analysis and Machine Intelligence} \textbf{1985}, \emph{PAMI-7},
  187--202\relax
\mciteBstWouldAddEndPuncttrue
\mciteSetBstMidEndSepPunct{\mcitedefaultmidpunct}
{\mcitedefaultendpunct}{\mcitedefaultseppunct}\relax
\EndOfBibitem
\bibitem[Siddiqi and Pizer(2008)Siddiqi, and Pizer]{SiddiqiPizer2008}
Siddiqi,~K.; Pizer,~S. \emph{Medial Representations: Mathematics, Algorithms
  and Applications}, 1st ed.; Springer Publishing Company, Incorporated,
  2008\relax
\mciteBstWouldAddEndPuncttrue
\mciteSetBstMidEndSepPunct{\mcitedefaultmidpunct}
{\mcitedefaultendpunct}{\mcitedefaultseppunct}\relax
\EndOfBibitem
\bibitem[Wohlgemuth \latin{et~al.}(2001)Wohlgemuth, Yufa, Hoffman, and
  Thomas]{Wohlgemuth2001}
Wohlgemuth,~M.; Yufa,~N.; Hoffman,~J.; Thomas,~E.~L. Triply Periodic
  Bicontinuous Cubic Microdomain Morphologies by Symmetries.
  \emph{Macromolecules} \textbf{2001}, \emph{34}, 6083--6089\relax
\mciteBstWouldAddEndPuncttrue
\mciteSetBstMidEndSepPunct{\mcitedefaultmidpunct}
{\mcitedefaultendpunct}{\mcitedefaultseppunct}\relax
\EndOfBibitem
\bibitem[Matsen(2002)]{Matsen2002}
Matsen,~M.~W. {The standard Gaussian model for block copolymer melts}.
  \emph{Journal of Physics: Condensed Matter} \textbf{2002}, \emph{14},
  R21--R47\relax
\mciteBstWouldAddEndPuncttrue
\mciteSetBstMidEndSepPunct{\mcitedefaultmidpunct}
{\mcitedefaultendpunct}{\mcitedefaultseppunct}\relax
\EndOfBibitem
\bibitem[Arora \latin{et~al.}(2016)Arora, Qin, Morse, Delaney, Fredrickson,
  Bates, and Dorfman]{Arora2016}
Arora,~A.; Qin,~J.; Morse,~D.~C.; Delaney,~K.~T.; Fredrickson,~G.~H.;
  Bates,~F.~S.; Dorfman,~K.~D. {Broadly Accessible Self-Consistent Field Theory
  for Block Polymer Materials Discovery}. \emph{Macromolecules} \textbf{2016},
  \emph{49}, 4675--4690\relax
\mciteBstWouldAddEndPuncttrue
\mciteSetBstMidEndSepPunct{\mcitedefaultmidpunct}
{\mcitedefaultendpunct}{\mcitedefaultseppunct}\relax
\EndOfBibitem
\bibitem[Prasad \latin{et~al.}(2017)Prasad, Seo, Hall, and Grason]{Prasad2017}
Prasad,~I.; Seo,~Y.; Hall,~L.~M.; Grason,~G.~M. Intradomain Textures in Block
  Copolymers: Multizone Alignment and Biaxiality. \emph{Phys. Rev. Lett.}
  \textbf{2017}, \emph{118}, 247801\relax
\mciteBstWouldAddEndPuncttrue
\mciteSetBstMidEndSepPunct{\mcitedefaultmidpunct}
{\mcitedefaultendpunct}{\mcitedefaultseppunct}\relax
\EndOfBibitem
\bibitem[Matsen(2001)]{Matsen2001}
Matsen,~M.~W. {Testing strong-segregation theory against self-consistent-field
  theory for block copolymer melts}. \emph{The Journal of Chemical Physics}
  \textbf{2001}, \emph{114}, 10528--10530\relax
\mciteBstWouldAddEndPuncttrue
\mciteSetBstMidEndSepPunct{\mcitedefaultmidpunct}
{\mcitedefaultendpunct}{\mcitedefaultseppunct}\relax
\EndOfBibitem
\bibitem[Matsen(2010)]{Matsen2010}
Matsen,~M.~W. Strong-segregation limit of the self-consistent field theory for
  diblock copolymer melts. \emph{The European physical journal. E, Soft matter}
  \textbf{2010}, \emph{33}, 297--306, Place: France\relax
\mciteBstWouldAddEndPuncttrue
\mciteSetBstMidEndSepPunct{\mcitedefaultmidpunct}
{\mcitedefaultendpunct}{\mcitedefaultseppunct}\relax
\EndOfBibitem
\bibitem[Feng \latin{et~al.}(2019)Feng, Burke, Zhou, Guo, Yang, Reddy, Prasad,
  Ho, Avgeropoulos, Grason, and Thomas]{Feng2019}
Feng,~X.; Burke,~C.~J.; Zhou,~M.; Guo,~H.; Yang,~K.; Reddy,~A.; Prasad,~I.;
  Ho,~R.-M.; Avgeropoulos,~A.; Grason,~G.~M.; Thomas,~E.~L. Seeing mesoatomic
  distortions in soft-matter crystals of a double-gyroid block copolymer.
  \emph{Nature} \textbf{2019}, \emph{575}, 175–179\relax
\mciteBstWouldAddEndPuncttrue
\mciteSetBstMidEndSepPunct{\mcitedefaultmidpunct}
{\mcitedefaultendpunct}{\mcitedefaultseppunct}\relax
\EndOfBibitem
\bibitem[Hamm and Kozlov(1998)Hamm, and Kozlov]{Hamm1998}
Hamm,~M.; Kozlov,~M.~M. Tilt model of inverted amphiphilic mesophases.
  \emph{The European Physical Journal B - Condensed Matter and Complex Systems}
  \textbf{1998}, \emph{6}, 519--528\relax
\mciteBstWouldAddEndPuncttrue
\mciteSetBstMidEndSepPunct{\mcitedefaultmidpunct}
{\mcitedefaultendpunct}{\mcitedefaultseppunct}\relax
\EndOfBibitem
\bibitem[Chen and Jin(2017)Chen, and Jin]{Chen2017}
Chen,~H.; Jin,~C. Competition brings out the best: modelling the frustration
  between curvature energy and chain stretching energy of lyotropic liquid
  crystals in bicontinuous cubic phases. \emph{Interface Focus} \textbf{2017},
  \emph{7}\relax
\mciteBstWouldAddEndPuncttrue
\mciteSetBstMidEndSepPunct{\mcitedefaultmidpunct}
{\mcitedefaultendpunct}{\mcitedefaultseppunct}\relax
\EndOfBibitem
\bibitem[Li \latin{et~al.}(2014)Li, Hur, Sai, Higuchi, Takahara, Jinnai,
  Gruner, and Wiesner]{Li2014}
Li,~Z.; Hur,~K.; Sai,~H.; Higuchi,~T.; Takahara,~A.; Jinnai,~H.; Gruner,~S.~M.;
  Wiesner,~U. Linking experiment and theory for three-dimensional networked
  binary metal nanoparticle--triblock terpolymer superstructures. \emph{Nature
  Communications} \textbf{2014}, \emph{5}, 3247\relax
\mciteBstWouldAddEndPuncttrue
\mciteSetBstMidEndSepPunct{\mcitedefaultmidpunct}
{\mcitedefaultendpunct}{\mcitedefaultseppunct}\relax
\EndOfBibitem
\bibitem[Winey \latin{et~al.}(1991)Winey, Thomas, and Fetters]{Winey1991Macro}
Winey,~K.~I.; Thomas,~E.~L.; Fetters,~L.~J. Swelling of lamellar diblock
  copolymer by homopolymer: influences of homopolymer concentration and
  molecular weight. \emph{Macromolecules} \textbf{1991}, \emph{24},
  6182--6188\relax
\mciteBstWouldAddEndPuncttrue
\mciteSetBstMidEndSepPunct{\mcitedefaultmidpunct}
{\mcitedefaultendpunct}{\mcitedefaultseppunct}\relax
\EndOfBibitem
\bibitem[Winey \latin{et~al.}(1992)Winey, Thomas, and Fetters]{Winey1992}
Winey,~K.~I.; Thomas,~E.~L.; Fetters,~L.~J. The ordered bicontinuous
  double-diamond morphology in diblock copolymer/homopolymer blends.
  \emph{Macromolecules} \textbf{1992}, \emph{25}, 422--428\relax
\mciteBstWouldAddEndPuncttrue
\mciteSetBstMidEndSepPunct{\mcitedefaultmidpunct}
{\mcitedefaultendpunct}{\mcitedefaultseppunct}\relax
\EndOfBibitem
\bibitem[Winey \latin{et~al.}(1992)Winey, Thomas, and Fetters]{Winey1992b}
Winey,~K.~I.; Thomas,~E.~L.; Fetters,~L.~J. Isothermal morphology diagrams for
  binary blends of diblock copolymer and homopolymer. \emph{Macromolecules}
  \textbf{1992}, \emph{25}, 2645--2650\relax
\mciteBstWouldAddEndPuncttrue
\mciteSetBstMidEndSepPunct{\mcitedefaultmidpunct}
{\mcitedefaultendpunct}{\mcitedefaultseppunct}\relax
\EndOfBibitem
\bibitem[Takagi \latin{et~al.}(2017)Takagi, Takasaki, and Yamamoto]{Takagi2017}
Takagi,~H.; Takasaki,~T.; Yamamoto,~K. Phase Diagram for Block Copolymer and
  Homopolymer Blends: The Phase Boundary of the Ordered Bicontinuous Double
  Diamond Network Structure. \emph{Journal of Nanoscience and Nanotechnology}
  \textbf{2017}, \emph{17}, 9009--9014\relax
\mciteBstWouldAddEndPuncttrue
\mciteSetBstMidEndSepPunct{\mcitedefaultmidpunct}
{\mcitedefaultendpunct}{\mcitedefaultseppunct}\relax
\EndOfBibitem
\bibitem[Mueller \latin{et~al.}(2020)Mueller, Lindsay, Jayaraman, Lodge,
  Mahanthappa, and Bates]{Mueller2020}
Mueller,~A.~J.; Lindsay,~A.~P.; Jayaraman,~A.; Lodge,~T.~P.;
  Mahanthappa,~M.~K.; Bates,~F.~S. {Emergence of a C15 Laves Phase in Diblock
  Polymer/Homopolymer Blends}. \emph{ACS Macro Letters} \textbf{2020},
  \emph{9}, 576--582\relax
\mciteBstWouldAddEndPuncttrue
\mciteSetBstMidEndSepPunct{\mcitedefaultmidpunct}
{\mcitedefaultendpunct}{\mcitedefaultseppunct}\relax
\EndOfBibitem
\bibitem[Matsen(1995)]{Matsen1995PRL}
Matsen,~M.~W. Stabilizing New Morphologies by Blending Homopolymer with Block
  Copolymer. \emph{Phys. Rev. Lett.} \textbf{1995}, \emph{74}, 4225--4228\relax
\mciteBstWouldAddEndPuncttrue
\mciteSetBstMidEndSepPunct{\mcitedefaultmidpunct}
{\mcitedefaultendpunct}{\mcitedefaultseppunct}\relax
\EndOfBibitem
\bibitem[Grason and Thomas(2023)Grason, and Thomas]{GrasonThomas2023}
Grason,~G.~M.; Thomas,~E.~L. How does your gyroid grow? A mesoatomic
  perspective on supramolecular, soft matter network crystals. \emph{Phys. Rev.
  Mater.} \textbf{2023}, \emph{7}, 045603\relax
\mciteBstWouldAddEndPuncttrue
\mciteSetBstMidEndSepPunct{\mcitedefaultmidpunct}
{\mcitedefaultendpunct}{\mcitedefaultseppunct}\relax
\EndOfBibitem
\bibitem[Winey \latin{et~al.}(1991)Winey, Thomas, and Fetters]{Winey1991JCP}
Winey,~K.~I.; Thomas,~E.~L.; Fetters,~L.~J. {Ordered morphologies in binary
  blends of diblock copolymer and homopolymer and characterization of their
  intermaterial dividing surfaces}. \emph{The Journal of Chemical Physics}
  \textbf{1991}, \emph{95}, 9367--9375\relax
\mciteBstWouldAddEndPuncttrue
\mciteSetBstMidEndSepPunct{\mcitedefaultmidpunct}
{\mcitedefaultendpunct}{\mcitedefaultseppunct}\relax
\EndOfBibitem
\bibitem[Mayes \latin{et~al.}(1992)Mayes, Russell, Satija, and
  Majkrzak]{Mayes1992}
Mayes,~A.~M.; Russell,~T.~P.; Satija,~S.~K.; Majkrzak,~C.~F. Homopolymer
  distributions in ordered block copolymers. \emph{Macromolecules}
  \textbf{1992}, \emph{25}, 6523--6531\relax
\mciteBstWouldAddEndPuncttrue
\mciteSetBstMidEndSepPunct{\mcitedefaultmidpunct}
{\mcitedefaultendpunct}{\mcitedefaultseppunct}\relax
\EndOfBibitem
\end{mcitethebibliography}


\begin{thebibliography}{0}%
\makeatletter
\providecommand \@ifxundefined [1]{%
 \@ifx{#1\undefined}
}%
\providecommand \@ifnum [1]{%
 \ifnum #1\expandafter \@firstoftwo
 \else \expandafter \@secondoftwo
 \fi
}%
\providecommand \@ifx [1]{%
 \ifx #1\expandafter \@firstoftwo
 \else \expandafter \@secondoftwo
 \fi
}%
\providecommand \natexlab [1]{#1}%
\providecommand \enquote  [1]{``#1''}%
\providecommand \bibnamefont  [1]{#1}%
\providecommand \bibfnamefont [1]{#1}%
\providecommand \citenamefont [1]{#1}%
\providecommand \href@noop [0]{\@secondoftwo}%
\providecommand \href [0]{\begingroup \@sanitize@url \@href}%
\providecommand \@href[1]{\@@startlink{#1}\@@href}%
\providecommand \@@href[1]{\endgroup#1\@@endlink}%
\providecommand \@sanitize@url [0]{\catcode `\\12\catcode `\$12\catcode
  `\&12\catcode `\#12\catcode `\^12\catcode `\_12\catcode `\%12\relax}%
\providecommand \@@startlink[1]{}%
\providecommand \@@endlink[0]{}%
\providecommand \url  [0]{\begingroup\@sanitize@url \@url }%
\providecommand \@url [1]{\endgroup\@href {#1}{\urlprefix }}%
\providecommand \urlprefix  [0]{URL }%
\providecommand \Eprint [0]{\href }%
\providecommand \doibase [0]{http://dx.doi.org/}%
\providecommand \selectlanguage [0]{\@gobble}%
\providecommand \bibinfo  [0]{\@secondoftwo}%
\providecommand \bibfield  [0]{\@secondoftwo}%
\providecommand \translation [1]{[#1]}%
\providecommand \BibitemOpen [0]{}%
\providecommand \bibitemStop [0]{}%
\providecommand \bibitemNoStop [0]{.\EOS\space}%
\providecommand \EOS [0]{\spacefactor3000\relax}%
\providecommand \BibitemShut  [1]{\csname bibitem#1\endcsname}%
\let\auto@bib@innerbib\@empty
\end{thebibliography}%

\end{document}


\title{Supporting Information for ``Medial packing, frustration and competing network phases in strongly-segregated block copolymers''}

\author{Michael S. Dimitriyev\textsuperscript{1}}
\author{Abhiram Reddy\textsuperscript{2}}
\author{Gregory M. Grason\textsuperscript{1}}\email{grason@umass.edu}
\affiliation{\textsuperscript{1}Department of Polymer Science and Engineering, University of Massachusetts, Amherst, MA 01003}
\affiliation{\textsuperscript{2}Center for Computation \& Theory of Soft Materials, Northwestern University, Evanston, IL 60208}

\date{\today}

\maketitle 

\section*{Methods}

\subsection{Symmetry-adapted basis functions}
For DG, the four mode expansion is $\Psi_{\rm DG}(\mathbf{r}) = c_1 \psi_{110} + c_2 \psi_{211} + c_3 \psi_{220} + c_4\psi_{321}$, with each mode given by
\begin{equation}
    \begin{split}
        \psi_{110} &= \sin\left(\frac{2\pi x}{D}\right)\cos\left(\frac{2\pi y}{D}\right) + \sin\left(\frac{2\pi y}{D}\right)\cos\left(\frac{2\pi z}{D}\right) + \sin\left(\frac{2\pi z}{D}\right)\cos\left(\frac{2\pi x}{D}\right) \\
        \psi_{211} &= \sin\left(\frac{4\pi x}{D}\right)\cos\left(\frac{2\pi y}{D}\right)\sin\left(\frac{2\pi z}{D}\right) + \sin\left(\frac{4\pi y}{D}\right)\cos\left(\frac{2\pi z}{D}\right)\sin\left(\frac{2\pi x}{D}\right) \\
        &\mkern+16mu+ \sin\left(\frac{4\pi z}{D}\right)\cos\left(\frac{2\pi x}{D}\right)\sin\left(\frac{2\pi y}{D}\right) \\
        \psi_{220} &= \cos\left(\frac{4\pi x}{D}\right)\cos\left(\frac{4\pi y}{D}\right) + \cos\left(\frac{4\pi y}{D}\right)\cos\left(\frac{4\pi z}{D}\right) + \cos\left(\frac{4\pi z}{D}\right)\cos\left(\frac{4\pi x}{D}\right)\\
        \psi_{321} &= \cos\left(\frac{6\pi x}{D}\right)\sin\left(\frac{2\pi y}{D}\right)\sin\left(\frac{4\pi z}{D}\right) - \sin\left(\frac{6\pi x}{D}\right)\sin\left(\frac{4\pi y}{D}\right)\cos\left(\frac{2\pi z}{D}\right) \\
        &\mkern+16mu+ \cos\left(\frac{6\pi y}{D}\right)\sin\left(\frac{2\pi z}{D}\right)\sin\left(\frac{4\pi x}{D}\right) - \sin\left(\frac{6\pi y}{D}\right)\sin\left(\frac{4\pi z}{D}\right)\cos\left(\frac{2\pi x}{D}\right) \\
        &\mkern+16mu+ \cos\left(\frac{6\pi z}{D}\right)\sin\left(\frac{2\pi x}{D}\right)\sin\left(\frac{4\pi y}{D}\right) - \sin\left(\frac{6\pi z}{D}\right)\sin\left(\frac{4\pi x}{D}\right)\cos\left(\frac{2\pi y}{D}\right)
    \end{split}
\end{equation}
where $\mathbf{r} = (x,y,z)$.
For DD, the expansion is $\Psi_{\rm DD}(\mathbf{r}) = c_1 \psi_{111} + c_2 \psi_{220} + c_3 \psi_{311} + c_4 \psi_{222}$, with each mode given by
\begin{equation}
    \begin{split}
        \psi_{111} &= \cos\left(\frac{2\pi x}{D}\right)\cos\left(\frac{2\pi y}{D}\right)\cos\left(\frac{2\pi z}{D}\right) + \sin\left(\frac{2\pi x}{D}\right)\sin\left(\frac{2\pi y}{D}\right)\cos\left(\frac{2\pi z}{D}\right) \\
        &\mkern+16mu+ \sin\left(\frac{2\pi y}{D}\right)\sin\left(\frac{2\pi z}{D}\right)\cos\left(\frac{2\pi x}{D}\right) + \sin\left(\frac{2\pi z}{D}\right)\sin\left(\frac{2\pi x}{D}\right)\cos\left(\frac{2\pi y}{D}\right) \\
        \psi_{220} &= \sin\left(\frac{4\pi x}{D}\right)\sin\left(\frac{4\pi y}{D}\right) + \sin\left(\frac{4\pi y}{D}\right)\sin\left(\frac{4\pi z}{D}\right) + \sin\left(\frac{4\pi z}{D}\right)\sin\left(\frac{4\pi x}{D}\right) \\
        \psi_{311} &= \cos\left(\frac{6\pi x}{D}\right)\cos\left(\frac{2\pi (y-z)}{D}\right) - \sin\left(\frac{6\pi x}{D}\right)\sin\left(\frac{2\pi (y+z)}{D}\right) \\
        &\mkern+16mu+ \cos\left(\frac{6\pi y}{D}\right)\cos\left(\frac{2\pi (z-x)}{D}\right) - \sin\left(\frac{6\pi y}{D}\right)\sin\left(\frac{2\pi (z+x)}{D}\right) \\
        &\mkern+16mu+ \cos\left(\frac{6\pi z}{D}\right)\cos\left(\frac{2\pi (x-y)}{D}\right) - \sin\left(\frac{6\pi z}{D}\right)\sin\left(\frac{2\pi (x+y)}{D}\right) \\
        \psi_{222} &= \cos\left(\frac{4\pi x}{D}\right)\cos\left(\frac{4\pi y}{D}\right)\cos\left(\frac{4\pi z}{D}\right)
    \end{split}
\end{equation}
Finally, for DP, the expansion is $\Psi_{\rm DP} = c_1 \psi_{100} + c_2 \psi_{110} + c_3 \psi_{111} + c_4 \psi_{200}$, with each mode given by
\begin{equation}
    \begin{split}
        \psi_{100} &= \cos\left(\frac{2\pi x}{D}\right) + \cos\left(\frac{2\pi y}{D}\right) + \cos\left(\frac{2\pi z}{D}\right) \\
        \psi_{110} &= \cos\left(\frac{2\pi x}{D}\right)\cos\left(\frac{2\pi y}{D}\right) + \cos\left(\frac{2\pi y}{D}\right)\cos\left(\frac{2\pi z}{D}\right) + \cos\left(\frac{2\pi z}{D}\right)\cos\left(\frac{2\pi x}{D}\right) \\
        \psi_{111} &= \cos\left(\frac{2\pi x}{D}\right)\cos\left(\frac{2\pi y}{D}\right)\cos\left(\frac{2\pi z}{D}\right) \\
        \psi_{200} &= \cos\left(\frac{4\pi x}{D}\right) + \cos\left(\frac{4\pi y}{D}\right) + \cos\left(\frac{4\pi z}{D}\right)
    \end{split}
\end{equation}

\newpage

\section*{Supplementary Figures}

\begin{figure}[ht!]
\centering
\includegraphics[width=3in]{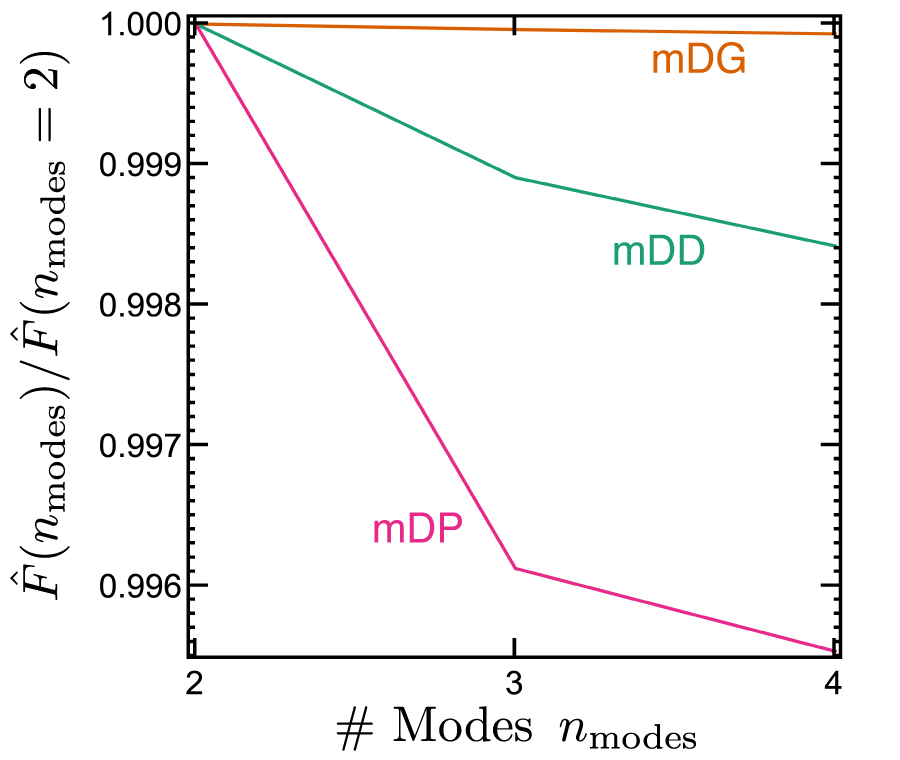}
\caption{\label{fig:fe_mode_comp} Average free energy per chain as a function of number of modes used for calculation, normalized by the free energy per chain calculated using 2 modes. These averages were taken over a collection of mSST variational calculations ($\epsilon = 0.5,\, 1.0$, 2.0 and $f = 0.10,\, 0.11,\, \dots,\, 0.50$), where the free energy of a given number of modes is the minimum reported by a Nelder-Mead minimization algorithm.}
\end{figure}


\begin{figure}[ht!]
\centering
\includegraphics[width=\textwidth]{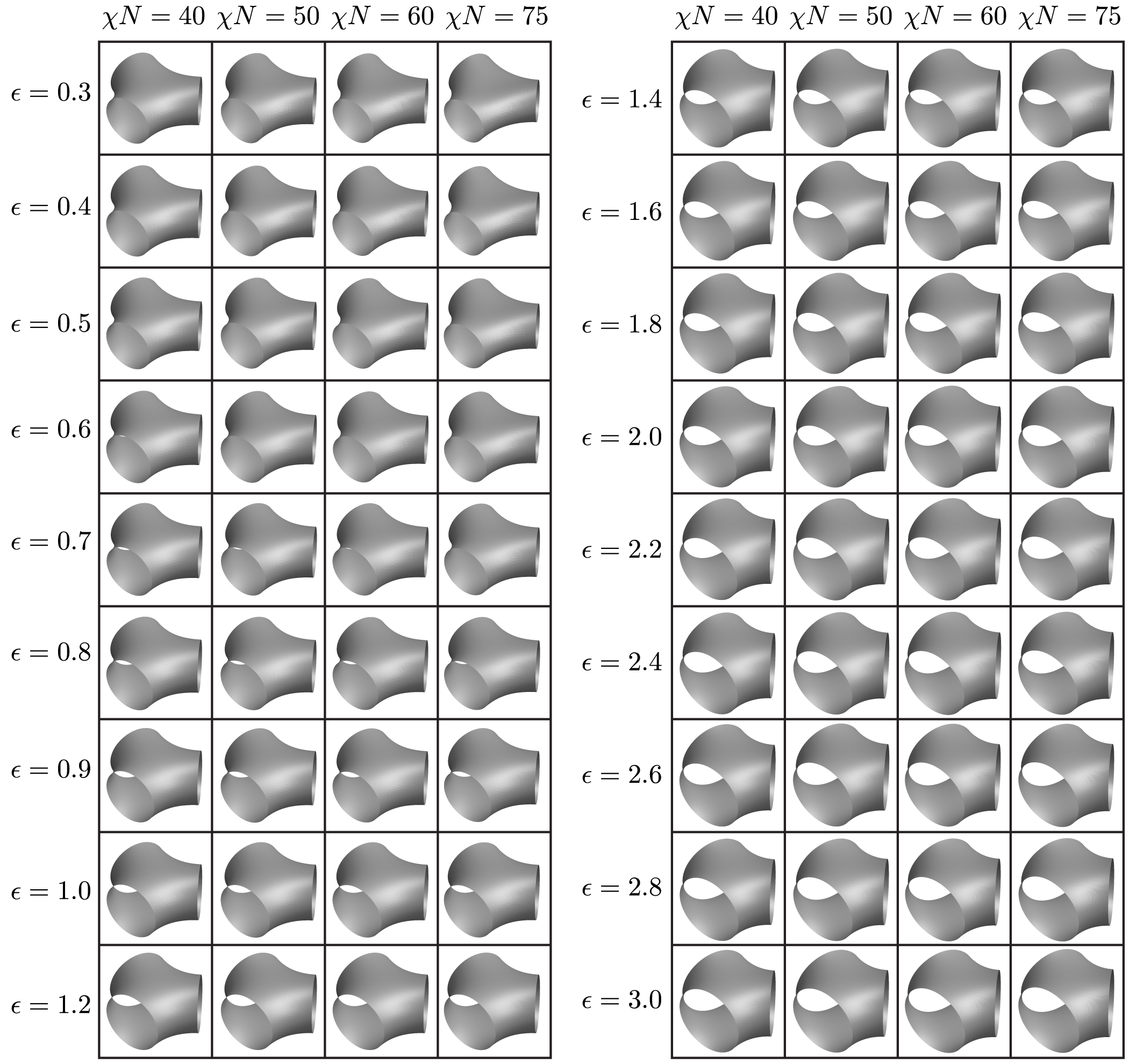}
\caption{\label{fig:dg_imds} Rendered nodal IMDSs for DG structures computed at finite segregation along the Lam/Hex boundary.}
\end{figure}

\begin{figure}[ht!]
\centering
\includegraphics[width=\textwidth]{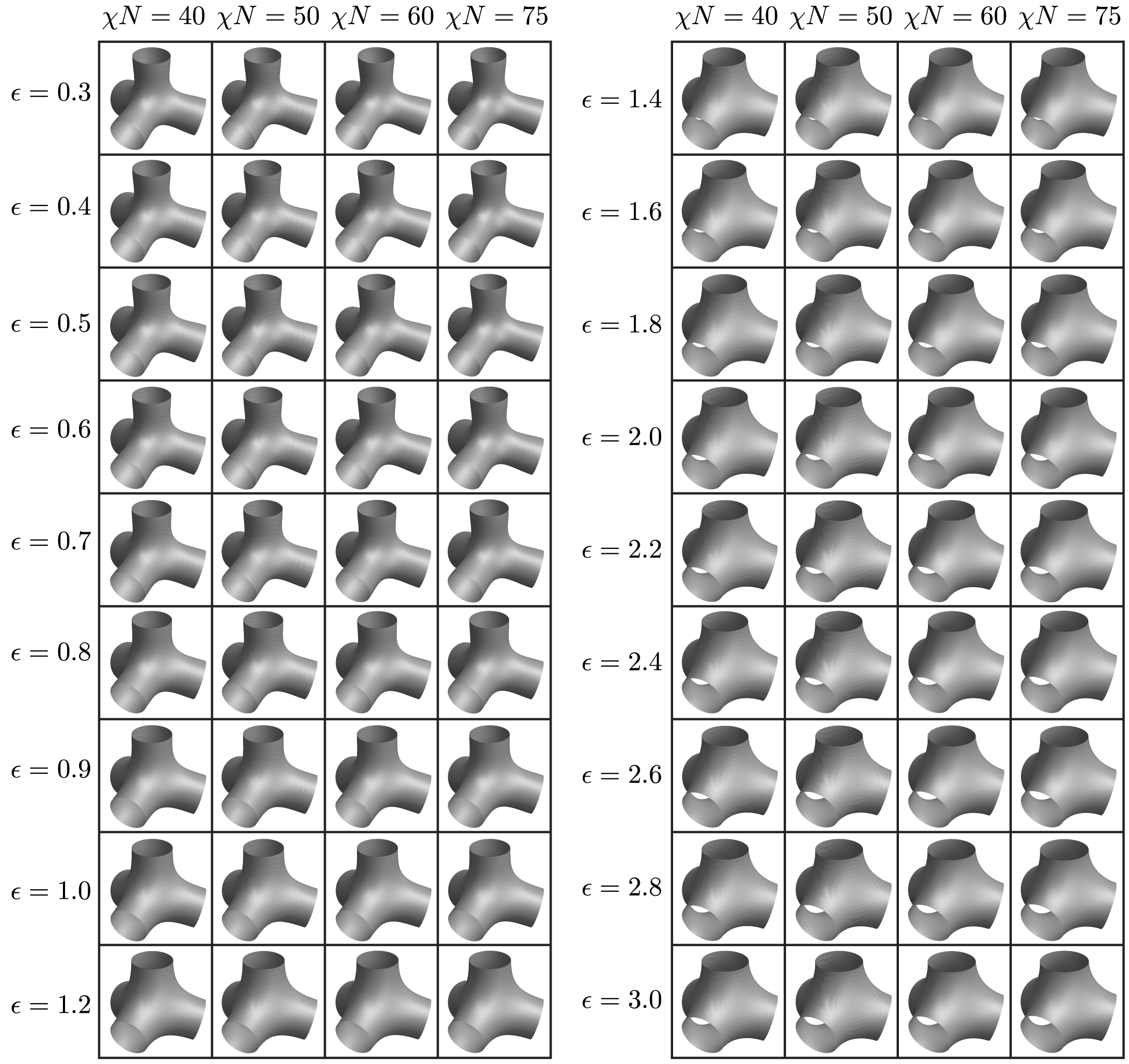}
\caption{\label{fig:dd_imds} Rendered nodal IMDSs for DD structures computed at finite segregation along the Lam/Hex boundary.}
\end{figure}

\begin{figure}[ht!]
\centering
\includegraphics[width=\textwidth]{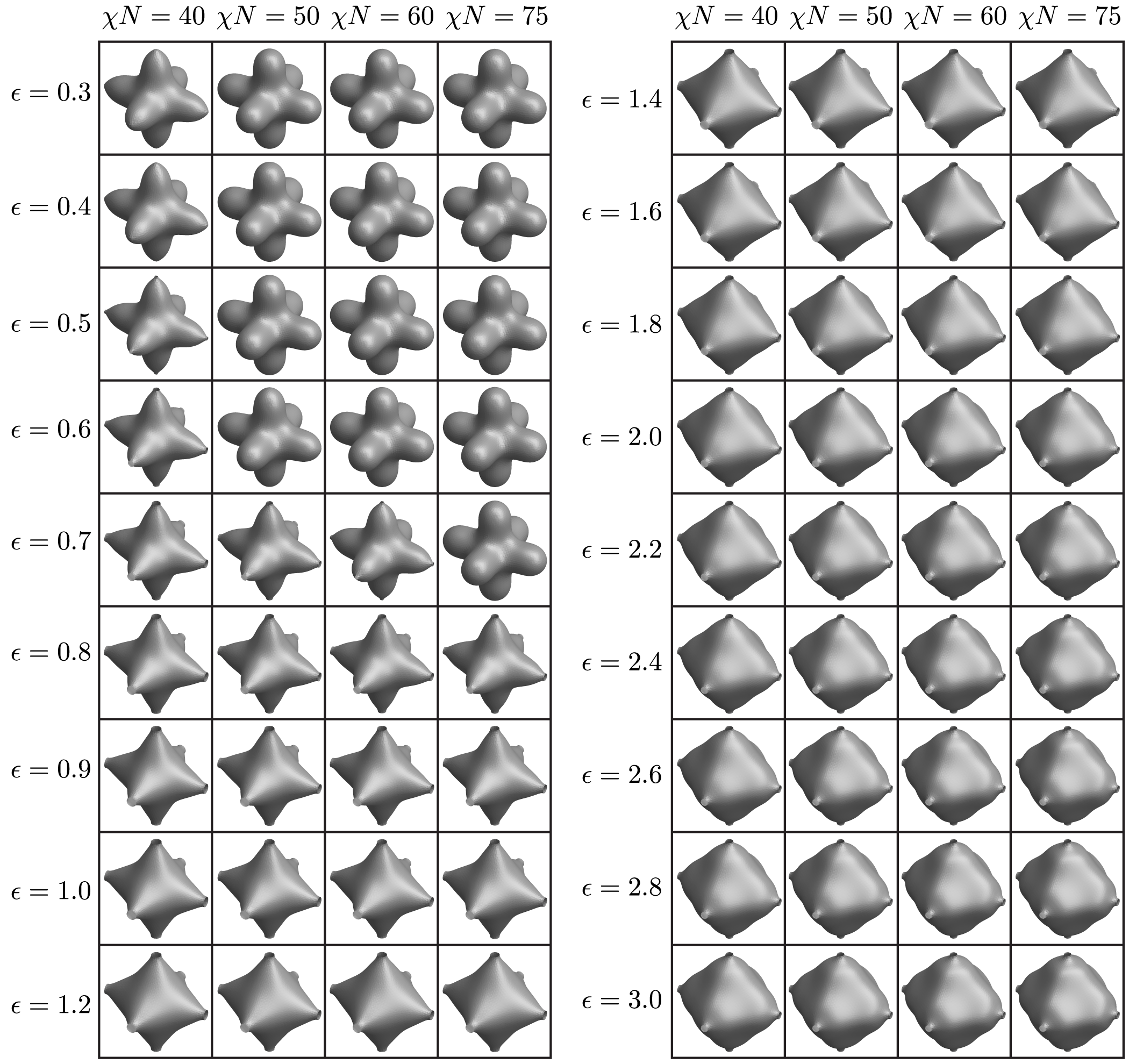}
\caption{\label{fig:dp_imds} Rendered nodal IMDSs for DP structures computed at finite segregation along the Lam/Hex boundary. Note that for $\epsilon = 2.8$ and $\epsilon = 3.0$, we show IMDSs computed at $\chi N = 72.5$ due to difficulties with numerical convergence at $\chi N = 75$.}
\end{figure}


\begin{figure}[ht!]
\centering
\includegraphics[width=4in]{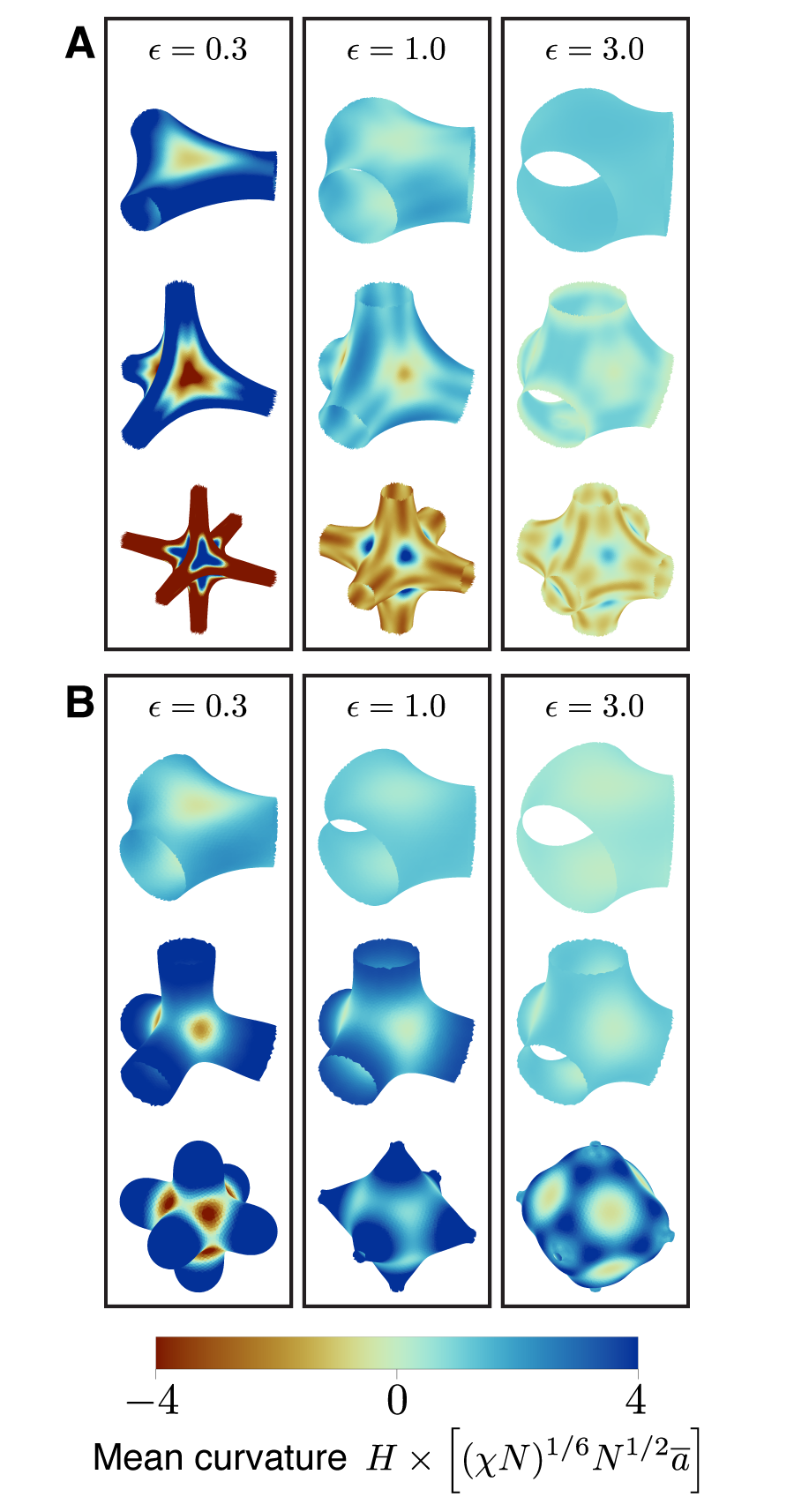}
\caption{\label{fig:H_maps} IMDS mean curvature distributions for $\epsilon = 0.3,\, 1.0$, and 3.0 along the Lam-Hex boundary. (A) shows mSST results and (B) shows SCFT results at $\chi N = 75$ (with the exception of DP at $\epsilon = 3.0$, for which $\chi N = 72.5$ was used).}
\end{figure}

\begin{figure}[ht!]
\centering
\includegraphics[width=4in]{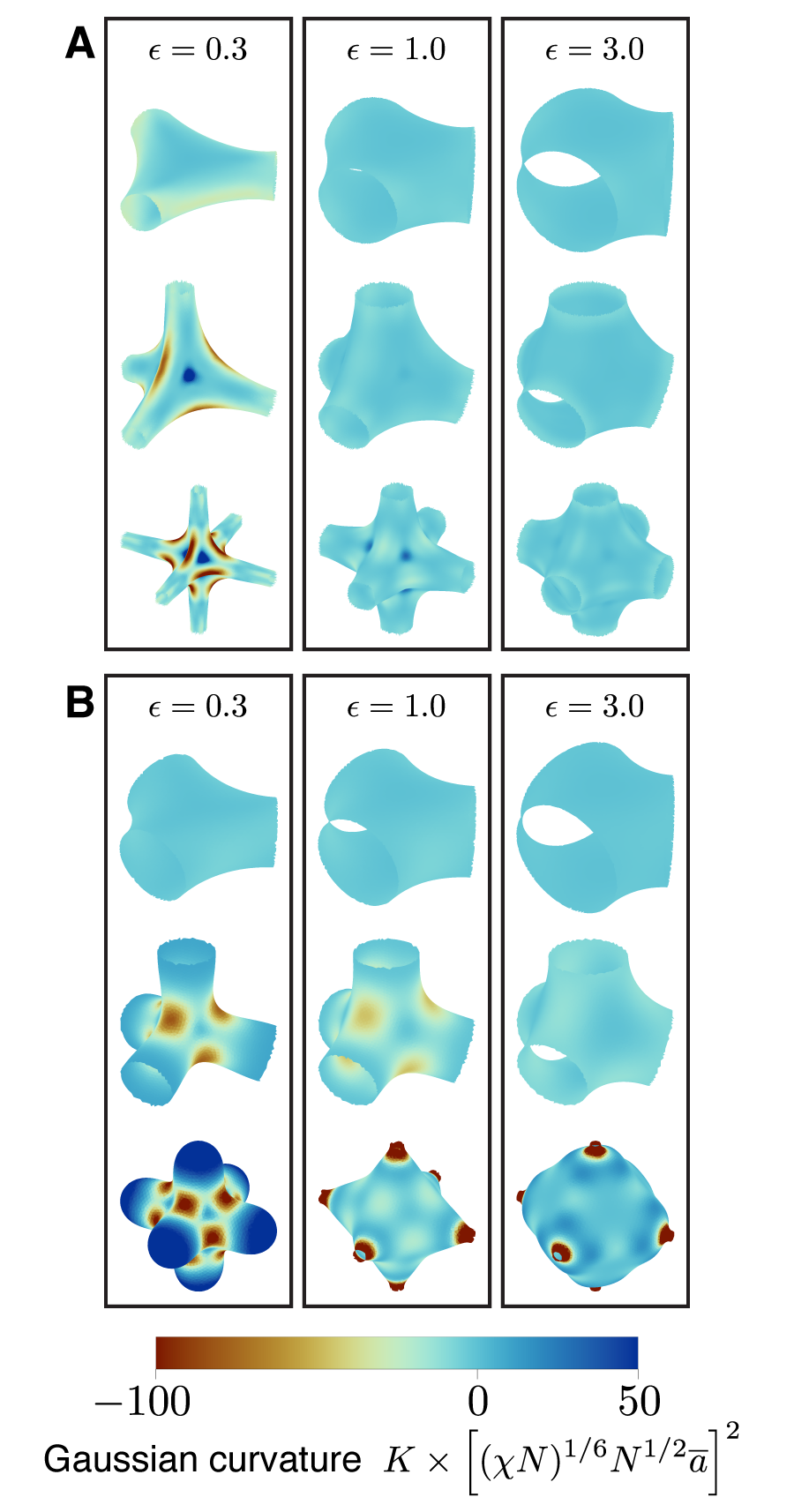}
\caption{\label{fig:H_maps} IMDS Gaussian curvature distributions for $\epsilon = 0.3,\, 1.0$, and 3.0 along the Lam-Hex boundary. (A) shows mSST results and (B) shows SCFT results at $\chi N = 75$ (with the exception of DP at $\epsilon = 3.0$, for which $\chi N = 72.5$ was used).}
\end{figure}

\begin{figure}[ht!]
\centering
\includegraphics[width=3in]{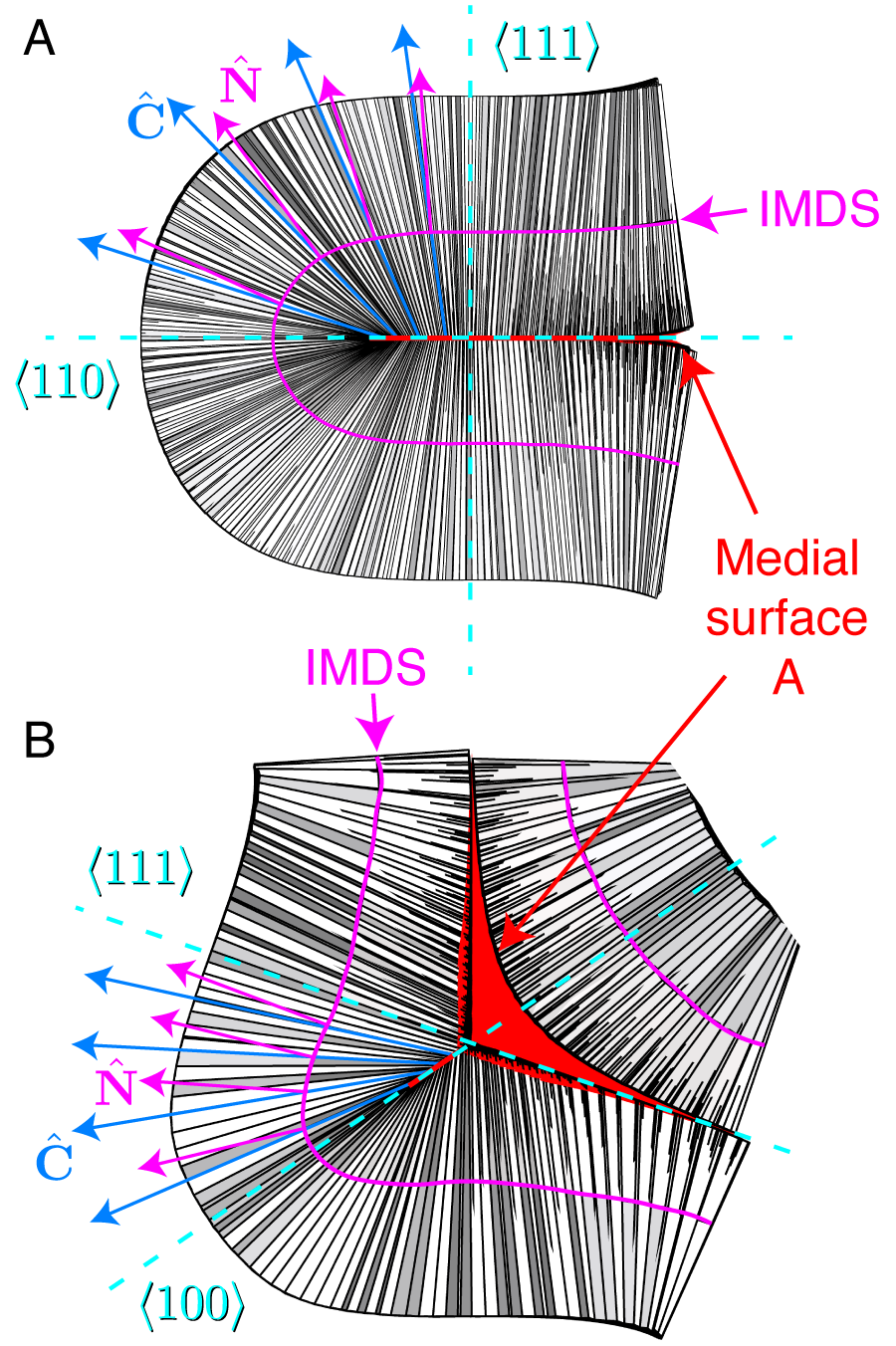}
\caption{\label{fig:medial_tilt} Cutaways of the (A) DG and (B) DD mesoatoms showing wedges whose long edges are aligned with the local surface normals $\hat{\mathbf{C}}$ (blue and orange lines) of a given generating surface, which terminate on a medial surface (red). The volume-balanced IMDSs with normals $\hat{\mathbf{N}}$ (magenta) closely conform to the shapes of the A-block medial surfaces.}
\end{figure}


\begin{figure}[ht!]
\centering
\includegraphics[width=5in]{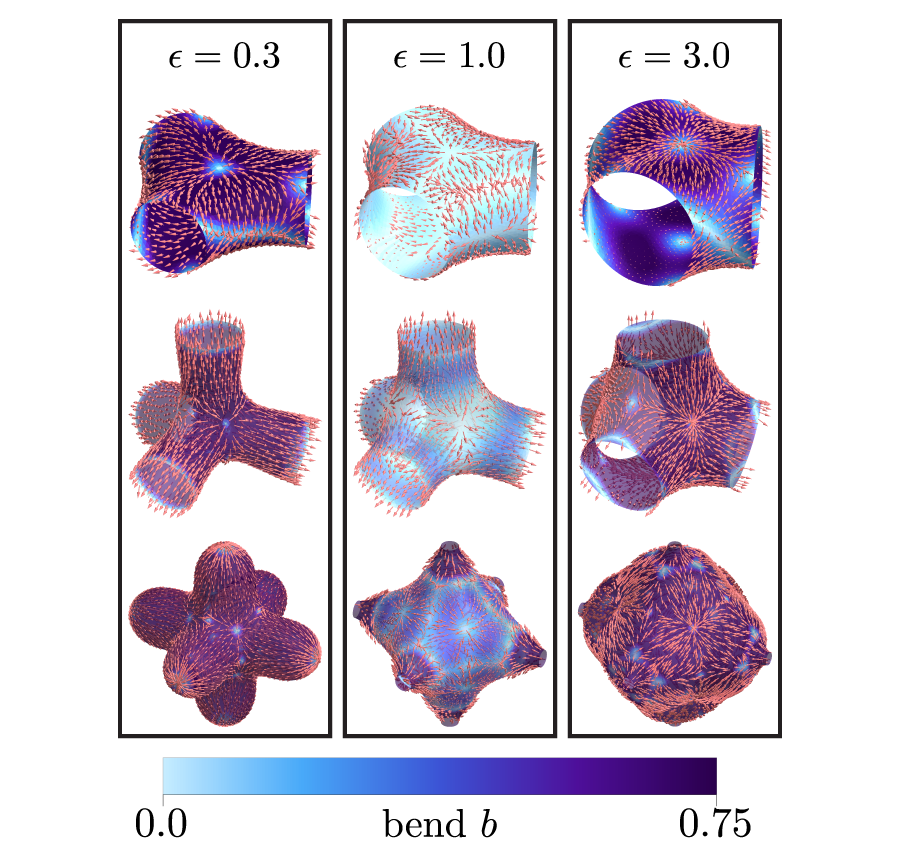}
\caption{\label{fig:add_bending} Chain bending distributions calculated from SCFT at $\chi N = 75$ for $\epsilon = 0.3,\, 1.0$, and 3.0 along the Lam-Hex boundary. Note that $\chi N = 72.5$ is used for DP at $\epsilon = 3.0$.}
\end{figure}


\begin{figure}[ht!]
\centering
\includegraphics[width=\textwidth]{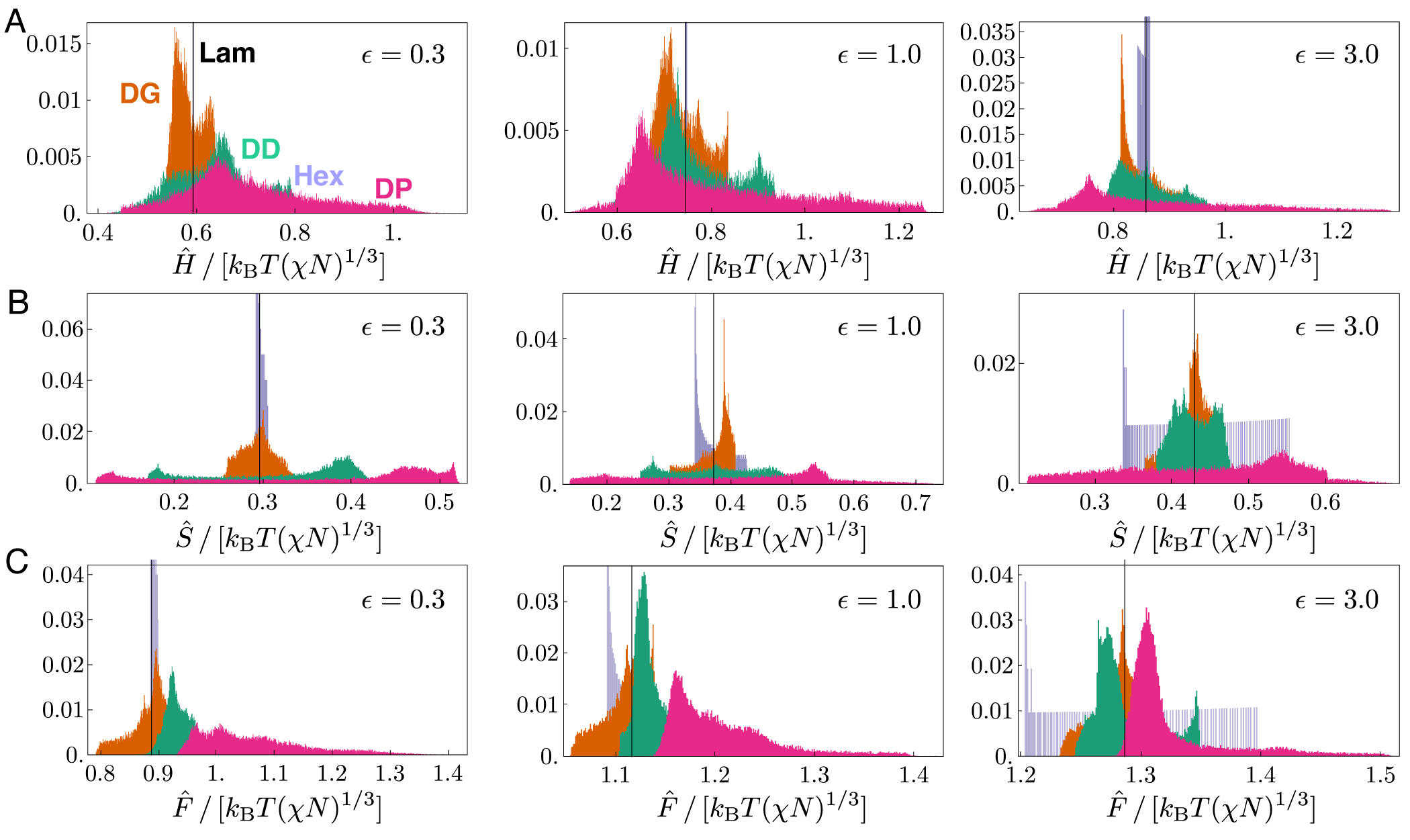}
\caption{\label{fig:ext_histograms} Histograms showing the distribution of (A) enthalpy per chain $\hat{H}$, (B) entropic cost of stretching per chain $\hat{S}$, and (C) total free energy per chain $\hat{F}$, taken over the collection of wedges used in mSST calculations for the network phases for $\epsilon = 0.3,\, 1.0$, and 3.0. The distribution representing the singular packing environment for the lamellar phase is shown as a black line. Hex phase packing environments were obtained using the kinked path ansatz. The histograms are all normalized so that the sum over bins equals 1 for each data set independently and all bin widths represent an energy interval of $10^{-3} k_{\rm B}T(\chi N)^{1/3}$.}
\end{figure}


\begin{figure}[ht!]
\centering
\includegraphics[width=4in]{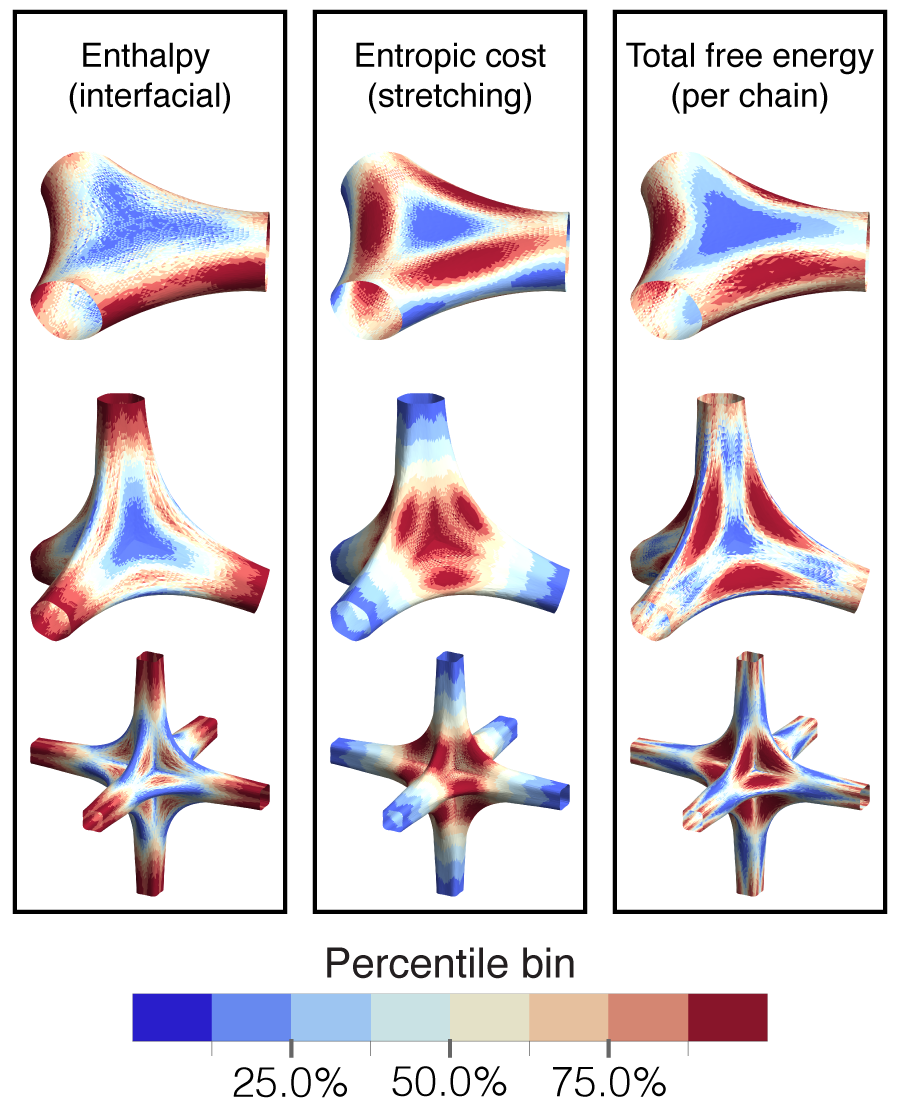}
\caption{\label{fig:fe_maps_03} Local free energy per chain contributions, mapped onto the IMDSs of the network phases for $\epsilon = 0.3$, $f \approx 0.07$. (A) shows the local interfacial enthalpy, (B) shows the local stretching free energy, (C) shows the total free energy per chain. Regions are binned according to the percentile rank of the free energy per chain distribution, such that the dark red corresponds to the regions containing the 12.5\% chains with the highest free energy and the dark blue corresponds to the 12.5\% lowest free energy chains.}
\end{figure}

\begin{figure}[ht!]
\centering
\includegraphics[width=4in]{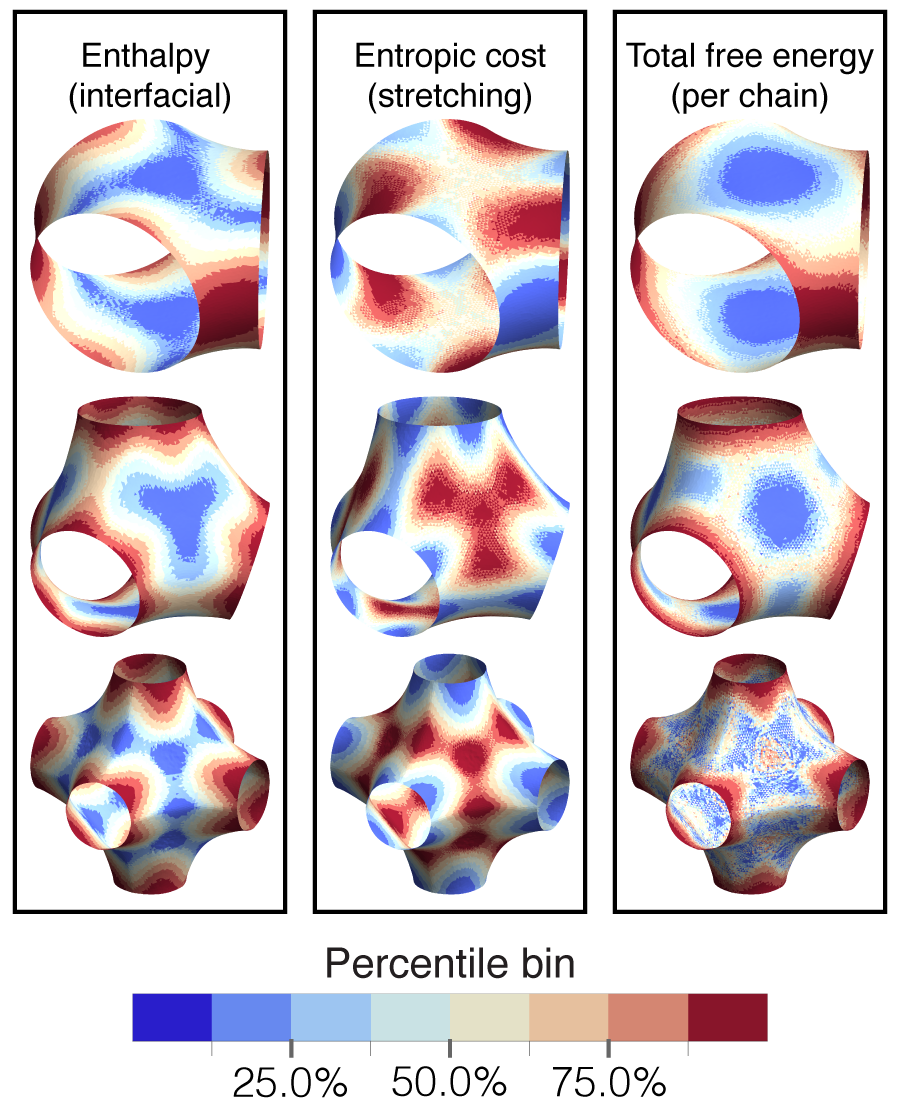}
\caption{\label{fig:fe_maps_30} Local free energy per chain contributions, mapped onto the IMDSs of the network phases for $\epsilon = 3.0$, $f \approx 0.55$. (A) shows the local interfacial enthalpy, (B) shows the local stretching free energy, (C) shows the total free energy per chain. Regions are binned according to the percentile rank of the free energy per chain distribution, such that the dark red corresponds to the regions containing the 12.5\% chains with the highest free energy and the dark blue corresponds to the 12.5\% lowest free energy chains.}
\end{figure}